\documentclass[final,3p,times]{elsarticle}
\usepackage{graphicx}
\usepackage{amssymb,amsmath}
\biboptions{sort,compress}
\journal{Annals of Physics}

\begin{document}

\begin{frontmatter}

\title{Implicit and explicit renormalization: two complementary views
  of effective interactions}

\author[1]{E. Ruiz Arriola}
\author[2]{S. Szpigel\corref{cor1}}
\ead{szpigel@mackenzie.br}

\author[3]{V. S. Tim\'oteo}

\cortext[cor1]{Corresponding author.}

\address[1]{Departamento de F\'isica At\'omica, Molecular y Nuclear
  and Instituto Carlos I de Fisica Te\'orica y Computacional \\ Universidad de Granada, E-18071 Granada, Spain}

\address[2]{Centro de R\'adio-Astronomia e Astrof\'\i sica Mackenzie, Escola de Engenharia,
Universidade Presbiteriana Mackenzie \\ 01302-907, S\~ao Paulo, SP, Brasil}

\address[3]{Grupo de \'Optica e Modelagem Num\'erica - GOMNI, Faculdade de Tecnologia - FT, Universidade Estadual de Campinas - UNICAMP \\ 13484-332, Limeira, SP, Brasil}

\begin{abstract}
We analyze quantitatively the interplay between explicit and implicit
renormalization in Nuclear Physics. By explicit renormalization we
mean to integrate out higher energy modes below a given cutoff scale
using the similarity renormalization group (SRG) with a block-diagonal
evolution generator, which separates the total Hilbert-space into a
model space and its complementary. In the implicit renormalization we
impose given conditions at low energies for a cutoff theory. In both
cases we compare the outcoming effective interactions as functions of
the cutoff scale. We carry out a comprehensive analysis of a toy-model
which captures the main features of the nucleon-nucleon ($NN$) $S$-wave
interaction at low energies. We find a wide energy region where both
approaches overlap.  This amounts to a great simplification in the
determination of the effective interaction. Actually, the outcoming
scales are within the expected ones relevant for the physics of light
nuclei.
\end{abstract}


\end{frontmatter}

\section{Introduction}

The idea of renormalization group from a Wilsonian point of view is quite
intuitive and appealing~\cite{Wilson:1973jj}. A truncated Hilbert
space is considered below some given maximal energy where the relevant
physical degrees of freedom are taken into account explicitly. All states above that maximal energy are integrated out and contribute to the structure of operators and their couplings in the reduced Hilbert state via scale-dependent effective interactions. The renormalization group equations arise from the requirement that
physical results ought to be independent on the chosen numerical
maximal energy value. While one may identify a fundamental underlying
theory with the corresponding elementary degrees of freedom, the
so-called {\it ab initio} calculations may not necessarily be the most
efficient way to pose the quantum mechanical many-body problem of
composite and extended interacting constituent particles. Actually,
for a self-bound system its compositeness vs the elementary character
depends on the shortest de Broglie wavelength involved in the physical
process under consideration as compared to the typical length scales
characterizing the interaction among constituents. For {\it known}
interactions the Wilsonian renormalization group approach proves a convenient
computational strategy to tackle the many-body problem.

In Nuclear Physics the interaction among nucleons is {\it unknown}
fundamentally and precisely except at long distances where one-pion
Exchange (OPE) dominates. At shorter distances the interaction may be
constrained from fits to nucleon-nucleon ($NN$) scattering data up to a given maximum
energy and with a given accuracy (see
Refs.~\cite{Perez:2013mwa,Perez:2013jpa,Perez:2013oba,Perez:2014yla}
for the most recent upgrade in the elastic regime for np and pp
data). Thus, the particular status of the nuclear force makes
renormalization group methods an ideal tool to address the problem of
nuclear binding.

In the case of atomic nuclei glued together by the $NN$ force the most
troublesome issue for nuclear structure calculations is the appearance
of an inner large core of about $a_c=0.5-0.6~{\rm fm}$ which becomes
visible for $NN$ scattering at pion production
threshold~\cite{Wiringa:1994wb} (see however
Ref.~\cite{Kukulin:2013oya} for an alternative interpretation). This
distinct feature generates a strong short-range repulsion which
complicates enormously the solution of the multinucleon problem
limiting the maximal number of nucleons in {\it ab initio}
calculations~\cite{Pieper:2001mp}. On the other hand, because the net
effect of the core is to prevent particles to come too close in the
nucleus ground-state the net contribution to the binding-energy
stemming from distances smaller than the core is tiny. From this point
of view one may equally assume weakly interacting particles at short
distances, thus under these circumstances the core may be replaced by
a suitable soft-core short-distance interaction which keeps invariant
the scattering information.

With this perspective in mind, the idea of effective interactions has
been developed after the early proposals of
Goldstone~\cite{Goldstone:1957zz}, Moshinsky~\cite{Moshinsky195819}
and Skyrme~\cite{Skyrme:1959zz} and Moszkowski and
Scott~\cite{1960AnPhy..11...65M} as a way to cut the gordian knot of
the Nuclear Many-Body Problem represented by strong short-range
repulsion. This allowed to take advantage of the much simpler mean
field framework based on those effective
interactions~\cite{Vautherin:1971aw} (for a review see
e.g.~\cite{Bender:2003jk}). The main problem of the effective
interaction approach is both the proliferation of independent
parameters as well as their huge numerical diversity (see e.g. the
recent compilation of parameters~\cite{Dutra:2012mb}).  This reflects
both the lack of a unambiguous link to the fundamental two-body
interaction as well as the quite disparate finite nuclei and nuclear
matter observables which have been used to fix the effective
Hamiltonian parameters. An effort has been made~\cite{Arriola:2010hj}
(see e.g. Ref.~\cite{Harada:2005tw} for a similar setup and
\cite{NavarroPerez:2013iwa,Perez:2014kpa} for alternative views) in
order to understand the origin of the two-body effective interactions
from free space $NN$ scattering without invoking
finite nuclei nor nuclear matter properties. This point of view
corresponds to what will be called here as {\it implicit}
renormalization.

These somewhat intuitive considerations have been made more precise by
a novel re-interpretation of the Nuclear Many-Body Problem from the
Wilsonian renormalization group point of view. The novel insight,
dubbed as $V_{\rm low \; k}$, is to provide an alternative approach to
the determination of effective interactions {\it directly} from the $NN$
{\it bare} potentials fitted to the scattering
data~\cite{Bogner:2001gq,Bogner:2003wn,Bogner:2006pc,Bogner:2006vp}
(for reviews see
e.g.~\cite{Coraggio:2008in,Bogner:2009bt,Furnstahl:2012fn,Furnstahl:2013oba}
and references therein) and their characterization as finite cutoff
counterterms~\cite{Holt:2003rj}. This point of view corresponds to
what will be called here as the {\it explicit} renormalization.  The
basis of the whole framework has been to recognize the relevance of
choosing the proper physical scale resolution in the formulation of
the problem. This amounts to a great simplification since at the
relevant scales the many-body problem is posed in terms of effective
degrees of freedom and hence the interaction decreases and
softens. Thus, a mean field solution can be used as a reliable zeroth
order approximation, from where corrections can perturbatively be
computed. Moreover, when the maximum energy is taken at about pion
production threshold or below, the $V_{\rm low \; k}$ interaction does
not depend on what particular {\it bare} potential was used to fit $NN$
scattering data. This universal character of model independent
effective interactions constitutes the main appeal of the approach.

It should be kept in mind that this $V_{\rm low \; k}$ method has not yet
been applied to long-range interactions such as Coulomb or van der
Waals type~\footnote{It should be noted that for current
  state-of-the-art many-body calculations of nuclei the Coulomb
  contributions are now routinely included in the SRG evolution (see
  e.g.,\cite{Jurgenson:2010wy,Roth:2013fqa}.}. As it is well known,
potentials can and have been derived from field theory principles by
analyzing the scattering problem in perturbation theory (see
e.g. Ref.~\cite{Partovi:1969wd} for a comprehensive exposition). The
outcoming meson-exchange potentials correspond to Yukawa-like forms at
long distances~\cite{Machleidt:1989tm}, and hence provide a finite
range for the interaction. On the other hand, the same derivations
provide singular interactions when directly extrapolated to short
distances. In momentum-space these interactions lead to ultraviolet
divergencies. The interpretation of these singularities has been
intensively analyzed in the literature using field theoretical
renormalization group ideas (see e.g. \cite{Cordon:2009pj} for a
discussion within the One-Boson-Exchange picture and references
therein). From a practical point of view, these approaches can be
thought of as introducing a short-range potential (which acts as a
regulator) which is actually fixed by some scattering properties
(corresponding to renormalization conditions).  In fact, the
long-distance behavior turns out to be regulator-independent. In this
paper we are not concerned about how a finite-range interaction is
deduced from an underlying theory nor which procedure was used to deal
with the short-range behavior, and for our purposes we will assume
that no serious short-distance singularity is present in the
interaction. Rather, we want to analyze the behavior of the system
when the scale resolution changes, and more specifically how effective
interactions for the Nuclear Many-Body Problem do exhibit the
necessary scale-dependence. We point out that the most recent
formulation of the problem is via the similarity renormalization group
(SRG) method~\cite{Bogner:2006pc,Furnstahl:2007fr,Bogner:2009bt},
where tremendous simplifications arise which entitle to circumvent the
problem at the relevant scales needed for light nuclei.

In the present paper we want to analyze the SRG method with a
block-diagonal (BD) generator~\cite{Anderson:2008mu} as applied to the
two-body problem~\footnote{It is noteworthy that this block-diagonal SRG approach
  could but has not yet been applied to multinucleon problem after
  properly handling the CM motion.}.  This allows to implement,
through a continuous and unitary evolution of a system of coupled
differential equations, a block-diagonal separation of the
Hilbert-space in two orthogonal (decoupled) subspaces ${\cal H}= {\cal
  H}_P \oplus {\cal H}_Q$, which are below or above a given
momentum cutoff $\Lambda$ respectively. The SRG evolution is carried
out as function of a momentum-dimension parameter $\lambda$ referred
to as the SRG-cutoff, which runs from $\lambda=\infty$ (the
ultraviolet limit) to $\lambda=0$ (the infrared limit) and
interpolates between a {\it bare} Hamiltonian, $H_{\lambda=\infty}$,
and the block-diagonal one $H_{\lambda=0}$ in a unitary way
$H_{\lambda=0}=U H_{\lambda=\infty} U^\dagger$. This is a unitary
implementation~\cite{Anderson:2008mu} to all energies of the
previously proposed $V_{\rm low \; k}$ approach~\cite{Bogner:2001gq},
where the higher-energy states are missing, and in practice a free
theory is assumed above the energy determined by the momentum cutoff
$\Lambda$. For the rest of the paper we will refer to this $\Lambda$
as the $V_{\rm low \; k}$-cutoff to be identified with the block-diagonal
SRG one. We emphasize that a complete Hilbert-space separation
corresponds to the limit $\lambda \to 0$. The SRG method has been
applied to the two-~\cite{Bogner:2006pc}, three~\cite{Hebeler:2012pr}
and many-body problem~
\cite{Jurgenson:2009qs,Tsukiyama:2010rj,Jurgenson:2010wy,Launey:2012zz,
Hergert:2012nb,Tsukiyama:2012sm}. The role of effective and long-distance symmetries
has been analyzed only very recently~\cite{Timoteo:2011tt,Launey:2012mda,Arriola:2013nja}.
Toy models have also been used to understand relevant features of the
equations~\cite{Bogner:2007qb,Jurgenson:2008jp}.

As mentioned, the SRG flow-equations are differential equations in the
SRG-cutoff $\lambda$ for unbound operators defined on the
Hilbert-space, and they have only been solved exactly for very simple
cases~\cite{Szpigel:1999gf}. For more realistic cases, one has to
resort to numerical analysis; SRG flow-equations are solved on a finite
$N-$dimensional momentum grid, $p_n$, which introduces an infrared
resolution scale $\Delta p_n$ into the problem. Furthermore, an
auxiliary numerical cutoff $P_{\rm max}$ must also be introduced from
the very start, under the assumption that high-momentum states are
truly irrelevant and hence decouple. Thus, a finite dimensional
Hilbert-space of dimension $N$ remains. This discretization will be of
utmost relevance for our analysis as we discuss briefly below.

The evolution along the SRG trajectory strictly requires the
SRG-cutoff $\lambda$ to be a continuous variable. In practice,
however, some integration method is used to numerically solve the SRG
flow-equations which requires a further grid in the SRG-cutoff,
$\lambda_i$. This new discretization introduces an integration step
$\Delta \lambda_i$ which acts as an additional infrared resolution
scale and can efficiently be made dependent on the actual $\lambda_i$
value.  Obviously, we must take $\Delta \lambda_i $ small enough not
to jeopardize the unitarity of the SRG transformation, a feature which
has to be checked during the course of the evolution~\footnote{
  Algorithms which preserve unitarity exactly, regardless of the
  evolution step, will be discussed elsewhere.}.  It turns out that
evolution steps $\Delta \lambda_i $ have to become smaller as
$\lambda_i$ approaches the origin. This is partly due to the
non-linear character of the SRG flow-equations and the onset of stiffness
due to the large dimensionality of the model space. Of course, the
existence of two small momentum scales $\Delta p_n$ and $\Delta
\lambda_i$ suggests some critical slowing down of the
calculations. Furthermore, when the $V_{\rm low \; k}$ cutoff $\Lambda$
(which takes values on the momentum grid $p_n$) is also small,
$\Lambda \le \Delta p_n$, we expect some dynamics cross-over as the
discretization effects become relevant and hence large deviations from
the continuum are expected.  Within the finite dimensional reduction
of the problem, the meaning of convergence of the SRG evolution as the
SRG cutoff $\lambda$ goes from the $\lambda=\infty$ to $\lambda=0$ is
not particularly subtle from a mathematical viewpoint. Technically,
the Hamiltonian is defined as an operator on a Hilbert-space. However,
the SRG evolution deals with a family of unitarily equivalent
operators, and hence one must introduce some metric to measure the
distance between operators. As it is well known in finite dimensional
spaces all metrics are equivalent. That means that if convergence is
accomplished according to a given metric any other metric does also
exhibit convergence. In practice one can check individual
matrix-elements, for instance. This of course poses the problem of the
continuum limit, which to our knowledge has never been discussed in
detail within the present context, so some choice of how the limits
are taken must be made. In all our numerical implementations of the
SRG flow-equations we will assume that the SRG limit is taken first
and only then the continuum limit will be pursued.

The basic goal of the BD-SRG (or equivalently $V_{\rm low \; k}$) is to
choose $\Lambda$ small enough to filter out model-dependent
small-distance physics and make the many-body problem more
perturbative, but large enough to keep the relevant and
well-constrained low-energy degrees of freedom in the Hilbert space
for few- or many-body calculation. In this sense the typical energy
scale $\Lambda$ is set by the pion mass or, equivalently, the energy
scale up to where the phase shift analysis can be reliably be
performed. Nonetheless, the limit $\Lambda \to 0$ is requested to fix
the renormalization condition with lowest energy scattering. As we
will show, the implicit renormalization handles the continuum limit
for small $\Lambda$'s, a regime where the finite-grid proves extremely
inefficient. On the other hand, when we try to extend this SRG
solution to higher $\Lambda$'s the finite-range aspects of the
interaction are lost. We envisage a possibility of reducing the number
of grid points precisely when the finite-range becomes relevant.

We consider a simple separable gaussian potential toy-model for the
two-body nuclear force inspired by the $^1S_0$ and $^3S_1$
partial-wave channels and perform a complete study in the framework of
the SRG. These two cases illustrate the situation where either none or
just one bound-state (corresponding to the deuteron) are present.  The
idea is to investigate the infrared limit ($\lambda \rightarrow 0$) of
the SRG with the block-diagonal generators which is the unitary
version of the $V_{\rm low \; k}$ approach. Our toy-model is constructed
so that the main $S$-wave two-nucleon observables (the phase-shifts at
low-momenta and the deuteron binding-energy) are reasonably described
with a short-range interaction and makes the SRG evolution towards the
infrared limit much more practical. We compare the effective
interactions obtained in the explicit and implicit renormalization
approaches and analyze to what extent in terms of the corresponding
cutoff scales the $V_{\rm low \; k}$ potential can be implicitly
described by low-energy parameters without explicitly solving the SRG flow-equations.
Shorter accounts of the present work have already been published~\cite{Arriola:2013gya,Arriola:2013yca}.
Here we provide more details and further results.

\section{Implicit renormalization: contact theory with a momentum cutoff}

Implicit renormalization is defined by looking for a $NN$ interaction $V_\Lambda(p,p')$,
regulated by a sharp or smooth momentum cutoff $\Lambda$, which reproduces $NN$
scattering data up to a given center-of-mass (CM) momentum $p \le \Lambda$. This problem
has in fact no unique solution as scattering data above that CM momentum are not specified.
This is in spirit the idea behind the $V_{\rm low \; k}$ approach
~\footnote{There are of course important differences, as in $V_{\rm low \; k}$ the half-off shell
equivalence is also required. This spoils by construction self-adjointness of the potential,
and a subsequent transformation to a self-adjoint potential must be carried out.}.

\subsection{Effective range expansion for the $NN$ interaction}

Here and in what follows we use units such that $\hbar=c=M=1$, where
$M$ is the nucleon mass.  The transition matrix $T$ for the scattering of
two nucleons is given by the partial-wave Lippmann-Schwinger (LS)
equation
\begin{equation}
T(p,p';k^2)=V(p,p')+\frac{2}{\pi}\; \int_{0}^{\infty} \; dq \; q^2 \;
\frac{V(p,q)}{k^2-q^2+i \; \epsilon} \; T(q,p';k^2) \; ,
\end{equation}
\noindent
where $V(p,p')$ is the $NN$ potential in a given
partial-wave. At low energies, the on-shell $T$-matrix can be
represented by an effective range expansion (ERE)
\begin{eqnarray}
T^{-1}(k,k;k^2)=-\left[k~{\rm cot}~\delta(k)-i~k \right]=-\left[-\frac{1}{a_0}
+\frac{1}{2}~r_e~k^2 + v_2~k^4 + {\cal O}(k^6)-i~k \right] \; ,
\label{eq:ERE}
\end{eqnarray}
\noindent
where $k=\sqrt{E}$ is the on-shell momentum in the CM frame,
$\delta(k)$ is the phase-shift, $a_0$ is the scattering length, $r_e$
is the effective range and $v_2$ is a shape parameter. The
experimental values of the ERE parameters $a_0$ and $r_e$ for the
$S$-wave channels are given by
\begin{eqnarray}
^1S_0 \; {\rm channel}: a_0=-23.74~{\rm fm}; r_e=2.77~{\rm fm} ~ , \nonumber\\
^3S_1\;  {\rm channel}: a_0=+5.420~{\rm fm}; r_e=1.75~{\rm fm} ~ .
\end{eqnarray}
In order to avoid a numerical integration on the complex plane, which
depends on the contour chosen to perform the sum, we use the LS
equation for the partial-wave reactance matrix $R$ with the principal
value prescription,
\begin{equation}
K(p,p';k^2)=V(p,p')+\frac{2}{\pi}\; {\cal P}\int_{0}^{\infty} \; dq \; q^2 \;
\frac{V(p,q)}{k^2-q^2} \; K(q,p';k^2) \; ,
\label{LSKNN}
\end{equation}
\noindent
where ${\cal P}$ denotes the principal value. This matrix has the
advantage of being real and the relation to the on-shell $T$-matrix is given by
\begin{equation}
\frac{1}{K(k,k;k^2)} = \frac{1}{T(k,k;k^2)} \; - \; i \, k \; .
\label{KTrel}
\end{equation}

\subsection{Two-nucleon bound-state}

The deuteron bound-state energy $E_d = -\gamma^2$ can be
obtained from the pole of the on-shell $T$-matrix for the $^3S_1$ channel.
Using the ERE expansion to order ${\cal O}(k^2)$ we have
\begin{eqnarray}
\frac{1}{T(k,k;k^2)}=\frac{1}{a_0}-\frac{1}{2}~r_e~k^2 +i~k =0 \Rightarrow k ={\rm i}~\gamma = {\rm i}~\left[\frac{1}{r_e}-\left(\frac{1}{r_e^2}-\frac{2}{a_0~r_e}\right)^{1/2}\right] \; .
\label{Tinvexp}
\end{eqnarray}
\begin{figure*}[t]
\begin{center}
\includegraphics[width=7.0cm]{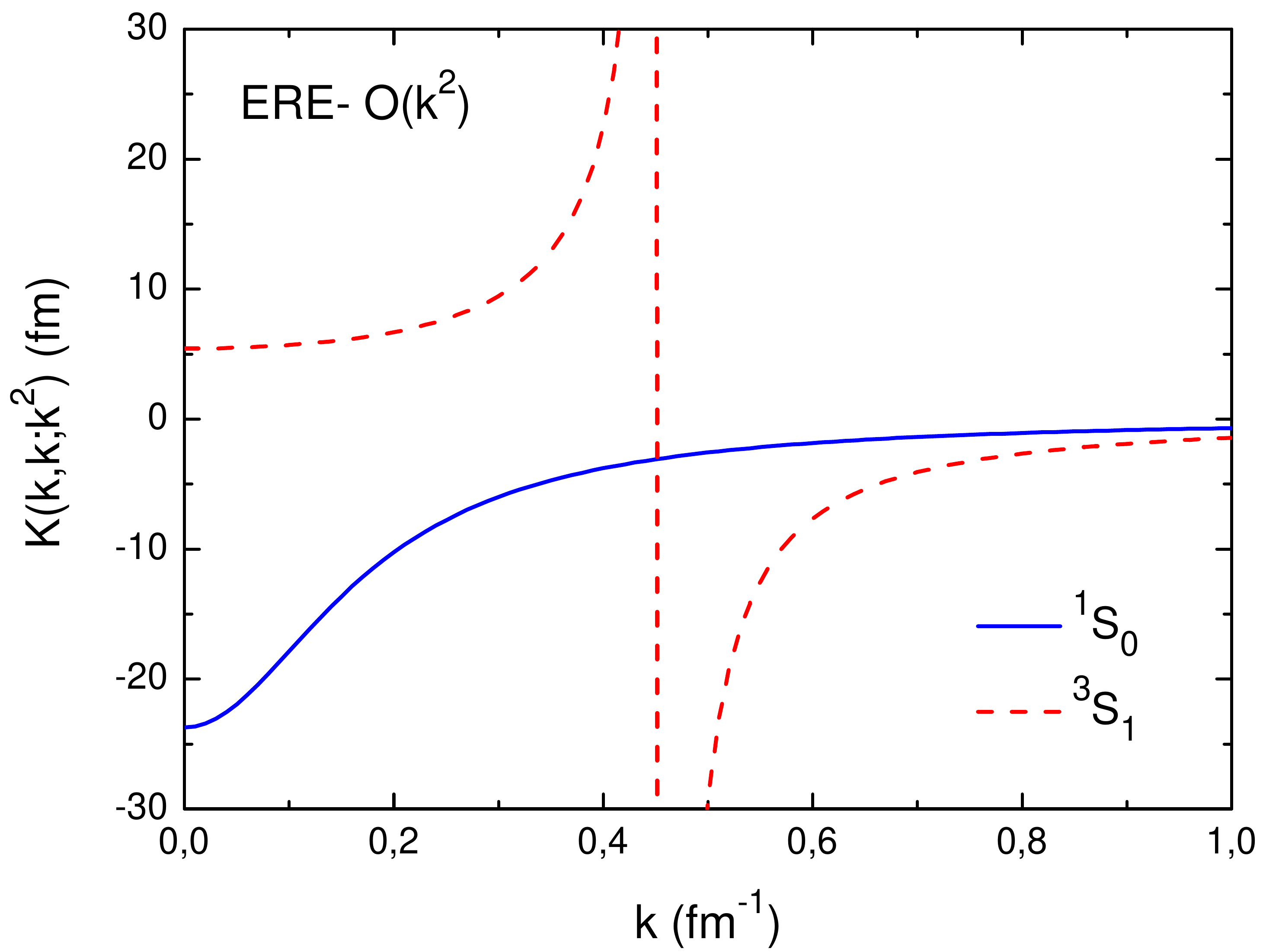}
\end{center}
\caption{On-shell $K$-matrix in the effective range expansion (ERE) to order ${\cal O}(k^2)$ for the $^1S_0$ channel and the $^3S_1$ channel $NN$ interactions.}
\label{fig:1}
\end{figure*}

One should note that the deuteron pole in the $T$-matrix for the
$^3S_1$ channel located at imaginary momentum $k=i~\gamma$ corresponds to a pole in the $K$-matrix located at a distinct real momentum $\beta$, as shown in Fig.~\ref{fig:1}. From Eq.~(\ref{KTrel}) we
have,
\begin{eqnarray}
\frac{1}{K(k,k;k^2)} = \frac{1}{a_0}-\frac{1}{2}~r_e~k^2=0
\Rightarrow k = \beta =\left[\frac{2}{a_0~r_e}\right]^{1/2} \;.
\label{Kinvexp}
\end{eqnarray}
\noindent
The relation between $\gamma$ and $\beta$ is then given by
\begin{eqnarray}
\gamma = \left[\frac{1}{r_e}-\left(\frac{1}{r_e^2}-\beta^2\right)^{1/2}\right] \; .
\end{eqnarray}

By replacing the experimental values of $a_0$ and $r_e$ for the $^3S_1$
channel in Eqs.~(\ref{Tinvexp}) and (\ref{Kinvexp}) we obtain $\gamma
\sim 0.2313~{\rm fm}{^{-1}}$ and $\beta \sim 0.4592~{\rm fm}{^{-1}}$.
Using $M=938.919~{\rm MeV} = 4.7581~{\rm fm}^{-1}$ we obtain the
deuteron binding energy $B_d = \gamma^2/M \sim 2.219~{\rm MeV}$.

\subsection{Contact theory regulated by a sharp momentum cutoff}

While the $NN$ potential is fairly general, we will analyze here
the simple case of hermitian polynomial potentials which correspond to contact
or zero-range interactions,
\begin{eqnarray}
 V(p,p')&=&V^{(0)}(p,p') + V^{(2)}(p,p') + V^{(4)}(p,p') + \dots \nonumber \\
&=& C_0 + C_2 (p^2 + p'^2) + C_4 (p^4+p'^4) + C_4' p^2 p'^ 2 + \dots  \, .
\label{eq:pot-mom}
\end{eqnarray}
In this case the LS equation is divergent, so we can endow the partial-wave $K$-matrix regulated by a sharp momentum cutoff $\Lambda$,
\begin{equation}
K_{\Lambda}(p,p';k^2)=V(p,p')+\frac{2}{\pi}\; {\cal P}\int_{0}^{\Lambda} \; dq \; q^2 \;
\frac{V(p,q)}{k^2-q^2} \; K_{\Lambda}(q,p';k^2) \; .
\label{LSKsharp}
\end{equation}
and determine the unknown $\Lambda$-dependent coefficients $C_0,C_2,
C_4, C_4', \dots$ through a renormalization procedure. This equation
has been solved in a number of occasions and the idea is, for a given
cutoff value $\Lambda$, to fix the unknown coefficients by fitting the
experimental values of the ERE parameters. We solve
Eq.~(\ref{LSKsharp}) analytically and match the expansion of the
inverse on-shell $K$-matrix in powers of $k^2$ to the ERE up to a
given order.

For the contact theory potential at leading-order ($LO$),
\begin{equation}
V_{\rm LO}(p,p')=V^{(0)}(p,p') = C_{0}^{(0)} \; ,
\label{vcLO}
\end{equation}
we obtain
\begin{eqnarray}
\frac{1}{a_0}=\frac{1}{C_{0}^{(0)}}+\frac{2\Lambda}{\pi} \; ,
\end{eqnarray}
and for the contact theory potential to next-to-leading-order ($NLO$),
\begin{equation}
V_{\rm NLO}(p,p')=V^{(0)}(p,p') + V^{(2)}(p,p') = C_{0}^{(2)} + C_{2}^{(2)}~(p^2+{p'}^2) \; ,
\label{vcNLO}
\end{equation}
a set of two coupled non-linear equations for $C_{0}^{(2)}$ and $C_{2}^{(2)}$ arise:
\begin{eqnarray}
\frac{1}{a_0}=\frac{45~ \pi^2 + 90~ C_{0}^{(2)} \pi~ \Lambda+60~ C_{2}^{(2)} \pi~ \Lambda^3-16 ~C_{2}^{(2)2} \Lambda^6}{45~ C_{0}^{(2)} \pi^2-18~ C_{2}^{(2)2}\pi~ \Lambda^5} \; ,
\label{NLO1}
\end{eqnarray}
\begin{eqnarray}
\frac{r_e}{2} =\frac{2~ (675~ C_{0}^{(2)2} \pi^2-540~ C_{0}^{(2)}~ C_{2}^{(2)2} \pi~ \Lambda^5+C_{2}^{(2)} \Lambda~ (675~ \pi^3+1125~ C_{2}^{(2)} \pi^2 \Lambda^3+600~ C_{2}^{(2)2} \pi ~\Lambda^6+208 ~C_{2}^{(2)3} \Lambda^9))}{27~ \pi ~\Lambda~ (5~ C_{0}^{(2)} \pi-2 ~C_{2}^{(2)2} \Lambda^5)^2} \; .
\label{NLO2}
\end{eqnarray}

\begin{figure*}[t]
\begin{center}
\includegraphics[width=5.3cm]{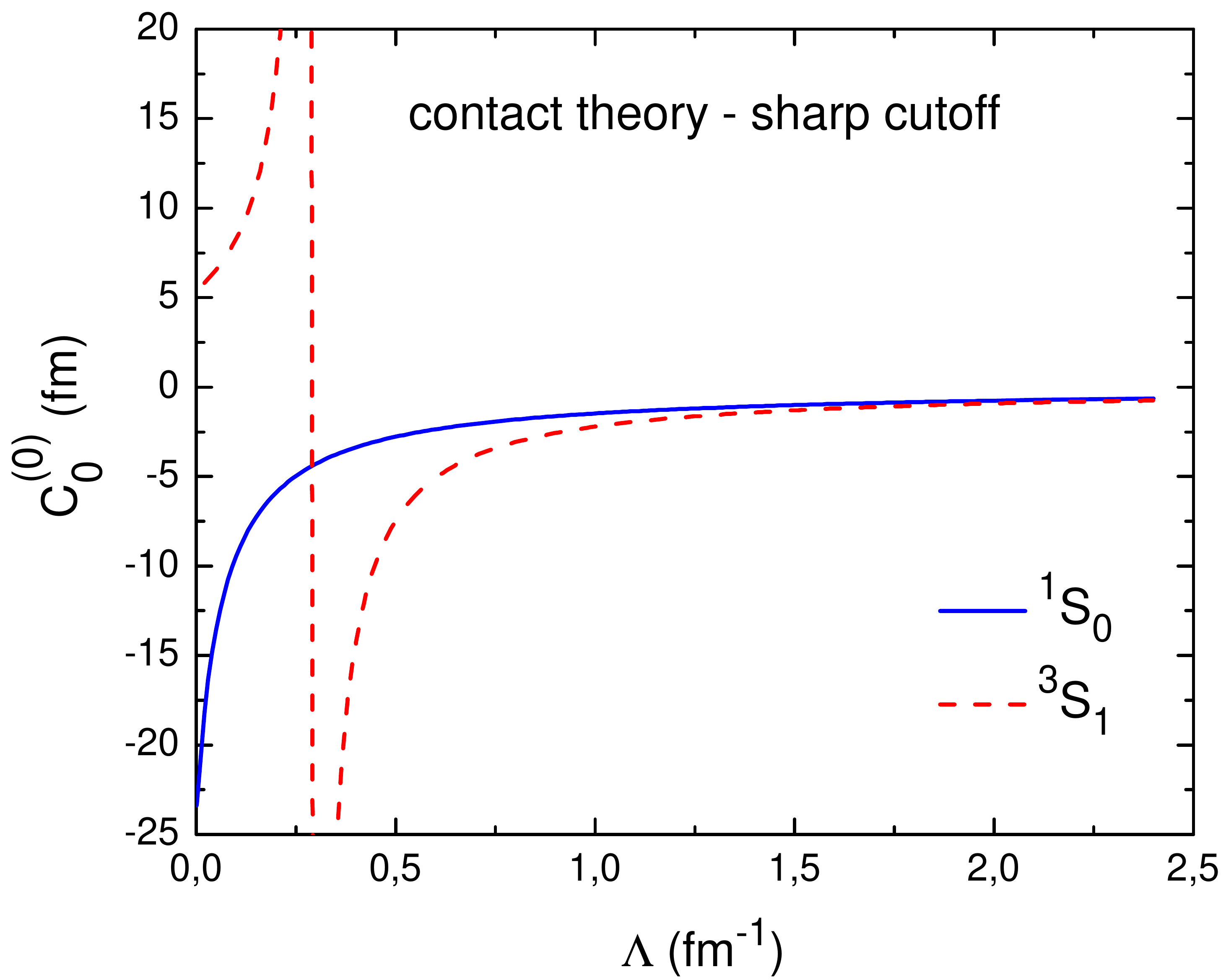}\hspace{0.1cm}
\includegraphics[width=5.3cm]{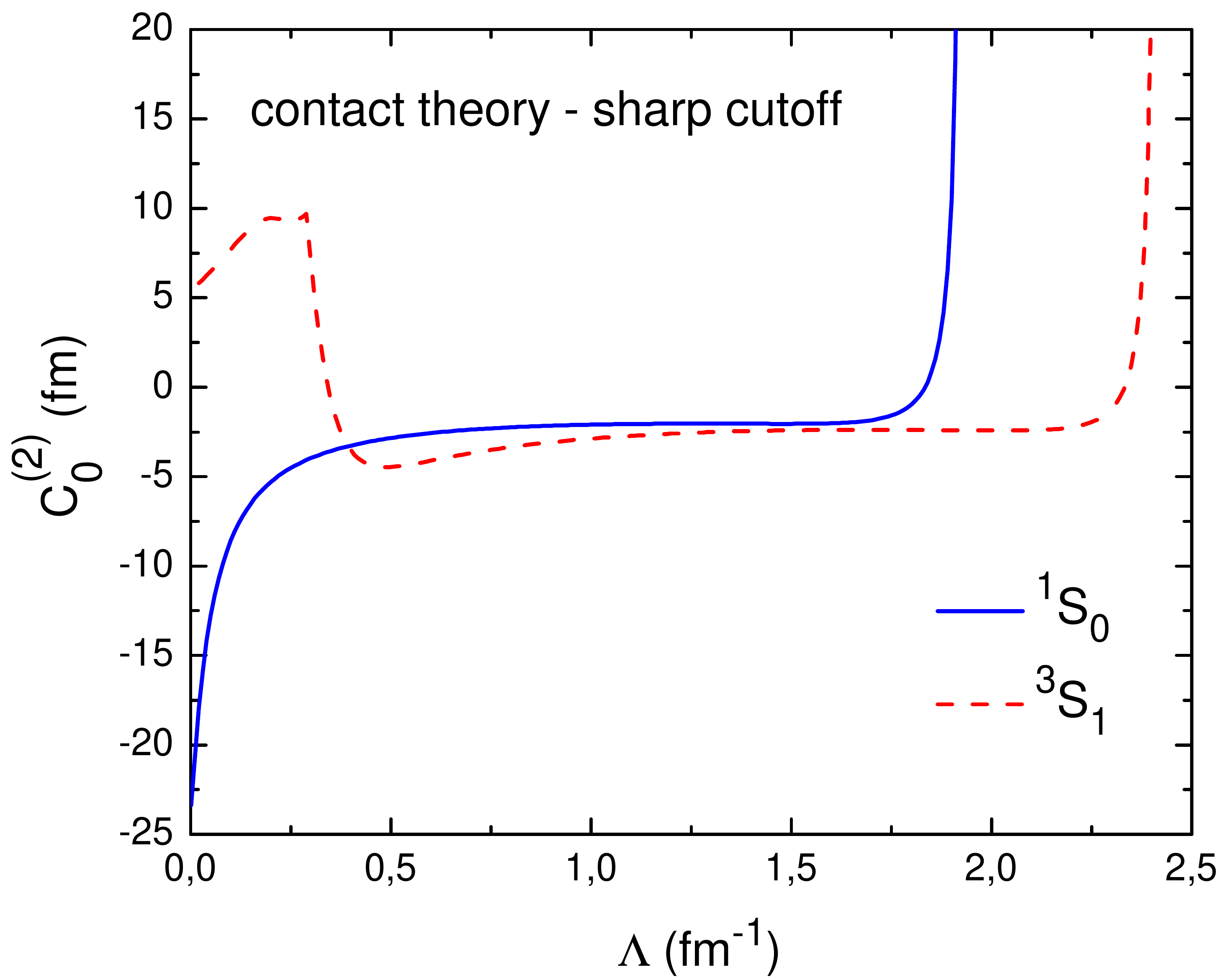}\hspace{0.1cm}
\includegraphics[width=5.3cm]{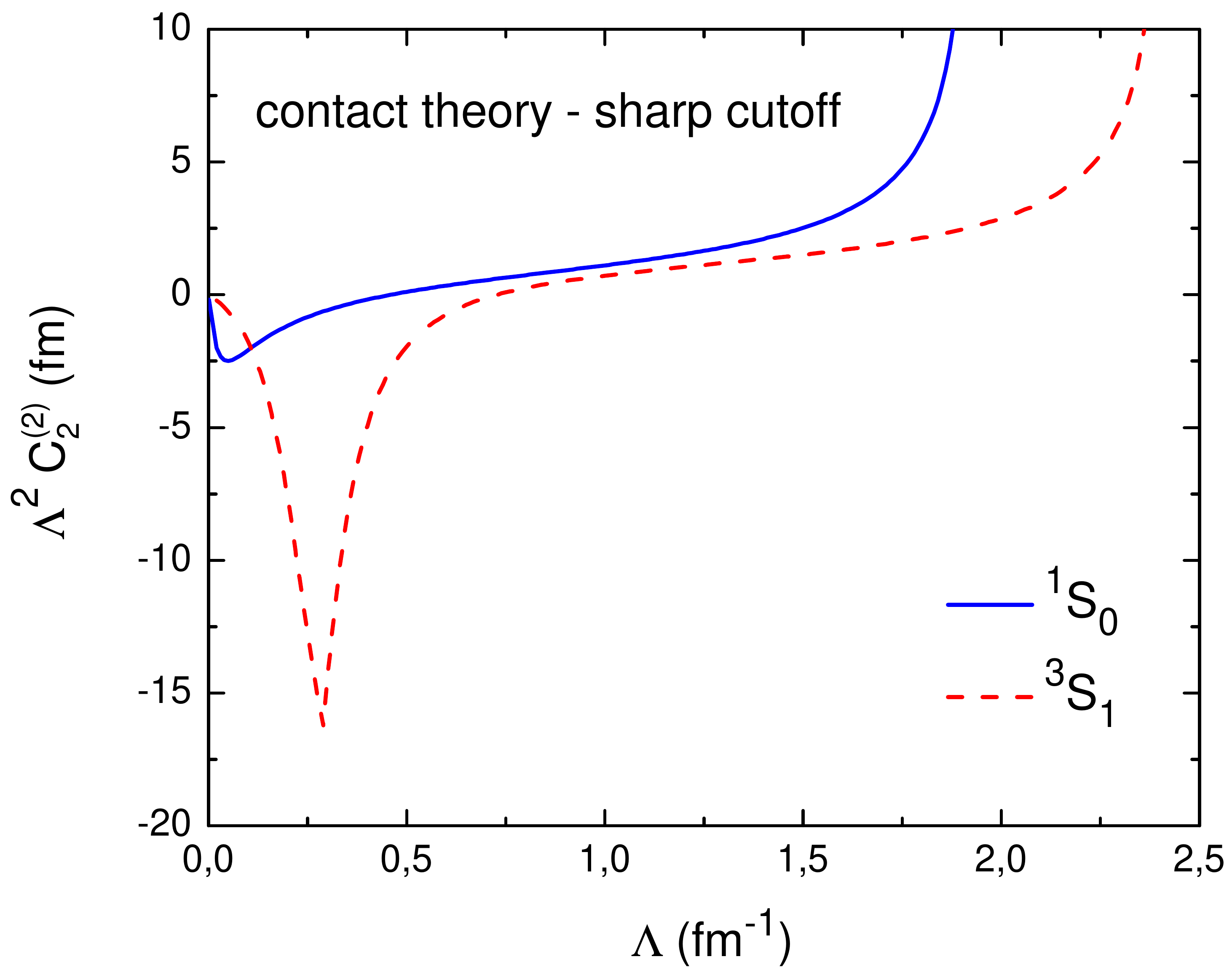}
\end{center}
\caption{Running of the coefficients $C_{0}^{(0)}$, $C_{0}^{(2)}$ and $C_{2}^{(2)}$ with the cutoff $\Lambda$ for the
  contact theory in the continuum regulated by a sharp momentum cutoff
  for the $^1S_0$ channel and the $^3S_1$ channel $NN$ potentials at $LO$ and $NLO$. The coefficients are
  determined from the analytical solution of the LS equation for the on-shell
  $K$-matrix by fitting the ERE parameters.}
\label{fig:2}
\end{figure*}

From the two possible solutions of Eqs.~(\ref{NLO1}) and (\ref{NLO2}),
we choose the one in which $C_{0}^{(2)}\rightarrow a_0$ and $\Lambda^2
~C_{2}^{(2)}\rightarrow 0$ in the limit $\Lambda \rightarrow 0$, as
shown in Fig.~\ref{fig:2}. One should note that in the case of the
$^3S_1$ channel the $LO$ potential coefficient $C_{0}^{(0)}$ is
singular and the derivatives of the $NLO$ potential coefficients
$C_{0}^{(2)}$ and $C_{2}^{(2)}$ are discontinuous at $\Lambda=\pi/2a_0
\sim 0.3~{\rm fm^{-1}}$, which is the momentum scale where the
deuteron bound-state appears. Moreover, as a consequence of Wigner's
causality bound, there is a maximum value $\Lambda_{\rm WB}$ for the
cutoff scale $\Lambda$ above which one cannot fix the potential
coefficients $C_{0}^{(2)}$ and $C_{2}^{(2)}$ by fitting the
experimental values of both the scattering length $a_0$ and the
effective range $r_e$ while keeping the renormalized potential
hermitian~\cite{Entem:2007jg,Szpigel:2010bj}. Indeed, for $\Lambda >
\Lambda_{\rm WB} \sim 1.9~\rm{fm}^{-1}$ in the case of the $^1S_0$
channel and $\Lambda > \Lambda_{\rm WB} \sim 2.4~\rm{fm}^{-1}$ in the
case of the $^3S_1$ channel, the coefficients $C_{0}^{(2)}$ and
$C_{2}^{(2)}$ diverge before taking complex values and hence violating
the hermiticity of the effective potential.

We can determine the running of the potential coefficients
$C_{0}^{(2)}$ and $C_{2}^{(2)}$ as a function of the cutoff $\Lambda$,
but the question is: when do we expect this running to be inaccurate?
While the $NLO$ provides a credible Wigner bound, in the $LO$ case
there is no reason to stop the evolution. One may of course try to
incorporate next-to-next-to-leading-order ($NNLO$) corrections. The
problem is that there are two such terms~\cite{Entem:2007jg}
\begin{eqnarray}
V^{(4)}(p,p') = C_4 ~(p^4+p'^4) + C_4' ~(p^2 \times p'^2) \; ,
\end{eqnarray}
but there is only {\it one} low-energy parameter in the ERE at order
${\cal O}(k^4)$, the shape parameter $v_2$ in Eq.~(\ref{eq:ERE}).
This is so because scattering does not depend just on the on-shell
potential. Thus, the implicit renormalization is manifestly not unique
beyond $NLO$.  This is just a manifestation of the existing
ambiguities in the inverse scattering problem \footnote{Actually, from
  a dimensional point of view the two-body operators with four
  derivatives are suppressed as compared to contact three-body
  operators. The off-shellness of the two-body problem can
  equivalently be translated into some three-body properties.}. The
ambiguity is genuine and inherent to the NN interaction; the off-shell
part is not uniquely fixed by the scattering data.

Clearly, and even for the coefficients
$C_{0}^{(2)}$ and $C_{2}^{(2)}$, increasing the $\Lambda$ values one
starts seeing more high-energy details of the theory. In the next
section we approach the problem by using a simple toy-model potential
which has the main features of the two-nucleon $S$-waves, namely no
(real) bound-states in the $^1S_0$ channel and one (real) bound-state
in the $^3S_1$ state (which is identified with the deuteron).

\subsection{Contact theory regulated by a smooth momentum cutoff}

We aim to compare the effective interactions obtained in the implicit
renormalization approach to those obtained in the explicit approach
implemented within the framework of the SRG method. Since the SRG
flow-equations have to be solved numerically on a finite momentum
grid, in order to perform a consistent comparison we consider the
implicit renormalization of a contact theory potential regulated by a
smooth momentum cutoff on the same grid. The LS equation for the
partial-wave $K$-matrix with a smooth momentum cutoff is given by
\begin{equation}
K_{\Lambda}(p,p';k^2)=V_{\Lambda}(p,p')+\frac{2}{\pi}\; {\cal P}\int_{0}^{\infty} \; dq \; q^2 \;
\frac{V_{\Lambda}(p,q)}{k^2-q^2} \; K_{\Lambda}(q,p';k^2) \; ,
\label{LSKsmooth}
\end{equation}
where $V_\Lambda (p,p')$ is the contact theory potential given in Eq.~(\ref{eq:pot-mom}) regulated by an exponential function
$f_{\Lambda}(p)=\exp[-(p/\Lambda)^{2n}]$
\begin{equation}
V_{\Lambda}(p, p') \equiv \exp[-(p/\Lambda)^{2n})]~ V(p, p')~ \exp[-(p'/\Lambda)^{2n}] \; ,
 \label{CR}
\end{equation}
\noindent
and $n=1, 2, \ldots$ determines the sharpness of the regulating function $f_{\Lambda}(p)$.

In the case of a finite momentum grid, the contact theory potential coefficients $C_{i}^{(j)}$ are determined from the numerical solution of the LS equation for the $K$-matrix by fitting the experimental values of the ERE
parameters. Following the method described by Steele and
Furnstahl~\cite{Steele:1998un}, we solve Eq.~(\ref{LSKsmooth}) on a gaussian grid with $N$ momentum points in the range
$[0,P_{\rm max}]$ and fit the difference between the corresponding
inverse on-shell $K_{\Lambda}$-matrix and the inverse on-shell
$K_{{\rm ERE}}$-matrix to an interpolating polynomial in $k^2/\Lambda^2$
to highest possible degree for a spread of very small on-shell momenta
$k_{{\rm low}}~\le 0.1~{\rm fm^{-1}}$:
\begin{eqnarray}
\Delta\left(\frac{1}{K}\right)=\frac{1}{K_{\Lambda}(k,k;k^2)} - \frac{1}{K_{{\rm ERE}}(k,k;k^2)} =
A_0 + A_2\;\frac{k^2}{\Lambda^2} + \ldots \; .
\end{eqnarray}
\noindent
Then we minimize the coefficients $A_i$ with respect to the variations of the coefficients $C_{i}^{(j)}$.

\begin{figure*}[t]
\begin{center}
\includegraphics[width=7.0cm]{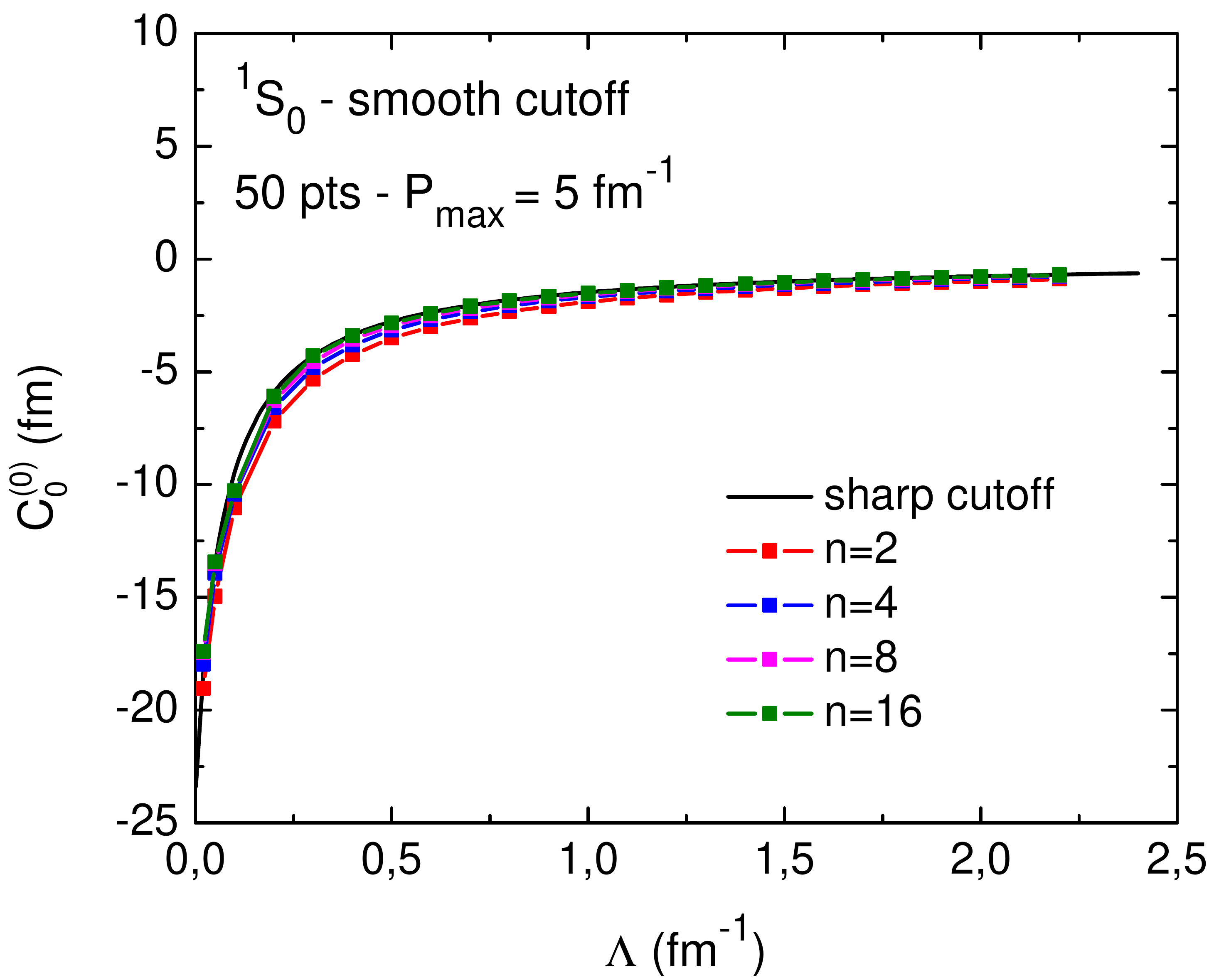}\hspace{0.5cm}
\includegraphics[width=7.0cm]{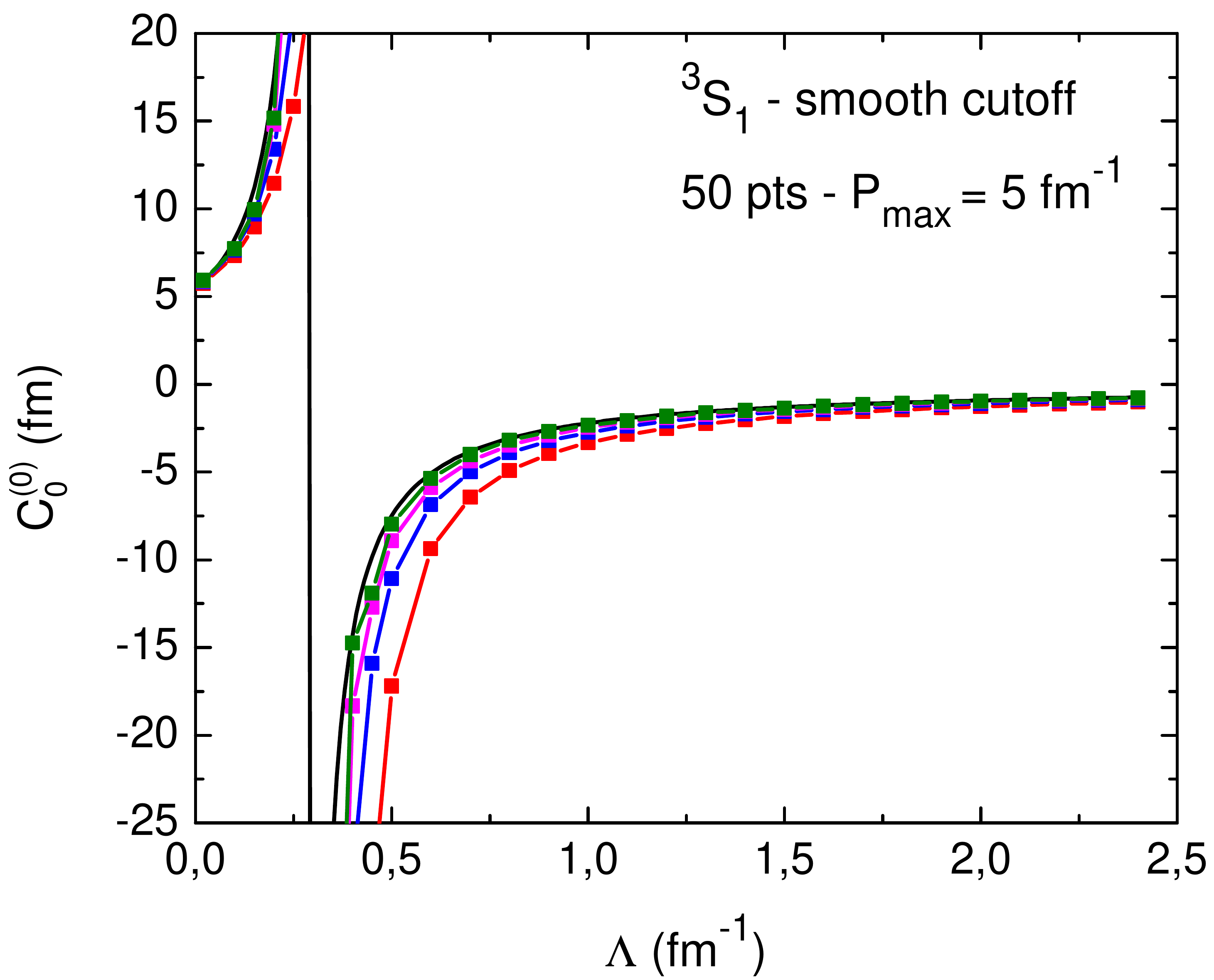}\\\vspace{0.5cm}
\includegraphics[width=7.0cm]{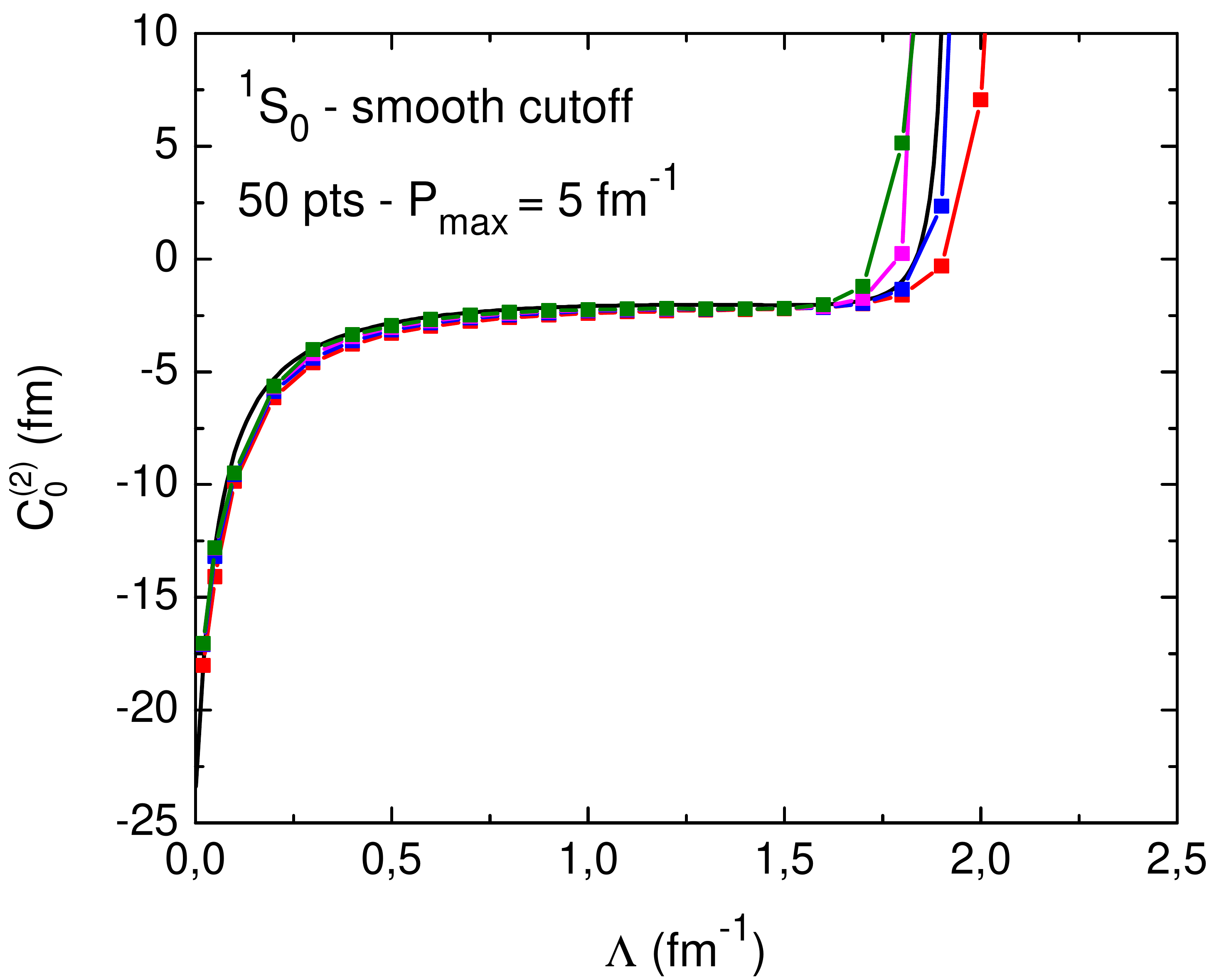}\hspace{0.5cm}
\includegraphics[width=7.0cm]{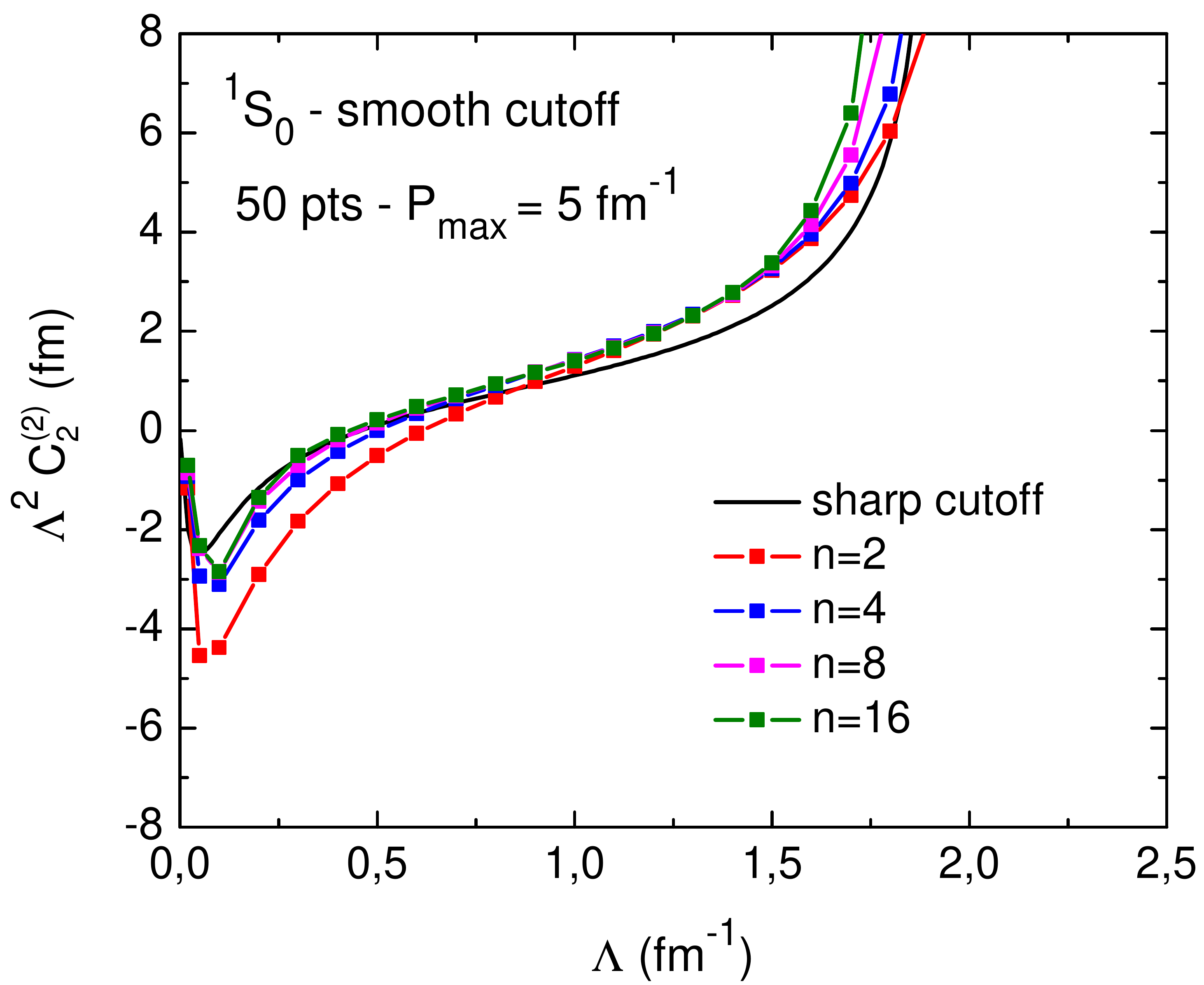}\\\vspace{0.5cm}
\includegraphics[width=7.0cm]{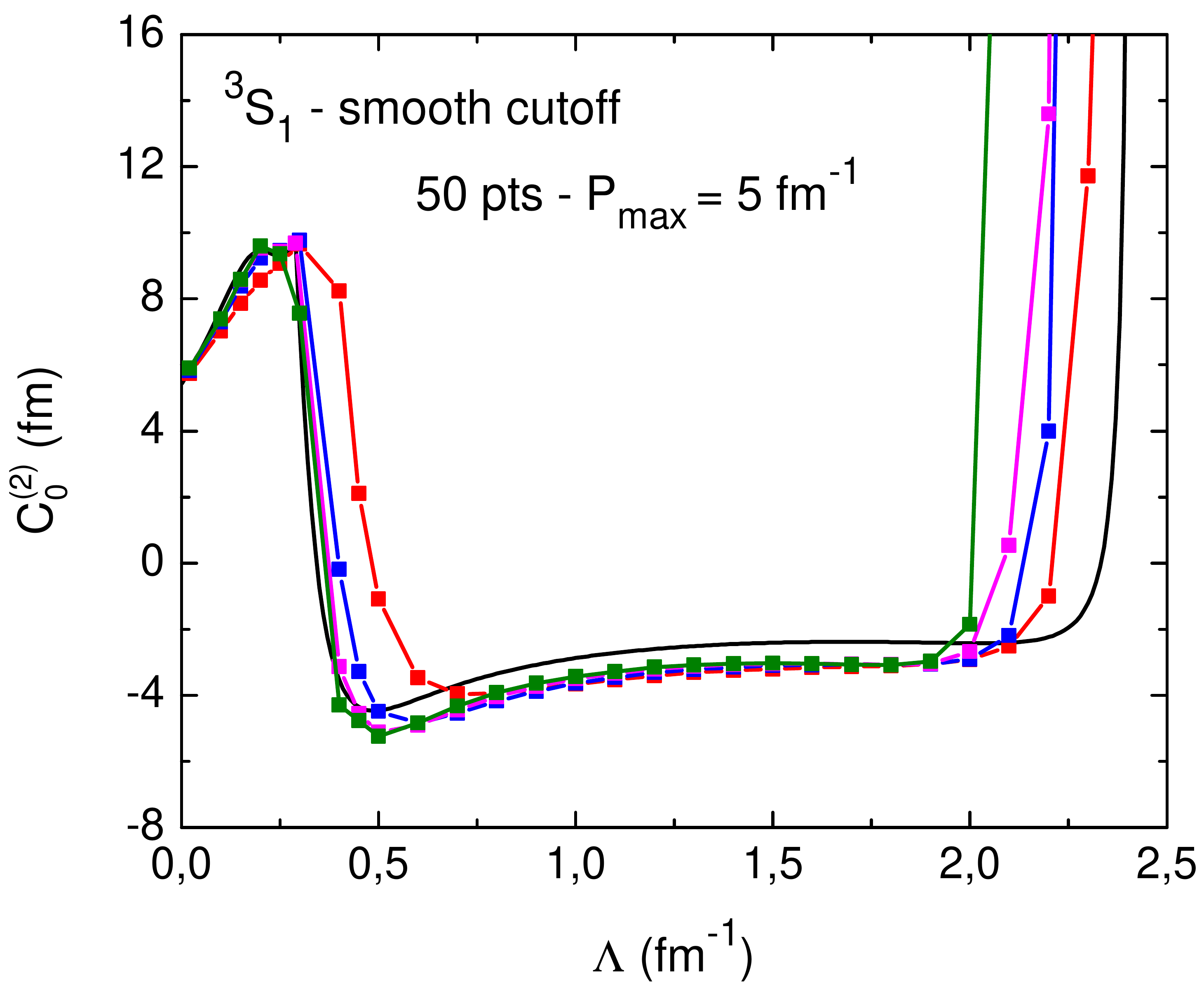}\hspace{0.5cm}
\includegraphics[width=7.0cm]{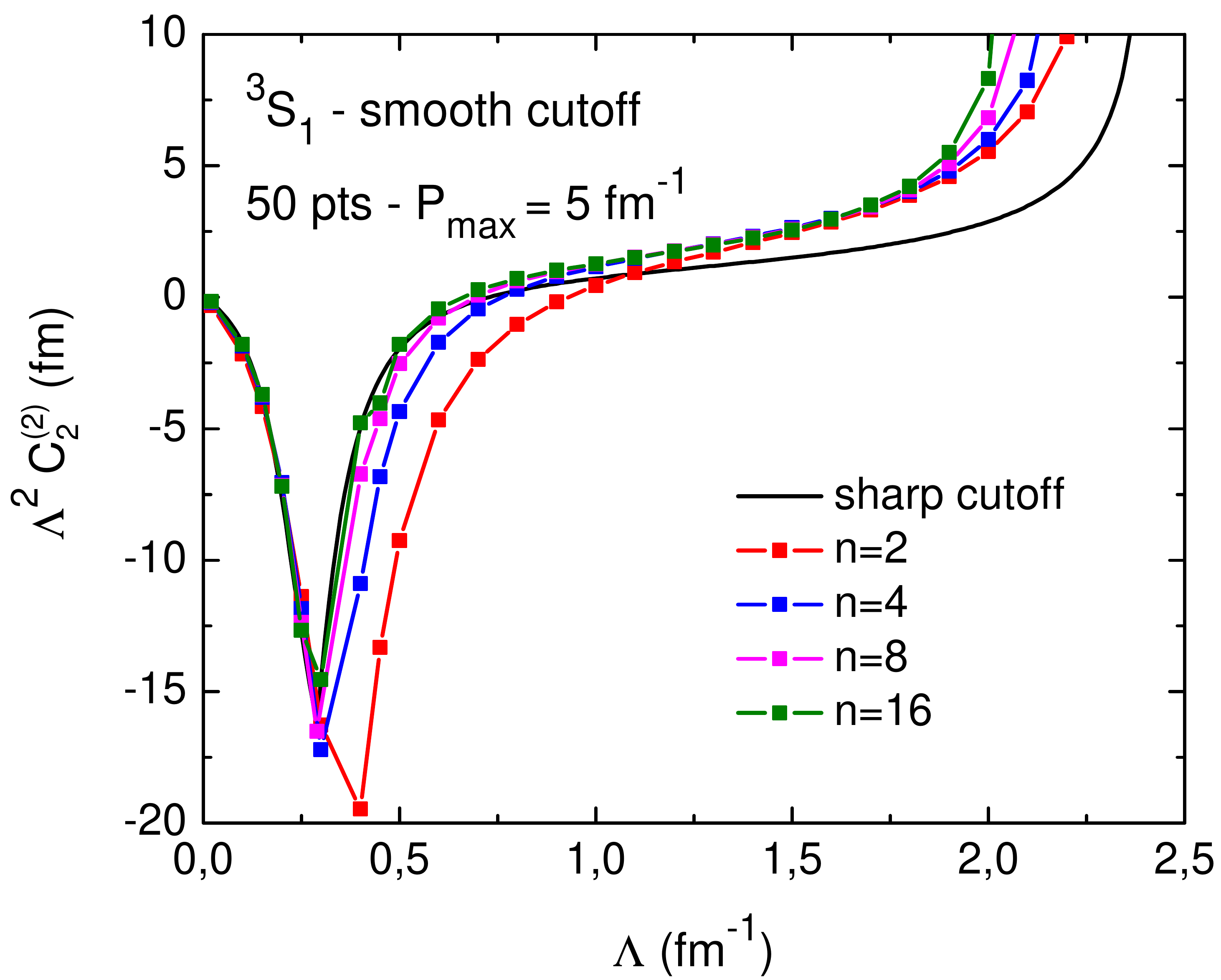}
\end{center}
\caption{Running of the strength $C_{0}^{(0)}$ with the cutoff $\Lambda$ for the contact theory on a finite momentum grid regulated by a smooth
cutoff for the $^1S_0$ and $^3S_1$ channels at $LO$ (upper panels) and at $NLO$ (middle panels: $^1S_0~$, lower panels: $^3S_1$). For comparison, we also show the corresponding $C_{0}^{(0)}$ for the contact theory in the continuum regulated by a sharp momentum cutoff.
In both cases the strengths are determined from the solution of the $LS$ equation for the on-shell $K$-matrix by fitting the ERE parameters.}
\label{fig:3}
\end{figure*}

In Fig.~\ref{fig:3} we compare the running of the coefficients $C_{i}^{(j)}$ with the cutoff $\Lambda$ for the $LO$ and $NLO$ contact theory potentials on a finite momentum grid using regulating functions with sharpness parameter $n =2,~4,~8~{\rm and}~16$ for different grid sizes
$(P_{\rm max}; N)$. As one can observe, when the sharpness parameter $n$ or the grid size increases we obtain a better agreement with the
results for the contact theory in the continuum regulated by a sharp cutoff.

\begin{figure*}[t]
\begin{center}
\includegraphics[width=7.0cm]{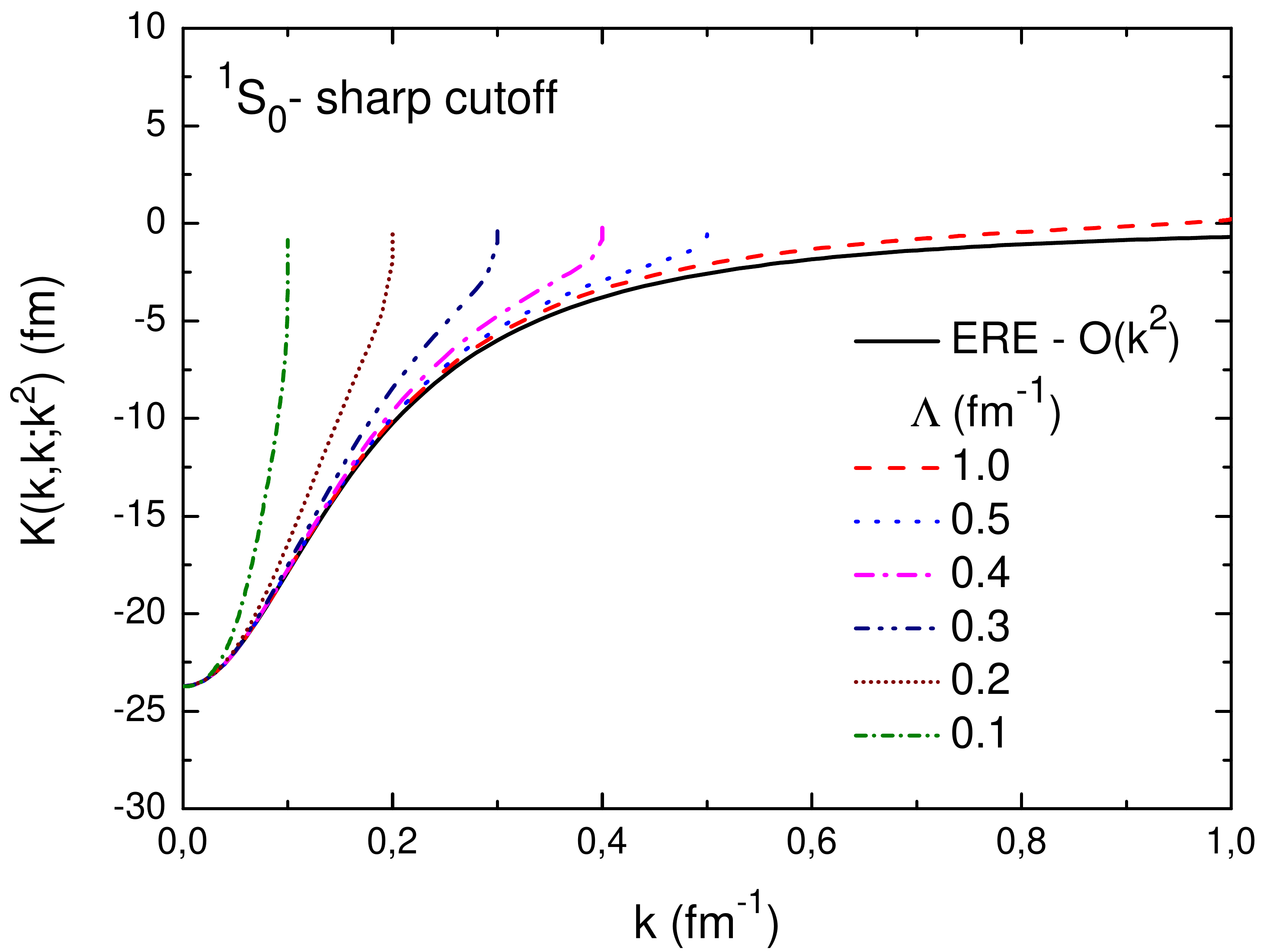}\hspace{0.5cm}
\includegraphics[width=7.0cm]{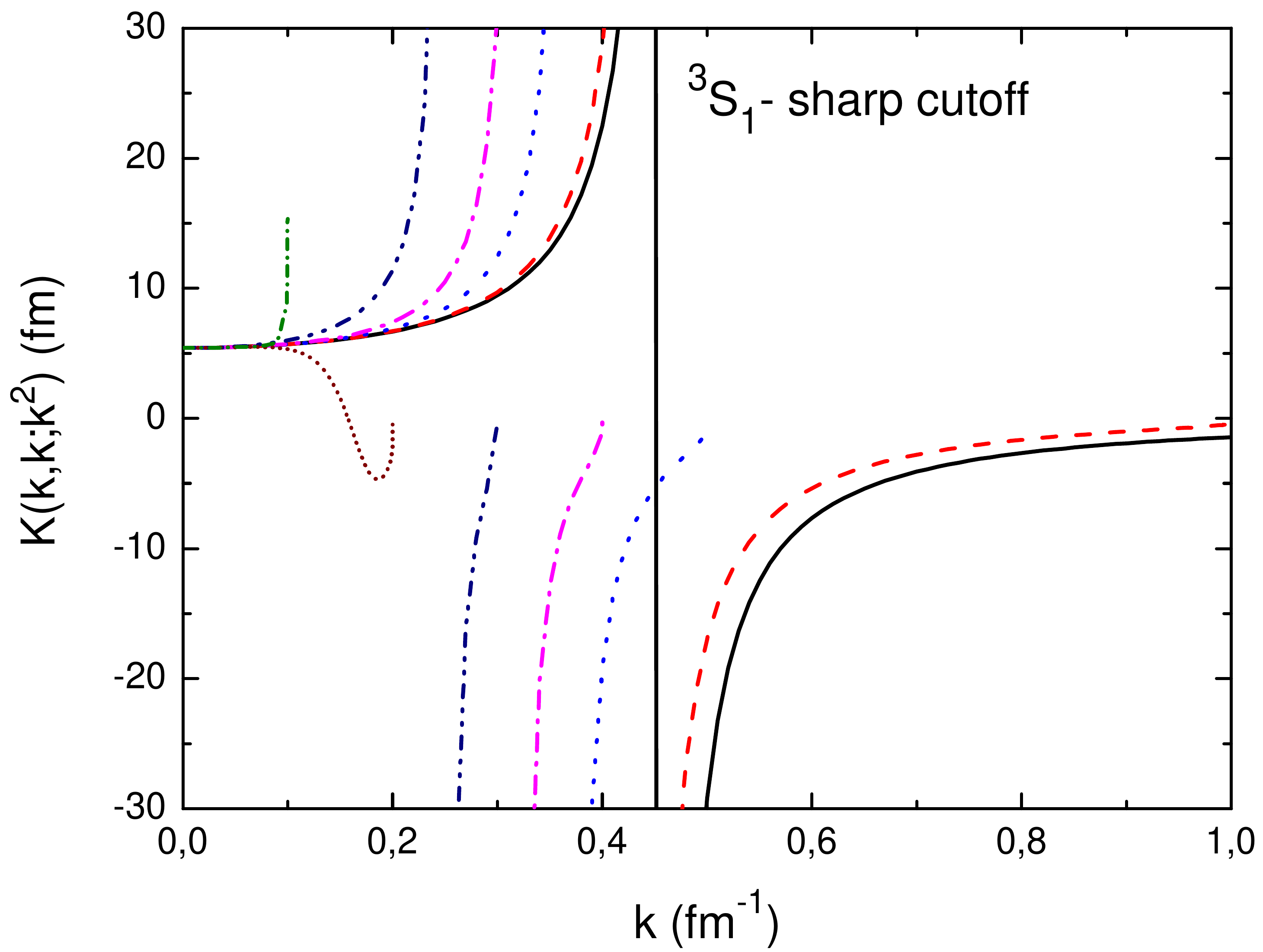}\\\vspace{0.5cm}
\includegraphics[width=7.0cm]{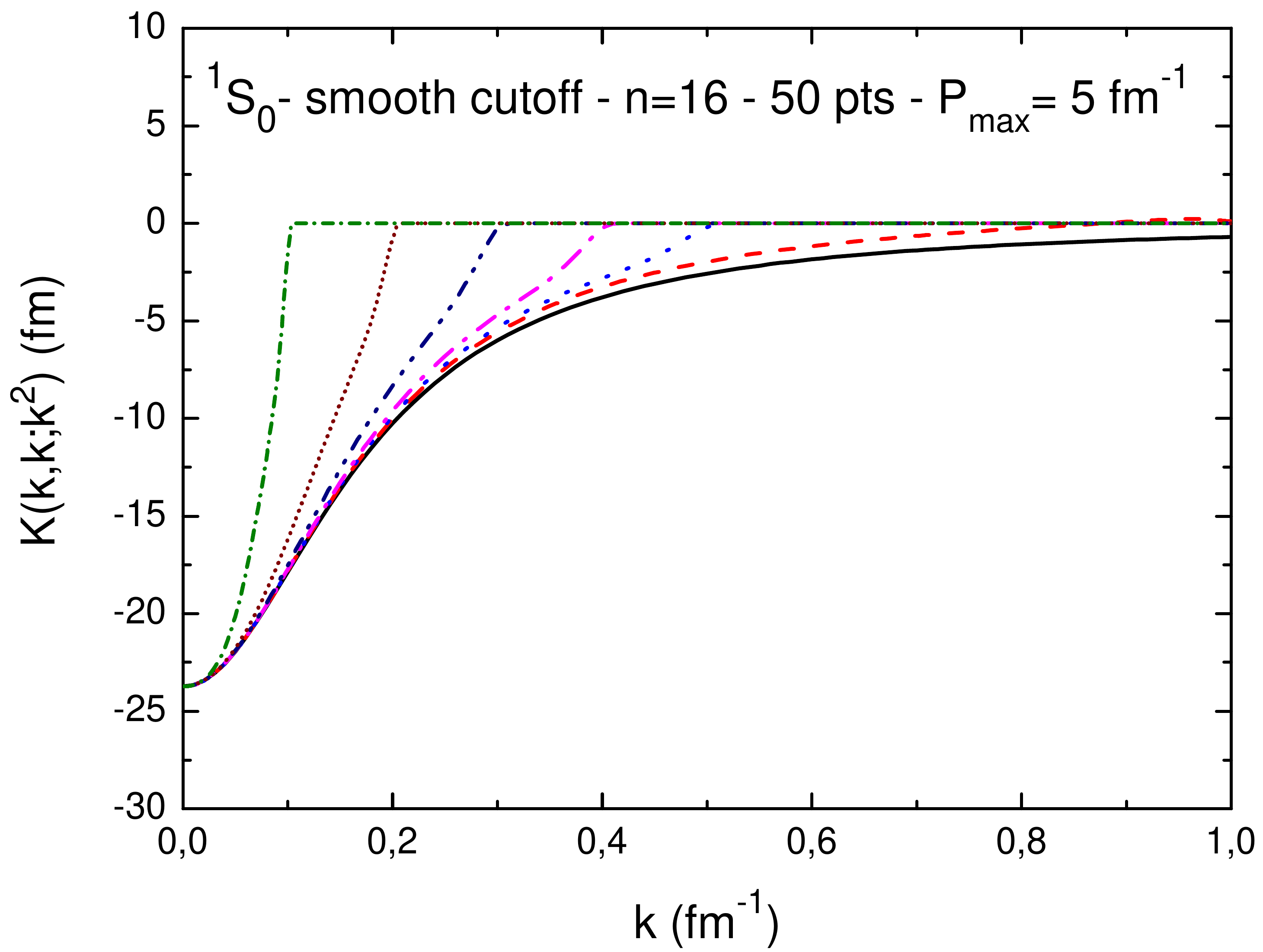}\hspace{0.5cm}
\includegraphics[width=7.0cm]{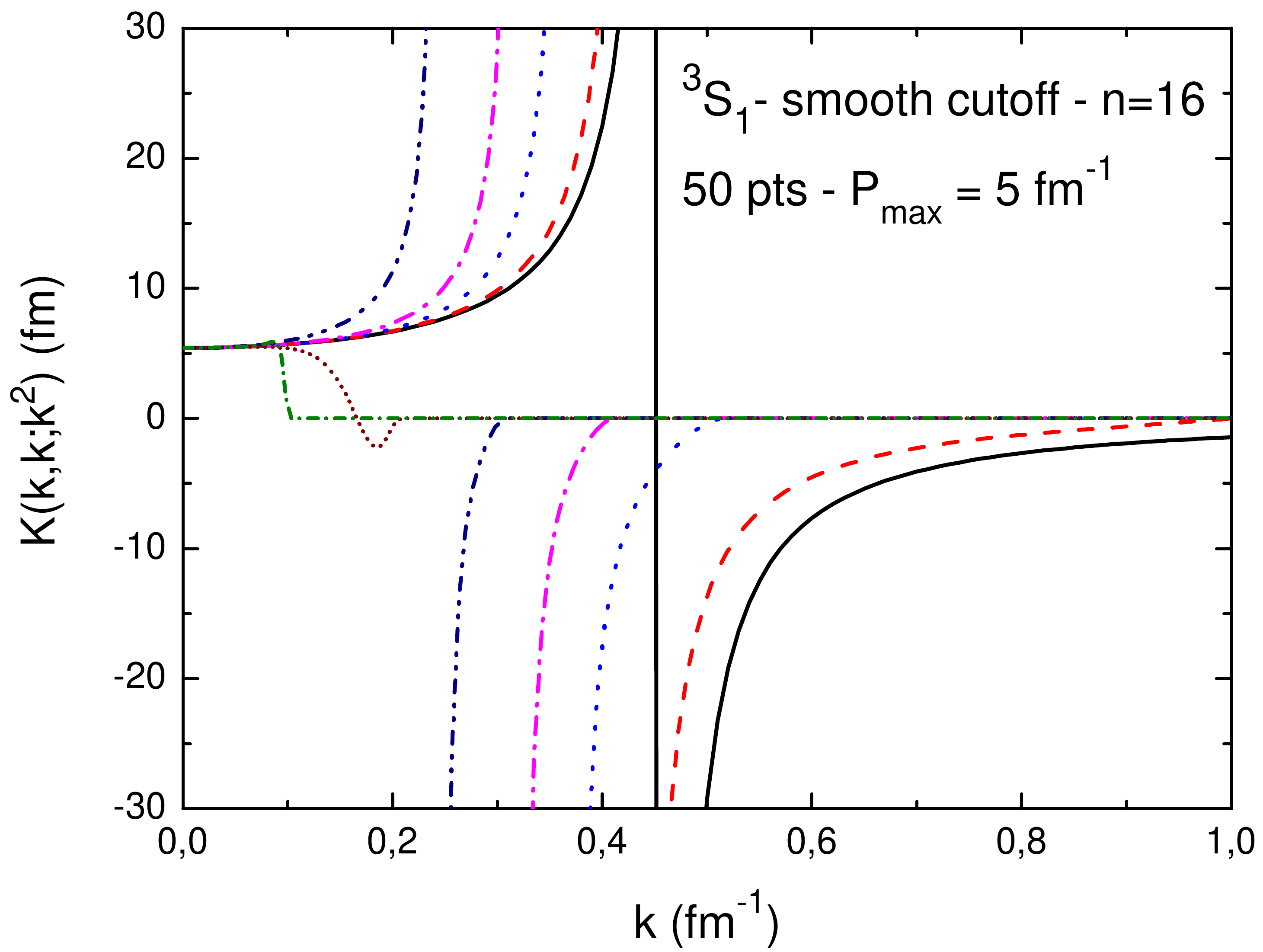}
\end{center}
\caption{On-shell $K$-matrix for the $^1S_0$ channel and the $^3S_1$
channel $NN$ potentials at $NLO$. Top panels: contact theory in the
continuum regulated by a sharp momentum cutoff; Bottom panels: contact theory
on a finite momentum grid with $N=50$ points regulated by a smooth exponential momentum cutoff. For
comparison, we also show the on-shell $K$-matrix in the ERE to order ${\cal O}(k^2)$.}
\label{fig:6}
\end{figure*}

In Fig.~\ref{fig:6} we show the on-shell $K$-matrix for the $^1S_0$
channel and the $^3S_1$ channel $NN$ potentials at $NLO$ as a function
of the momentum $k$ for several values of the cutoff scale $\Lambda$,
both in the contact theory with a sharp cutoff and in the contact
theory with a smooth cutoff. As one can observe, in both cases the
on-shell $K$-matrix changes with $\Lambda$, approaching the on-shell
$K$-matrix in the ERE to order ${\cal O}(k^2)$ as $\Lambda$
increases. One should also note that for the $^3S_1$ channel the pole
is shifted to the left as $\Lambda$ decreases and vanishes when
$\Lambda < 0.3~{\rm fm^{-1}}$, which is the threshold scale below which we
do not observe a bound-state in the $^3S_1$ channel.

\section{Explicit renormalization}

\subsection{The toy-model: separable gaussian potential}
\label{toymodel}

In the applications of the SRG method to Nuclear Physics, realistic potentials which fit $NN$ data up to the pion-production threshold ($\sqrt{m_\pi M_N} \sim 400~{\rm MeV}$) are usually taken as the initial $NN$ {\it bare} interaction. Due to the short-range repulsive core such potentials exhibit a long high-momentum tail, requiring the use of a large value for the auxiliary momentum cutoff $P_{\rm max}$ which complicates the numerical convergence when solving the SRG flow-equations \footnote{For the Argonne v18~\cite{Wiringa:1994wb} and the Nijmegen II~\cite{Stoks:1994wp} realistic potentials, for example, one needs $P_{\rm max} \sim 30~{\rm fm}^{-1}$ in order to ensure that the potential matrix-elements have vanished.}. For illustration purposes, in this work we consider as the $NN$ {\it bare} interaction a simple separable gaussian potential toy-model for the $S$-waves, given by
\begin{equation}
V(p, p') = C~g_L(p)~g_L(p') \; ,
 \label{gaupot}
\end{equation}
\noindent
with a gaussian form factor $g_L(p)=e^{-p^2/L^2}$. The potential parameters $(C, L)$ are determined
by fitting the experimental values of the ERE parameters, $a_0$ and $r_e$. For the toy-model potential in the continuum we obtain $C^{^1S_0}_{N=\infty} = -1.915884~{\rm fm}$ and $(1/L^2)^{1S0}_{N=\infty} =0.691269 ~{\rm fm^2}$ for the $^1S_0$ channel and $C^{^3S_1}_{N=\infty} = -2.300641~{\rm fm}$ and $(1/L^2)^{3S1}_{N=\infty} = 0.415154~{\rm fm^2}$ for the $^3S_1$ channel.

\begin{figure*}[t]
\begin{center}
\includegraphics[width=7.0cm]{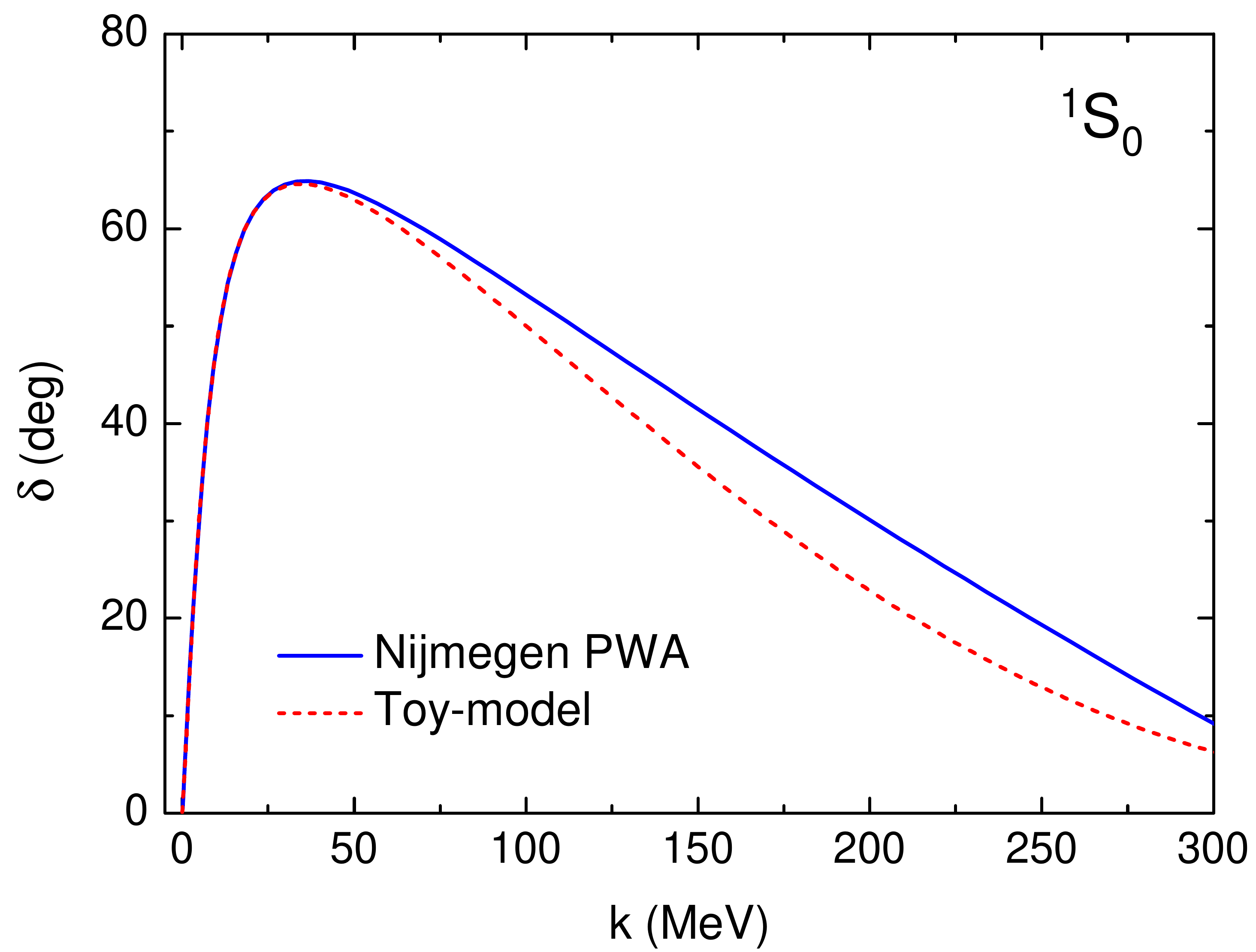}\hspace{0.5cm}
\includegraphics[width=7.0cm]{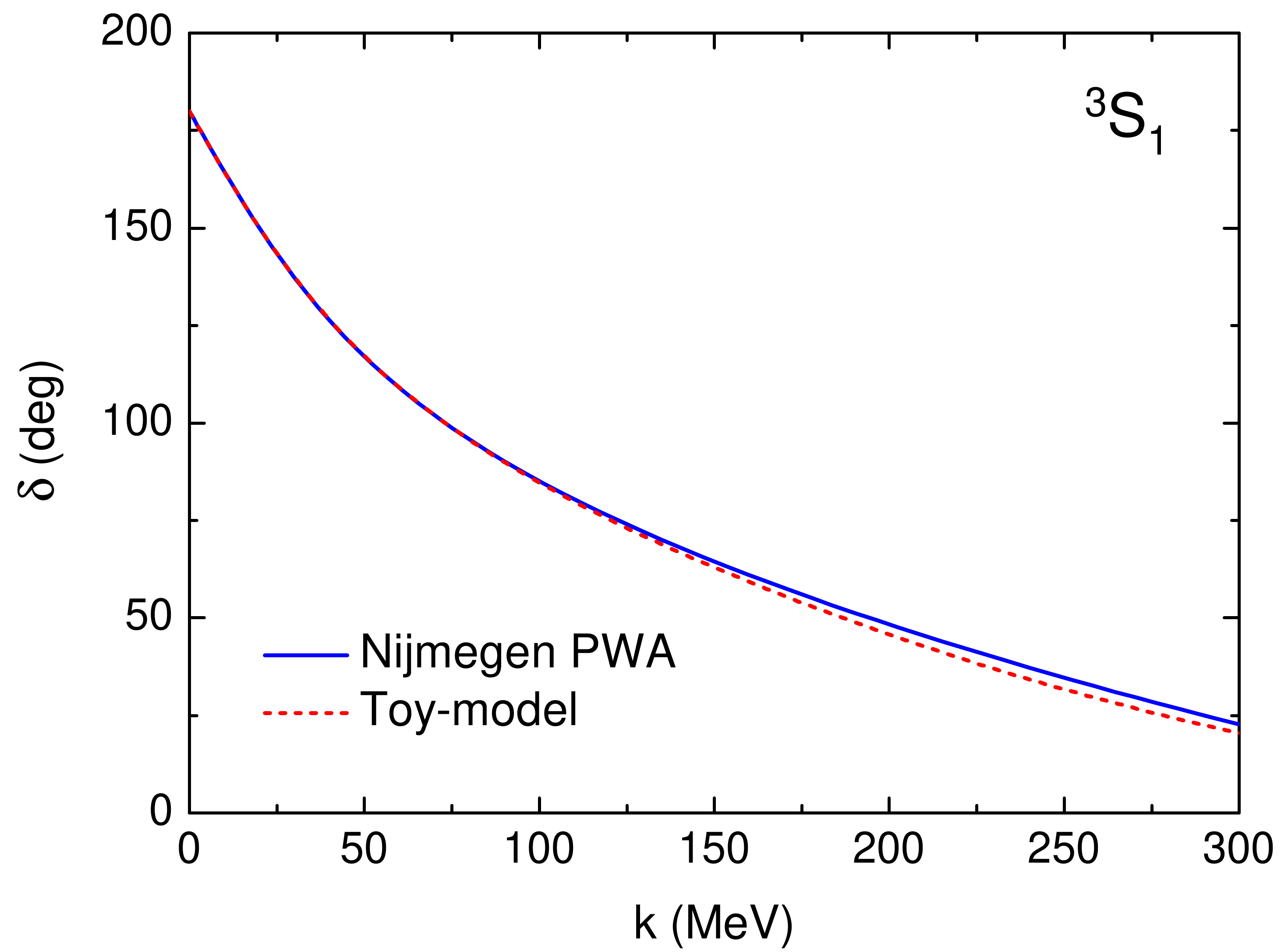}
\end{center}
\caption{$^1S_0$ and $^3S_1$ phase-shifts for the toy-model separable gaussian
potential compared to the results from the Nijmegen PWA \cite{Stoks:1993tb}.}
\label{fig:7}
\end{figure*}

In the case of a separable potential, it is straightforward to determine the phase-shifts using the {\it ansatz} for the $T$-matrix given by
\begin{eqnarray}
T(p,p';k^2) = g_L(p) ~t(k)~ g_L(p') \; ,
\end{eqnarray}
where $t(k)$ is called the reduced on-shell $T$-matrix. This leads to the relation
\begin{eqnarray}
 k \cot \delta(k) = - \frac{1}{V(k,k)} \left[1- \frac{2}{\pi}\;
{\cal P}\int_0^\infty dq~q^2 \frac{1}{k^2-q^2} V(q,q) \right] \; .
\label{phasetoy}
\end{eqnarray}
As shown in Fig.~\ref{fig:7}, the $^1S_0$ and $^3S_1$ phase-shifts computed from
Eq.~(\ref{phasetoy}) for the toy-model potential qualitatively
resemble the results from the much used 1993 Nijmegen partial-wave
analysis (PWA) \cite{Stoks:1993tb} or the more recent 2013
upgrades~\cite{Perez:2013mwa,Perez:2013jpa,Perez:2013oba,Perez:2014yla}. Moreover,
the on-shell $T$-matrix for the $^3S_1$ channel toy-model potential
has a pole located at $k= i~ 0.2314~{\rm fm^{-1}}$, corresponding to a
satisfactory deuteron binding-energy $B_d=2.22~{\rm MeV}$. The
deuteron wave function is obtained as
\begin{eqnarray}
\psi_d(p) = \frac{1}{\sqrt{\mathcal{N}}}  \frac{g_L(p)}{p^2+\gamma^2} \; .
\end{eqnarray}
The normalization condition,
\begin{eqnarray}
\frac2{\pi} \int_0^\infty dp~p^2 ~ |\psi_d(p)|^2 = 1 \; ,
\end{eqnarray}
implies that $\frac{1}{\sqrt{\mathcal{N}}} = 0.8548 ~{\rm fm}^{1/2}$. The matter radius is defined as
\begin{eqnarray}
\frac{1}{2\pi}  \int_0^\infty dp ~ |\psi_d(p)|^2 ~=~ r_m^2 \; ,
\end{eqnarray}
which gives $r_m~=~1.938~{\rm fm}$.

\subsection{SRG evolution of the toy-model potential}

The SRG, developed by Glazek and Wilson~\cite{Glazek:1993rc,Glazek:1994qc} and independently by
Wegner~\cite{Wegner200177} (for a review see e.g.~\cite{Kehrein:2006ti}), is a renormalization method based on a series of continuous unitary transformations that evolve hamiltonians with a cutoff on energy differences. Here we employ the formulation for the SRG developed by Wegner, which is based on a non-perturbative flow-equation that governs the unitary
evolution of a hamiltonian $H=T_{\rm rel}+V$ with a flow parameter $s$
that ranges from zero to infinity,
\begin{equation}
\frac{d H_s}{ds}=[\eta_s,H_s]\; ,
\end{equation}
\noindent
where $\eta_s=[G_s,H_s]$ is an anti-hermitian operator that generates
the unitary transformations. The flow parameter $s$ has dimensions of
$[{\rm energy}]^{-2}$ and in terms of a similarity cutoff $\lambda$
with dimension of momentum is given by the relation
$s=\lambda^{-4}$. The operator $G_s$ defines the anti-hermitian
generator $\eta_s$ and so specifies the flow of the hamiltonian. The
flow equation is to be solved with the boundary condition $H_s |_{_{s
    \rightarrow 0}} \equiv H_{0}$, where $H_{0}$ ($\equiv H_{\lambda=\infty}$) is the hamiltonian corresponding to the initial {\it bare} interaction.

Assuming that $T_{\rm rel}$ is independent of $s$, we obtain
\begin{equation}
\frac{d V_s}{ds}=[\eta_s,H_s]\; .
\end{equation}

We take the block-diagonal SRG generator ~\cite{Anderson:2008mu} given by

\begin{eqnarray}
 G_{s}=H_{s}^{\rm BD}
  \equiv \begin{pmatrix}
    PH_{s}P  & 0 \\ \\
    0  & QH_{s}Q
  \end{pmatrix} \; ,
\end{eqnarray}
\noindent
where $P$ and $Q=1-P$ are projection operators. In a partial-wave
relative momentum-space basis, the projection operators are determined
in terms of a momentum cutoff scale $\Lambda$ that divides the
momentum space into a low-momentum $P$-space ($p < \Lambda$) and a
high-momentum $Q$-space ($p > \Lambda$). Here we define the projection
operators just as step functions,
\begin{eqnarray}
P \equiv \theta(\Lambda - p) ; \; Q \equiv \theta(p -\Lambda) \; .
\end{eqnarray}

The full hamiltonian $H_s$ can be written as,

\begin{eqnarray}
 H_{s}
  \equiv \begin{pmatrix}
    PH_{s}P  & PH_{s}Q \\ \\
    QH_{s}P  & QH_{s}Q
  \end{pmatrix} \;.
\end{eqnarray}

The anti-hermitian operator $\eta_s$ is then given by
\begin{eqnarray}
 \eta_s=[G_s,H_s] =
   \begin{pmatrix}
    0  & P\eta_{s}Q \\ \\
    Q\eta_{s}P  & 0
  \end{pmatrix} \;,
\end{eqnarray}
\noindent
where
\begin{eqnarray}
&&P\eta_{s}Q = PH_{s}PH_{s}Q - PH_{s}QH_{s}Q \\
&&Q\eta_{s}P = QH_{s}QH_{s}P - QH_{s}PH_{s}P \;.
\end{eqnarray}
Thus, the SRG flow-equation with the block-diagonal generator can be written in matrix-form as
\begin{eqnarray}
   \begin{pmatrix}
    \frac{d}{ds}[PV_{s}P]  & \frac{d}{ds}[PV_{s}Q]\\ \\
    \frac{d}{ds}[QV_{s}P]  & \frac{d}{ds}[QV_{s}Q]
  \end{pmatrix} =
  \begin{pmatrix}
    P\eta_{s}QH_{s}P-PH_{s}Q\eta_{s}P  & P\eta_{s}QH_{s}Q-PH_{s}P\eta_{s}Q \\ \\
    Q\eta_{s}PH_{s}P-QH_{s}Q\eta_{s}P  & Q\eta_{s}PH_{s}Q-QH_{s}P\eta_{s}Q
  \end{pmatrix} \;.
  \label{flowBD}
\end{eqnarray}

The potential $V_s$ can be written as,
\begin{eqnarray}
 V_{s}
  \equiv \begin{pmatrix}
    PV_{s}P  & PV_{s}Q \\ \\
    QV_{s}P  & QV_{s}Q
  \end{pmatrix} \;.
\end{eqnarray}
By choosing the block-diagonal generator, the matrix-elements inside
the off-diagonal blocks $PV_{s}Q$ and $QV_{s}P$ are suppressed as the
flow parameter $s$ increases (or as the similarity cutoff $\lambda$
decreases), such that the hamiltonian is driven to a block-diagonal
form. In the limit $\lambda \rightarrow 0$ the $P$-space and the
$Q$-space become completely decoupled,
\begin{eqnarray}
\lim_{\lambda \to 0} V_\lambda = P V_{\rm low \, k} P + Q V_{\rm high \, k} Q =
\begin{pmatrix}
V_{\rm low \, k}      & 0 \\ \\
    0  & V_{\rm high\, k}
  \end{pmatrix}
\end{eqnarray}
Thus, while unitarity implies
$\delta_\lambda(p)= \delta(p)$ for any $\lambda$ one has
\begin{eqnarray}
\lim_{\lambda \to 0} \delta_\lambda (p) =
\delta_{\rm low \, k} (p)
+ \delta_{\rm high \, k} (p)
\end{eqnarray}
where $\delta_{\rm low \, k} (p)=\delta(p) \theta(\Lambda-p) $ and
$\delta_{\rm high \, k} (p)= \delta(p) \theta(p-\Lambda)$ are the
phase-shifts of the $V_{\rm low \, k} $ and $ V_{\rm high\, k} $
potentials respectively.

\begin{figure*}[t]
\begin{center}
\includegraphics[width=5.2cm]{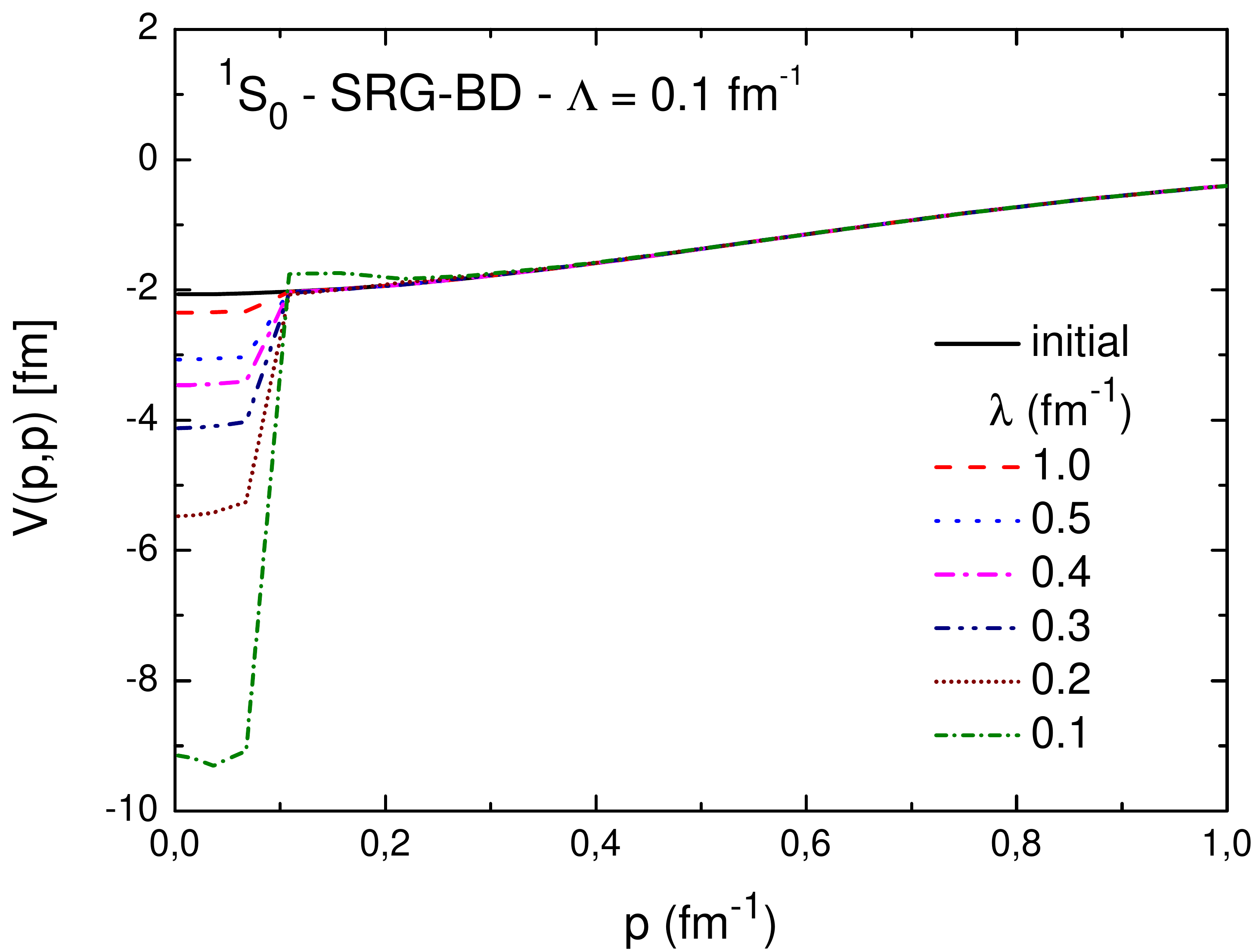}\hspace{0.2cm}
\includegraphics[width=5.2cm]{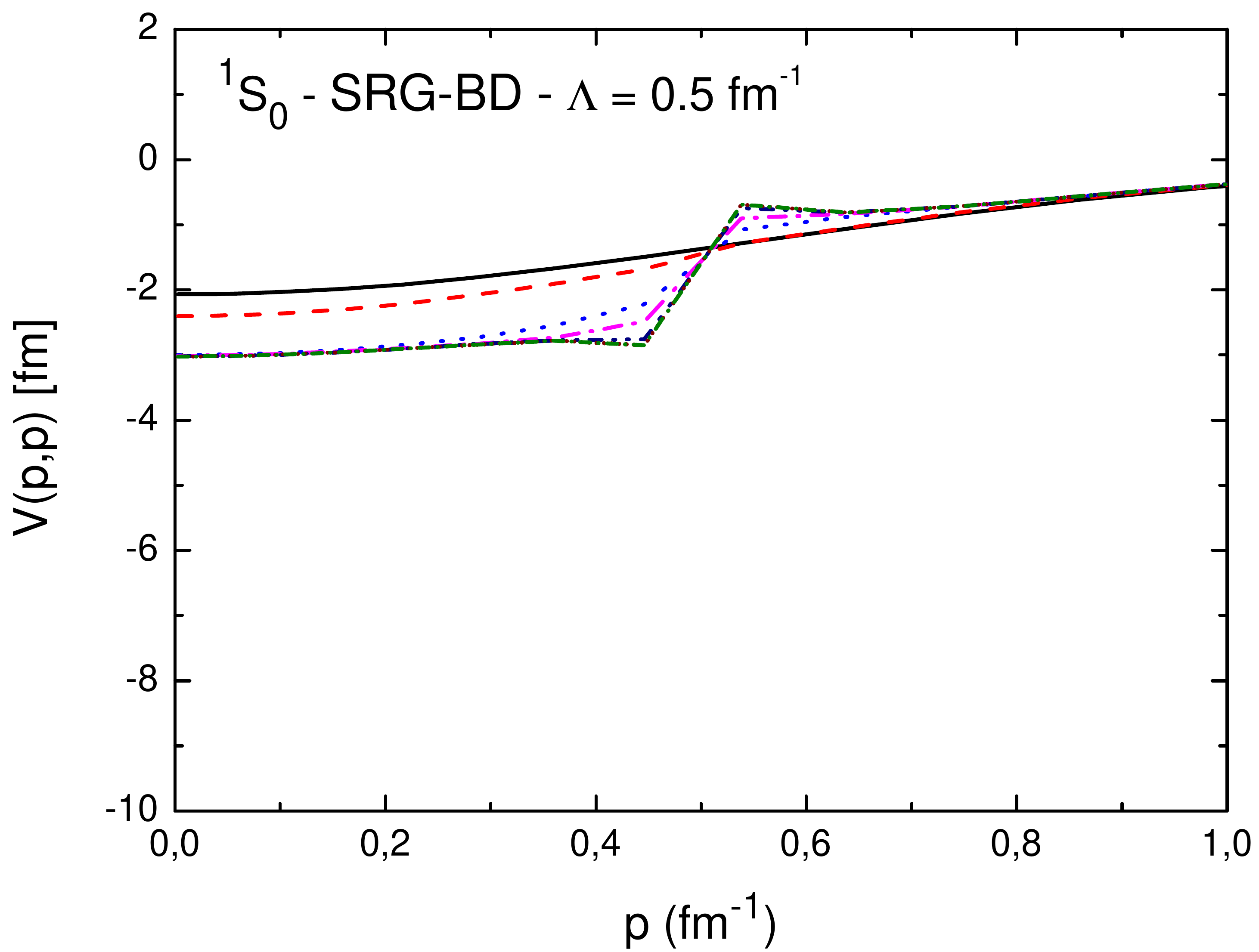}\hspace{0.2cm}
\includegraphics[width=5.2cm]{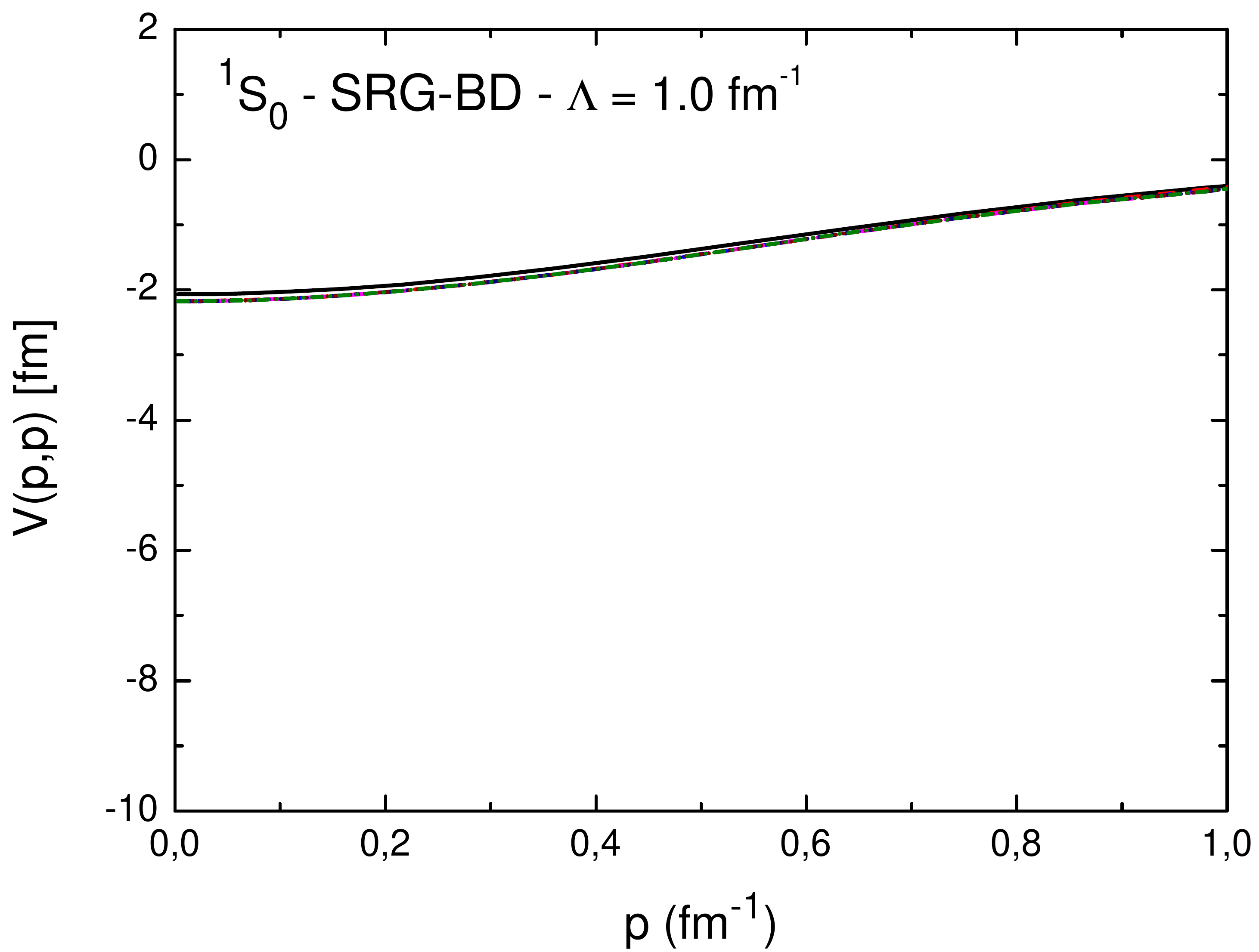}\\ \vspace{0.5cm}
\includegraphics[width=5.2cm]{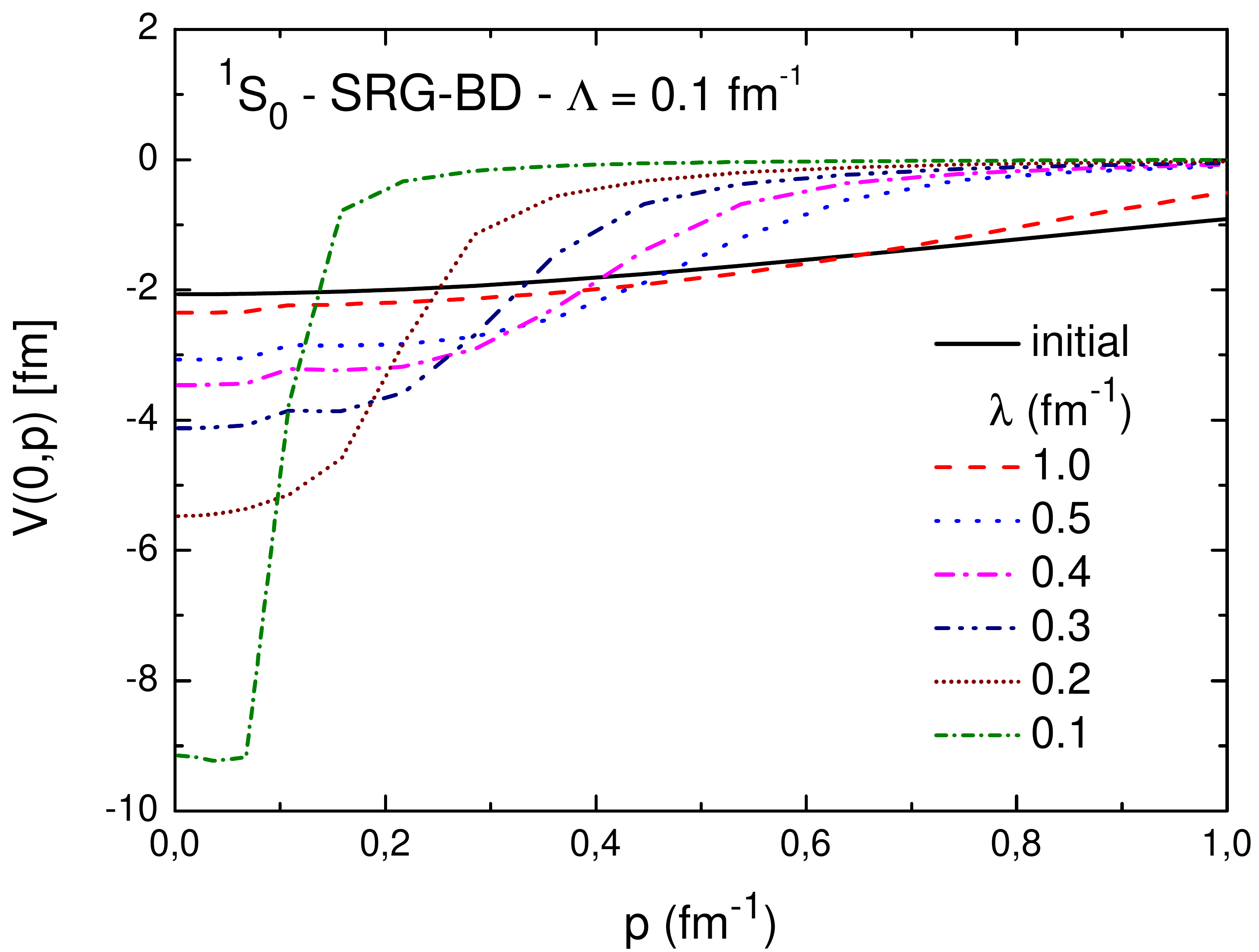}\hspace{0.2cm}
\includegraphics[width=5.2cm]{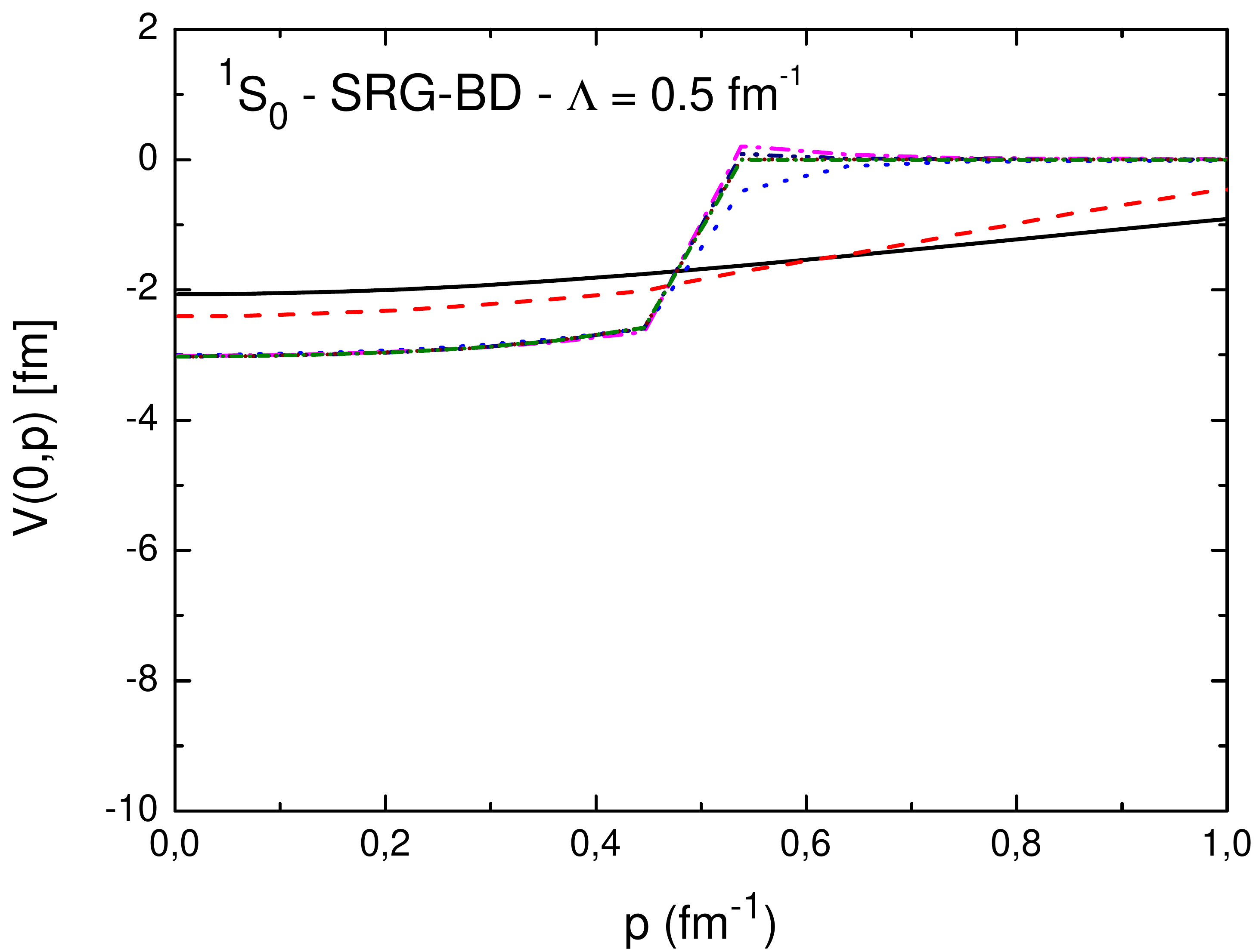}\hspace{0.2cm}
\includegraphics[width=5.2cm]{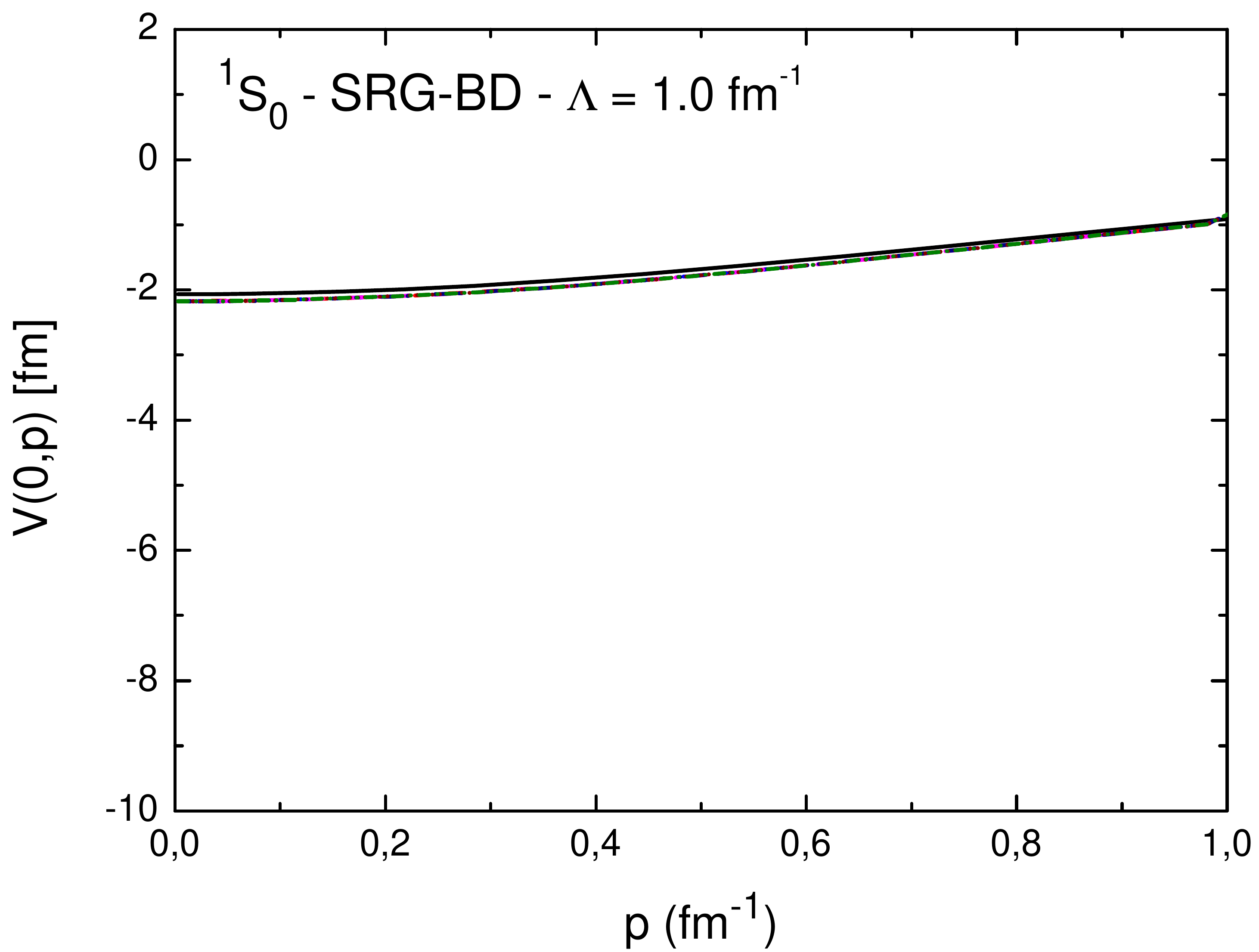}
\end{center}
\caption{Diagonal (upper panels) and fully off-diagonal (lower panels) matrix-elements for the $^1S_0$ channel toy-model potential on a gaussian grid (with $N=50$ momentum points and $P_{\rm max} = 5~{\rm fm}^{-1}$) evolved through the SRG transformation with the block-diagonal generator to different SRG cutoffs $\lambda$ for some values of the block-diagonal cutoff $\Lambda$. For comparison, we also show the diagonal matrix-elements for the initial ($\lambda \rightarrow \infty$) toy-model potential.}
\label{fig:8}
\end{figure*}

\begin{figure*}[t]
\begin{center}
\includegraphics[width=5.2cm]{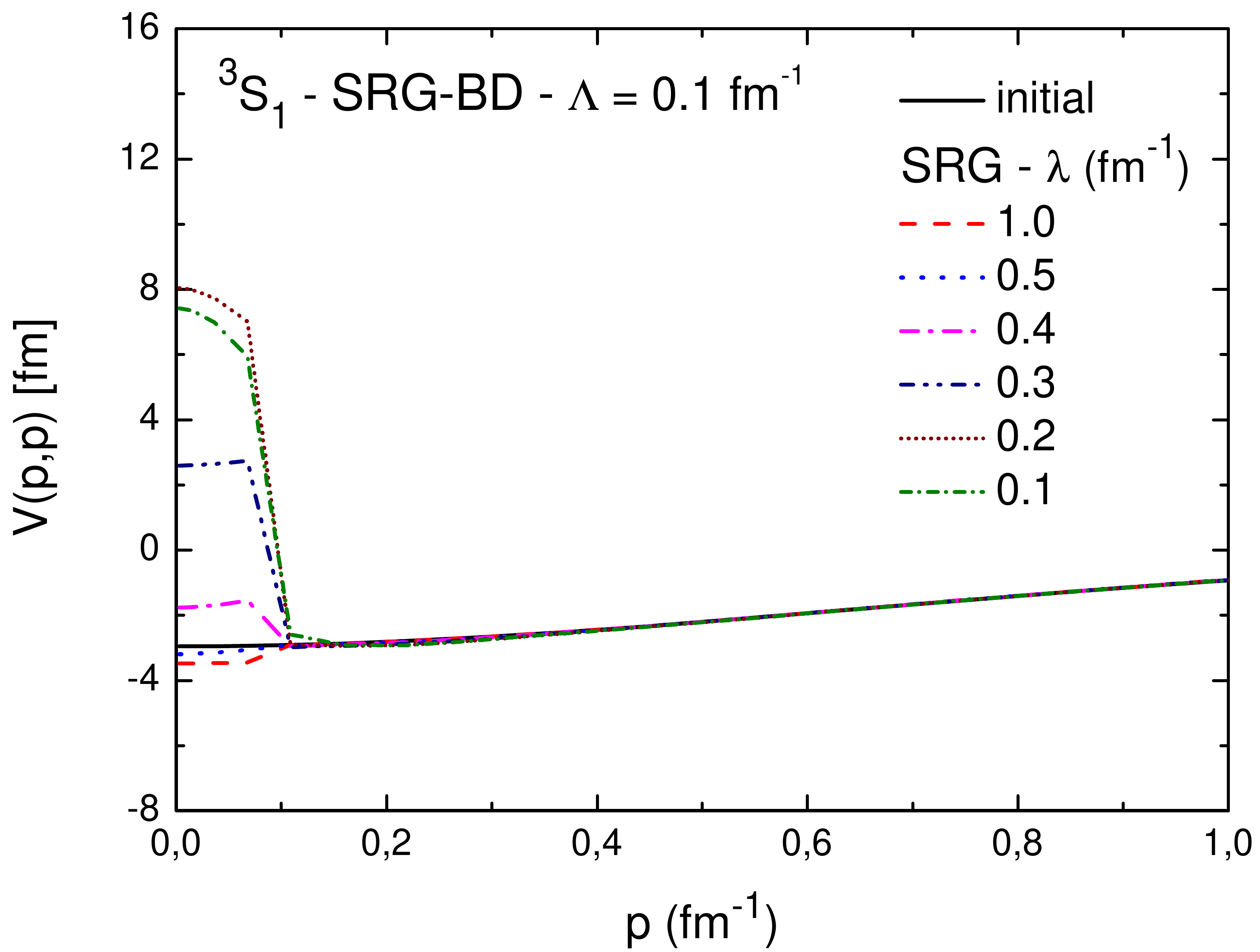}\hspace{0.2cm}
\includegraphics[width=5.2cm]{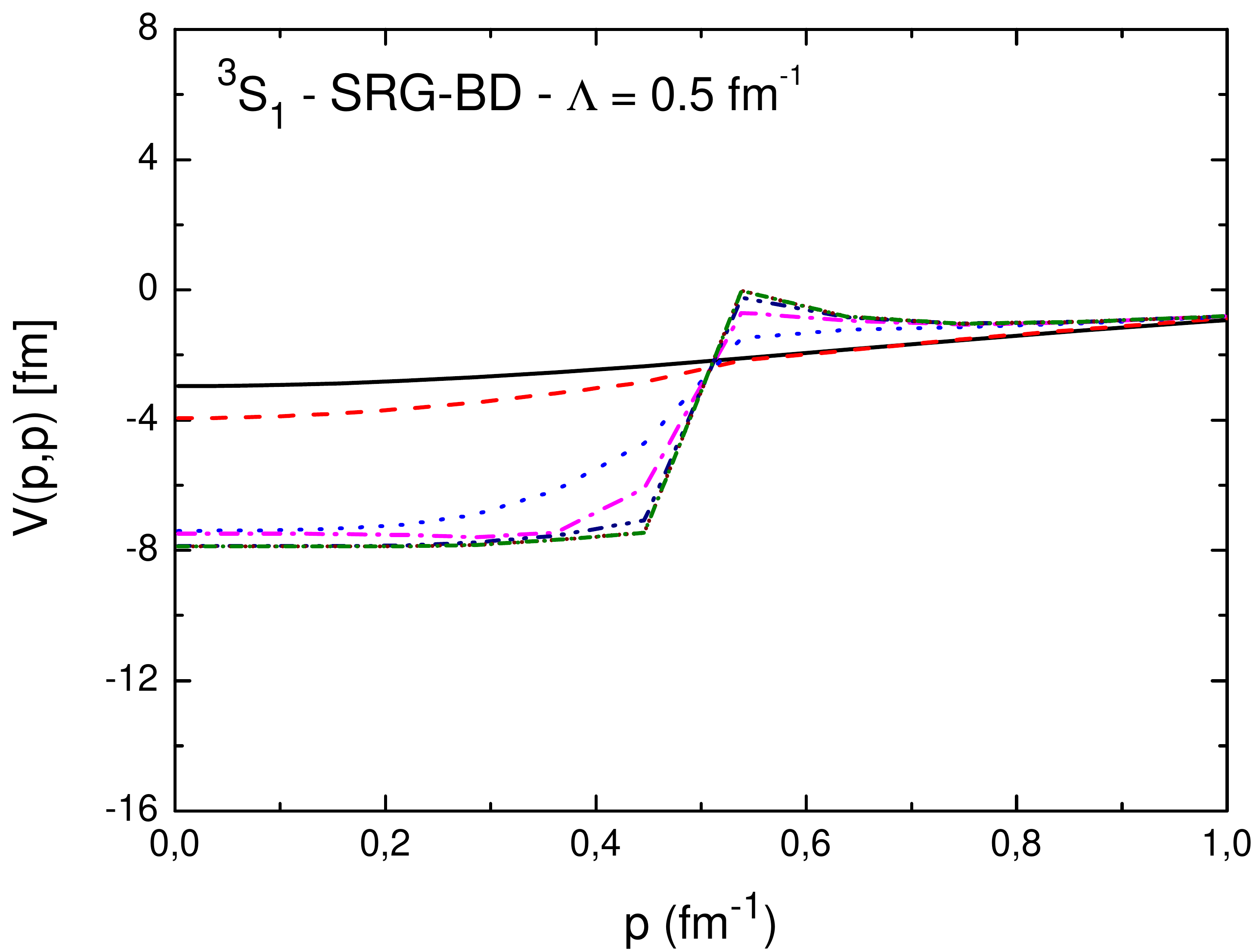}\hspace{0.2cm}
\includegraphics[width=5.2cm]{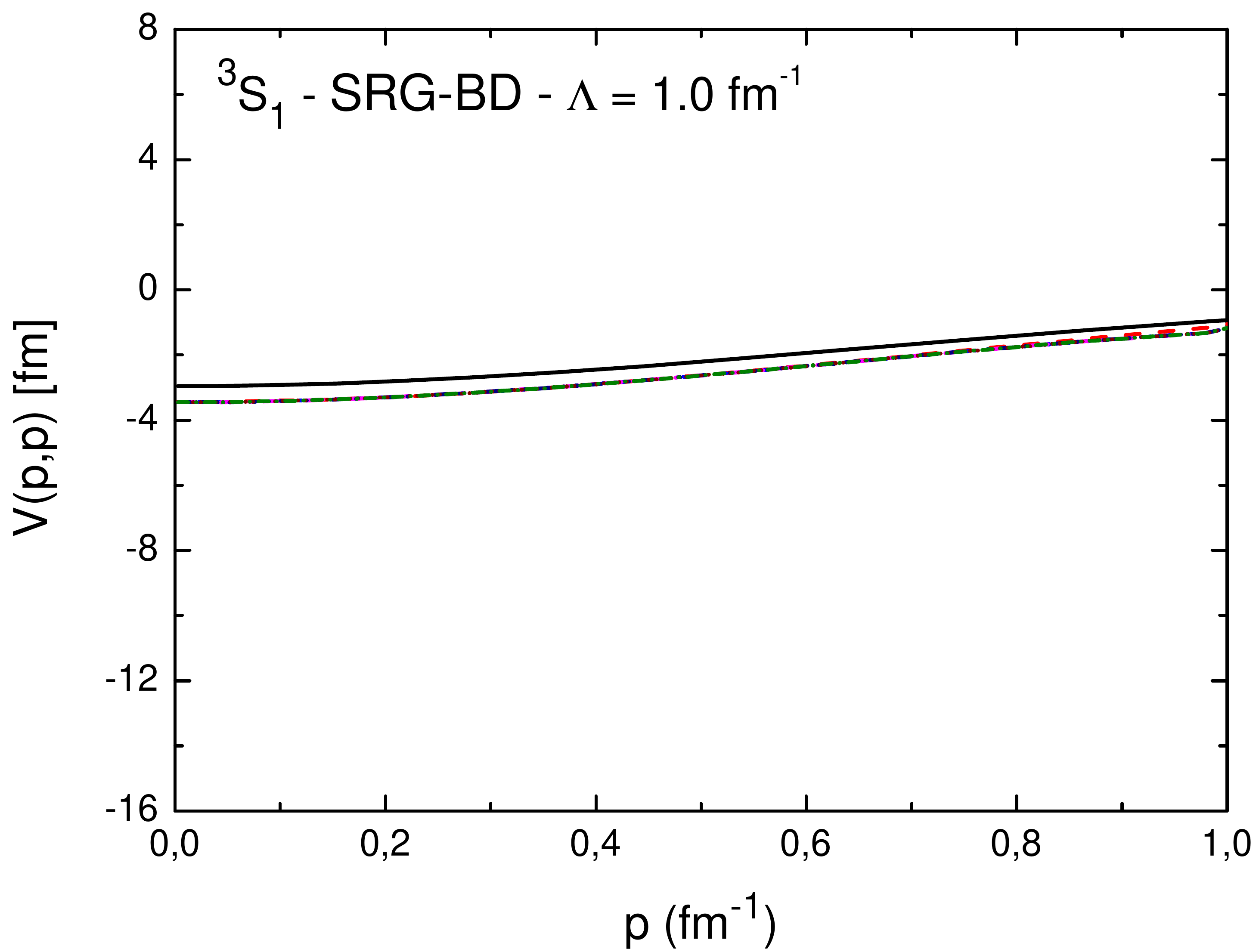}\\\vspace{0.5cm}
\includegraphics[width=5.2cm]{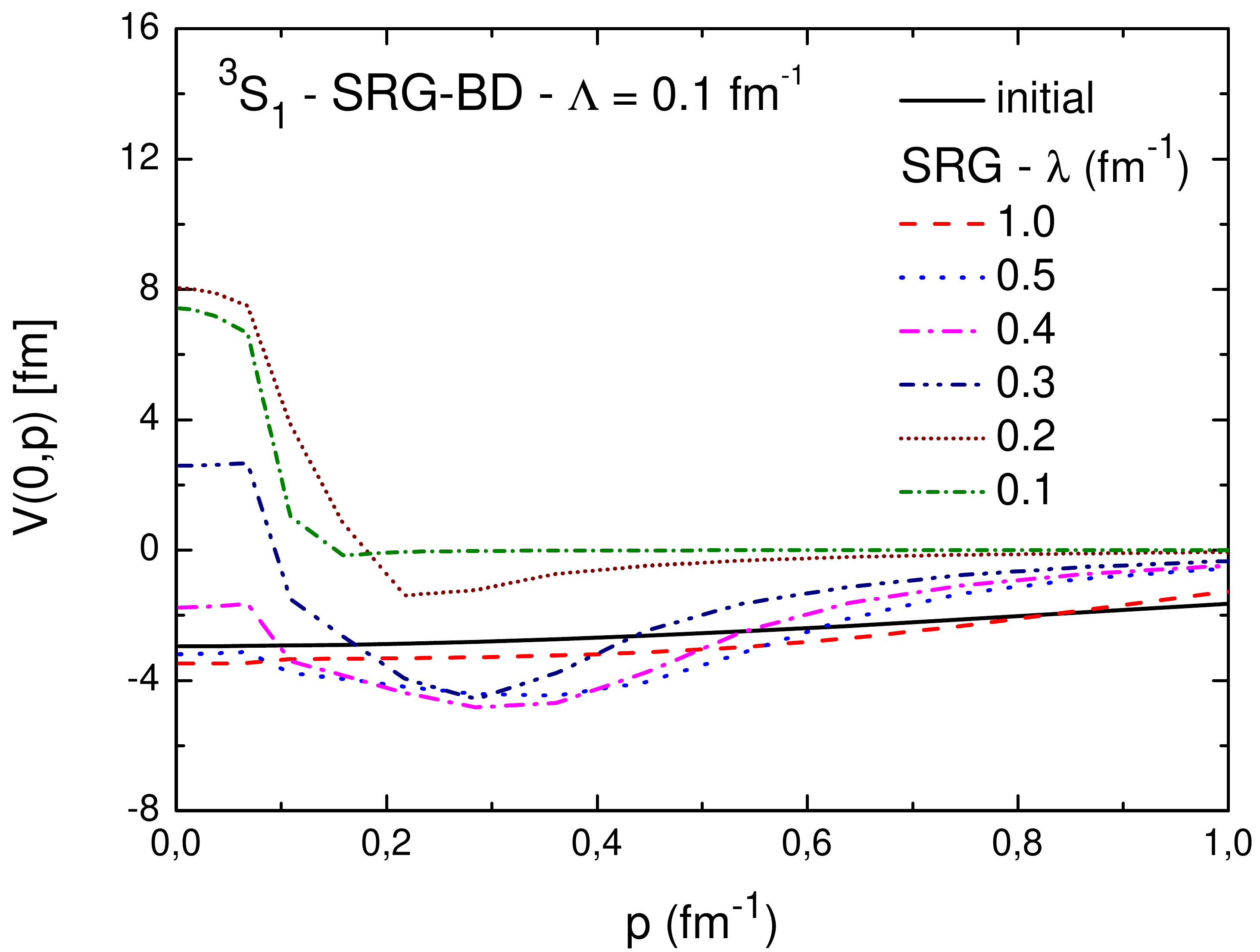}\hspace{0.2cm}
\includegraphics[width=5.2cm]{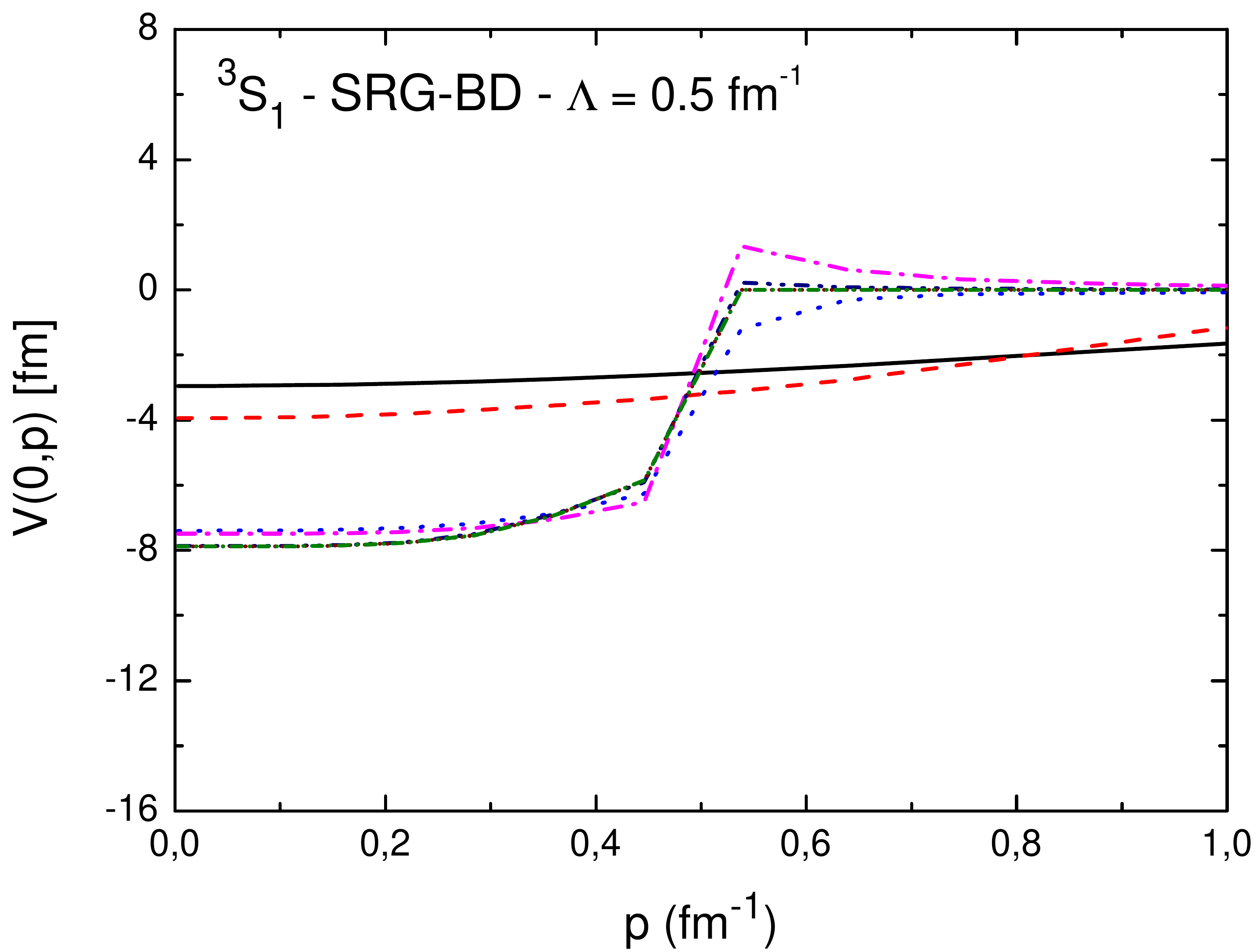}\hspace{0.2cm}
\includegraphics[width=5.2cm]{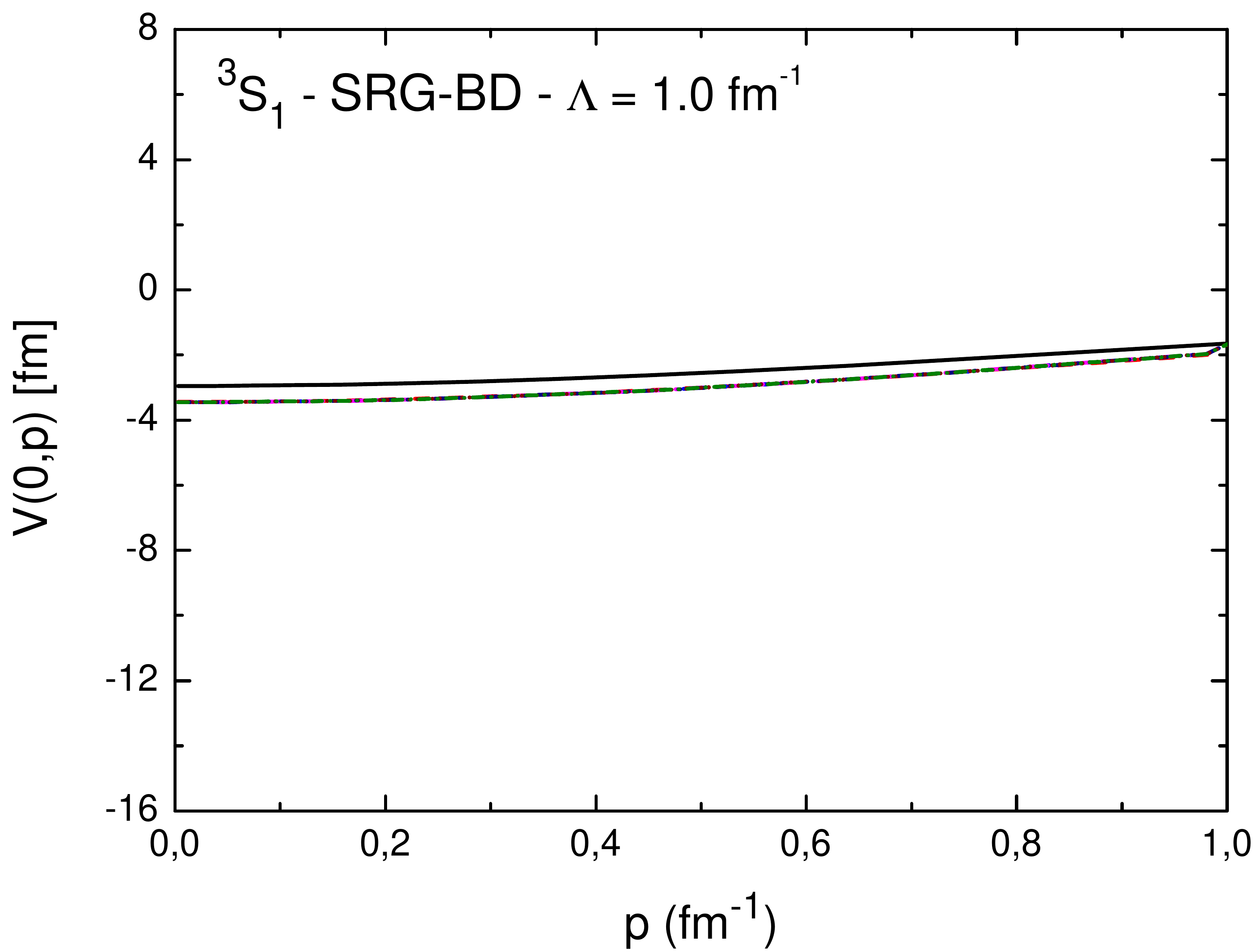}
\end{center}
\caption{Diagonal (upper panels) and fully off-diagonal (lower panels) matrix-elements for the $^3S_1$ channel toy-model potential on a gaussian grid (with $N=50$ momentum points and $P_{\rm max} = 5~{\rm fm}^{-1}$) evolved through the SRG transformation with the block-diagonal generator to different SRG cutoffs $\lambda$ for some values of the block-diagonal cutoff $\Lambda$. For comparison, we also show the diagonal matrix-elements for the initial ($\lambda \rightarrow \infty$) toy-model potential.}
\label{fig:9}
\end{figure*}

Here, we consider the SRG evolution of the toy-model separable
gaussian potential described in section \ref{toymodel}. The parameters
$(C, L)$ are determined from the numerical solution of the LS equation
for the $K$-matrix by fitting the experimental values of the ERE
parameters, $a_0$ and $r_e$. For a gaussian grid with $N=50$ momentum
points and $P_{\rm max}=5~{\rm fm}^{-1}$ we obtain
$C^{^1S_0}_{N=50}=-2.069625~{\rm fm}$ and
$(1/L^2)^{1S0}_{N=50}=0.818149~{\rm fm^2}$ for the $^1S_0$ channel and
$C^{^3S_1}_{N=50} = -2.952983~{\rm fm}$ and $(1/L^2)^{3S1}_{N=50}
=0.581541~{\rm fm^2}$ for the $^3S_1$ channel. One should note that
these values are slightly different from those obtained for the
toy-model potential in the continuum ($N=\infty$) with the same values
of $a_0$ and $r_e$.  We solve Eq.(\ref{flowBD}) numerically, obtaining
an exact (non-perturbative) solution for the SRG-evolved toy-model
potential (apart from numerical errors). The discretization of the
relative momentum space on a grid with $N$ points leads to a system of
$4N^2$ non-linear first-order coupled differential equations which is
solved by using a variable-step fifth-order Runge-Kutta algorithm. In
Figs.~\ref{fig:8} and \ref{fig:9} we show the SRG evolution of the
$^1S_0$ channel and the $^3S_1$ channel toy-model potentials. As one
can observe, the diagonal matrix-elements inside the $P$-space ($p <
\Lambda$) change significantly as the SRG cutoff $\lambda$
decreases. Such a change becomes smaller as the block-diagonal cutoff
$\Lambda$ increases. On the other hand, the diagonal matrix-elements
inside the $Q$-space ($p > \Lambda$) remain practically unchanged
(apart from a small change for momenta $p$ near $\Lambda$). The
evolution of the fully off-diagonal matrix-elements inside the
$P$-space is similar. As expected, the fully off-diagonal
matrix-elements inside the $PV_{s}Q$ block (and the $QV_{s}P$ block)
go to zero as the similarity cutoff $\lambda$ decreases. One should
also note that in the case of the $^3S_1$ channel potential, the
evolution follows a different pattern for $\Lambda \leq 0.3~{\rm
  fm^{-1}}$, which is the scale where the deuteron bound-state
appears.

\begin{figure*}[t]
\begin{center}
\includegraphics[width=7cm]{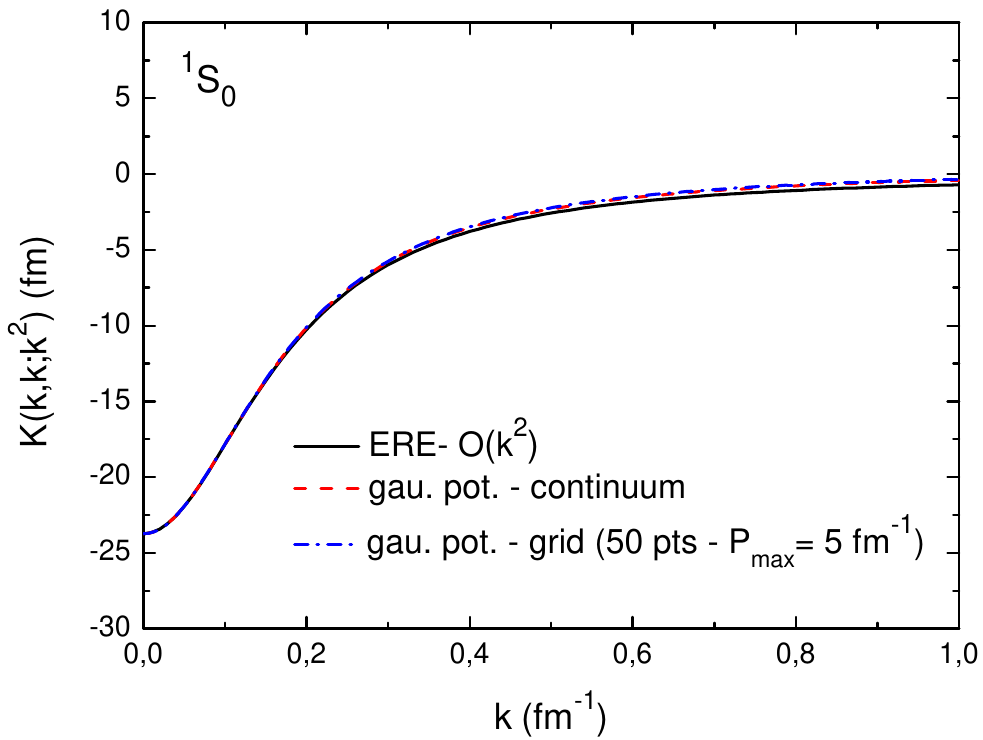}\hspace{0.5cm}
\includegraphics[width=7cm]{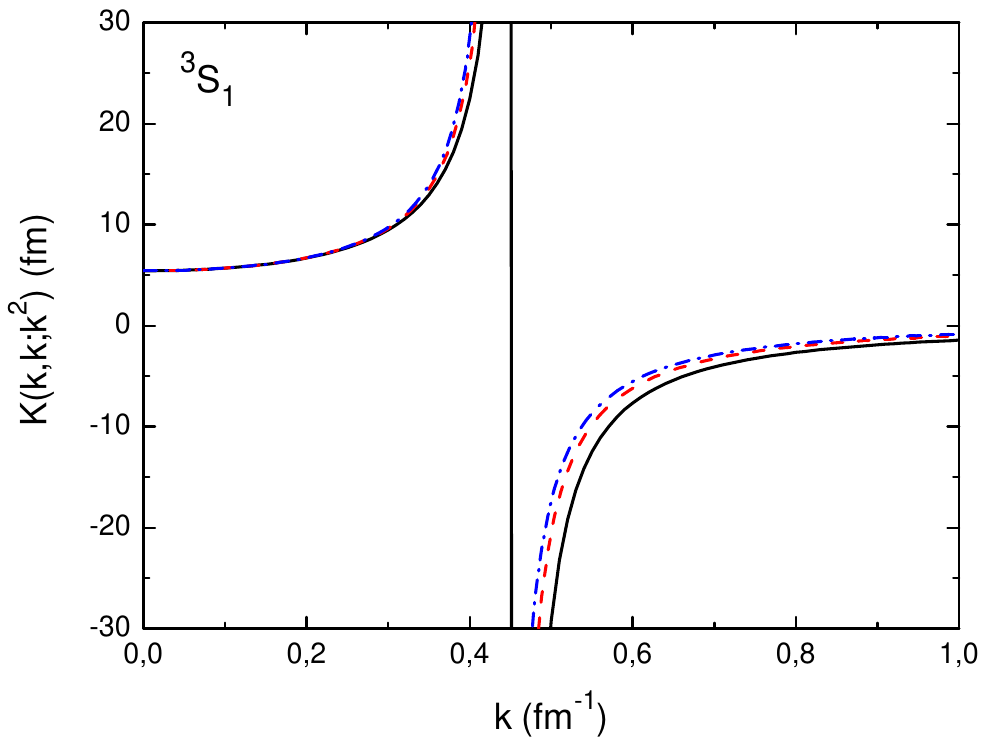}
\end{center}
\caption{On-shell $K$-matrices for the $^1S_0$ channel and the $^3S_1$ channel toy-model potentials in the continuum and on a gaussian grid (with $N=50$ momentum points and $P_{\rm max}=5~{\rm fm}^{-1}$). For comparison, we also show the on-shell $K$-matrices in the ERE to order ${\cal O}(k^2)$.}
\label{fig:12}
\end{figure*}

In Fig.~\ref{fig:12}, we show the on-shell $K$-matrices for the $^1S_0$ channel and the $^3S_1$ channel toy-model potentials in the continuum and on a gaussian grid with $N=50$ momentum points and $P_{\rm max}=5~{\rm fm}^{-1}$, compared with the corresponding on-shell $K$-matrices obtained from the ERE to order ${\cal O}(k^2)$. We have checked through explicit calculations that the on-shell $K$-matrices for the SRG-evolved toy-model potentials remain invariant under the change of both the similarity cutoff $\lambda$ and the block-diagonal cutoff $\Lambda$, as expected from the unitarity of the SRG transformation.

\begin{figure*}[ht]
\begin{center}
\includegraphics[width=5cm]{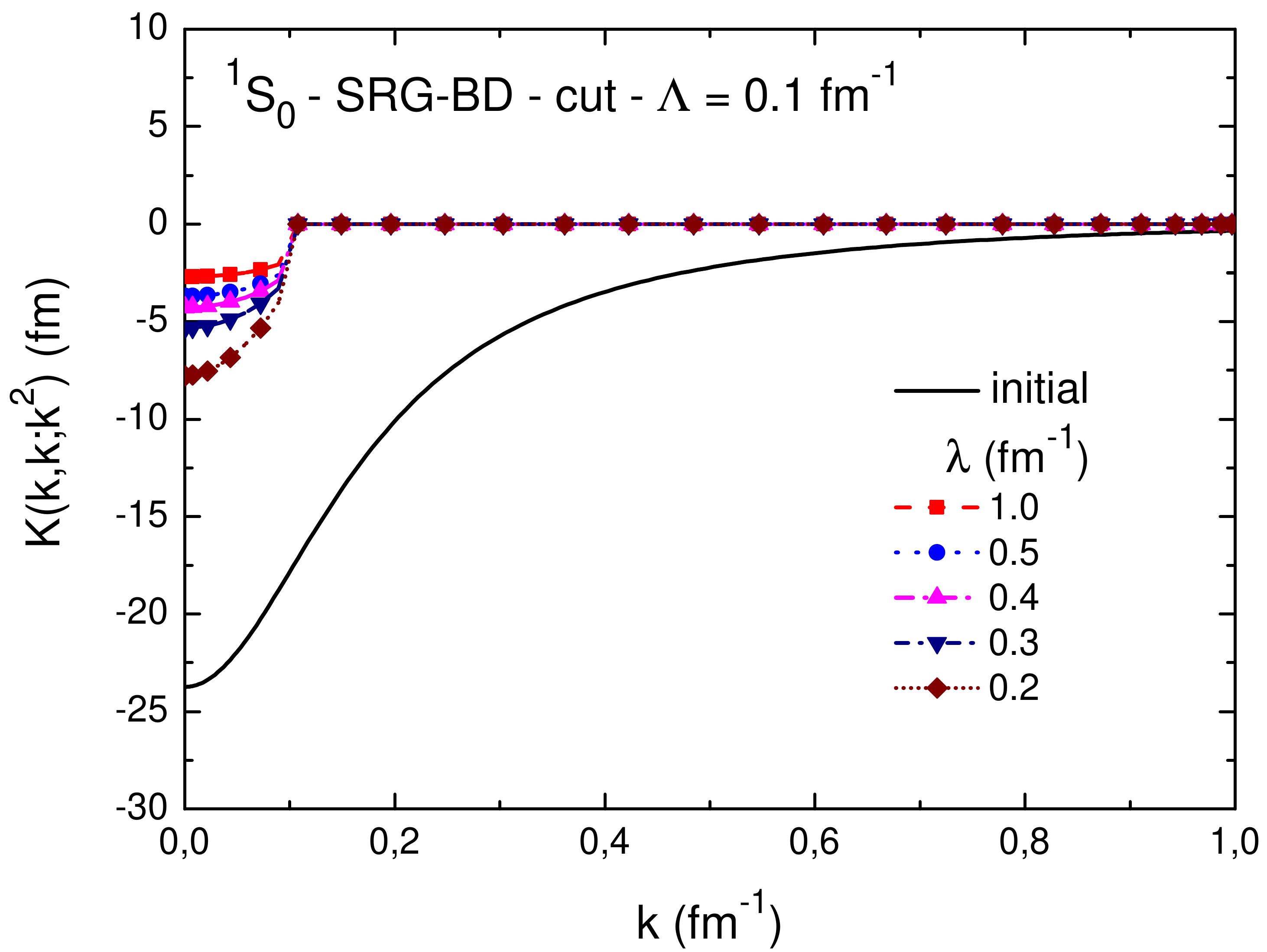}\hspace{0.5cm}
\includegraphics[width=5cm]{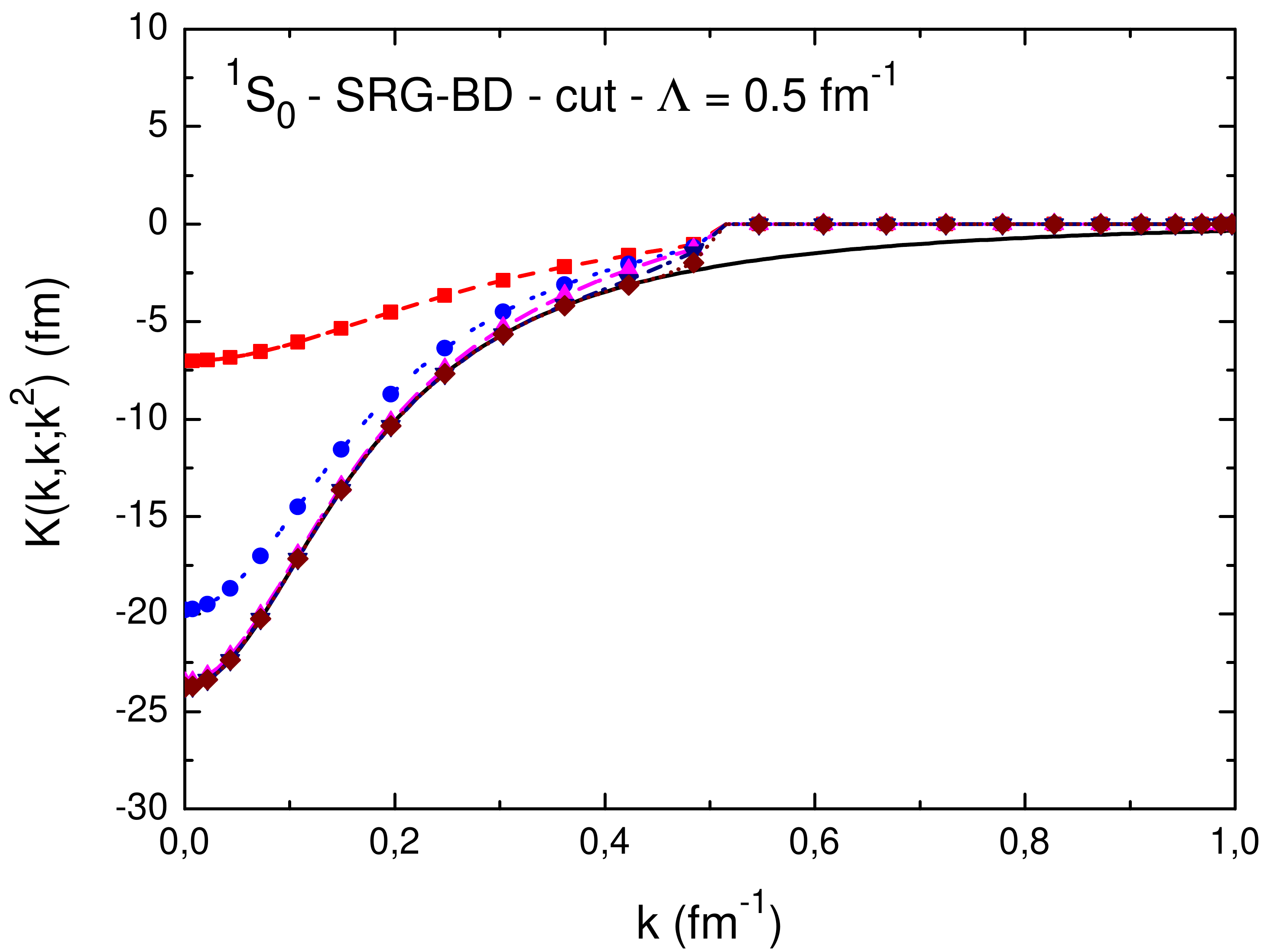}\hspace{0.5cm}
\includegraphics[width=5cm]{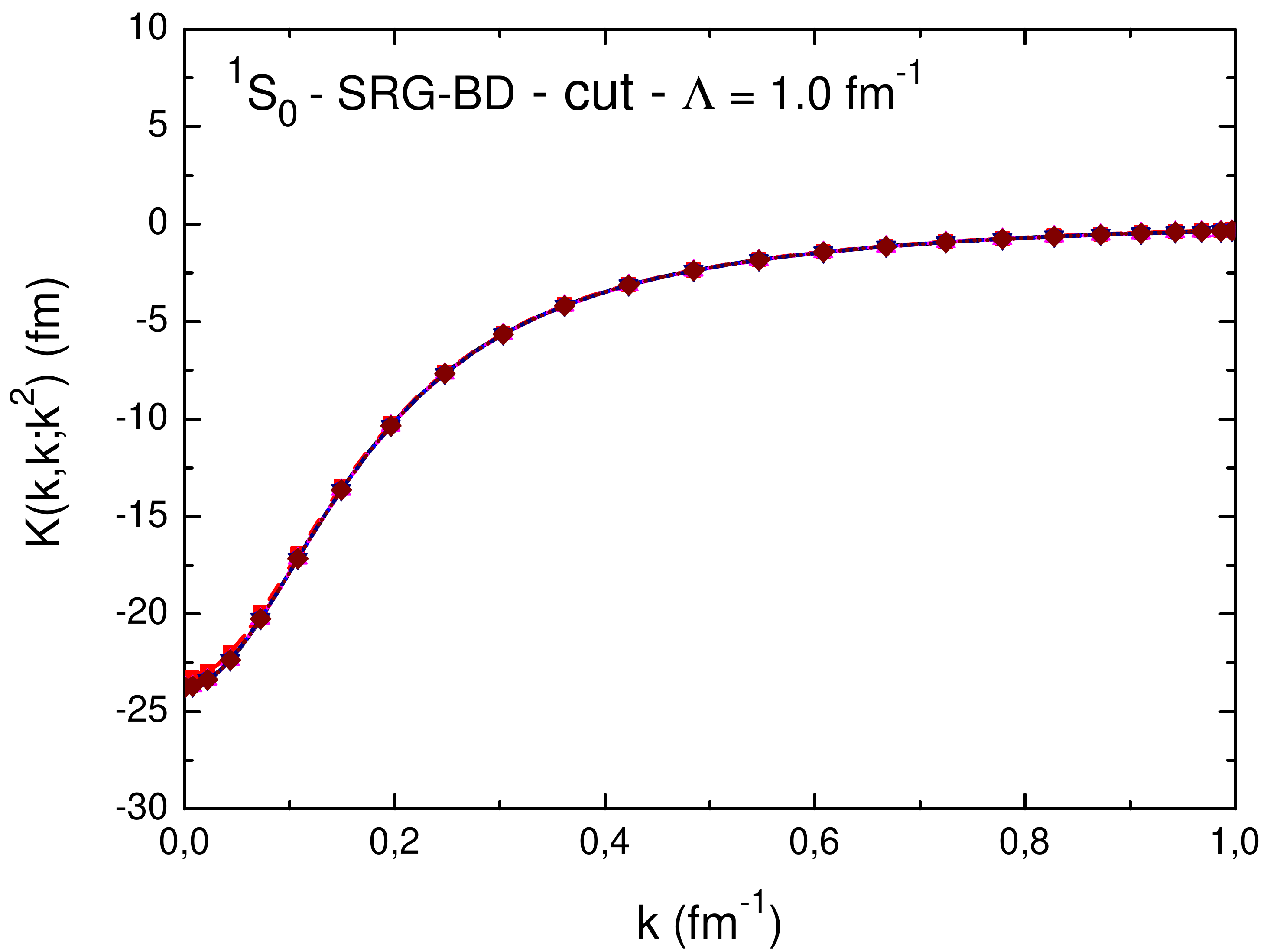}\\\vspace{0.5cm}
\includegraphics[width=5cm]{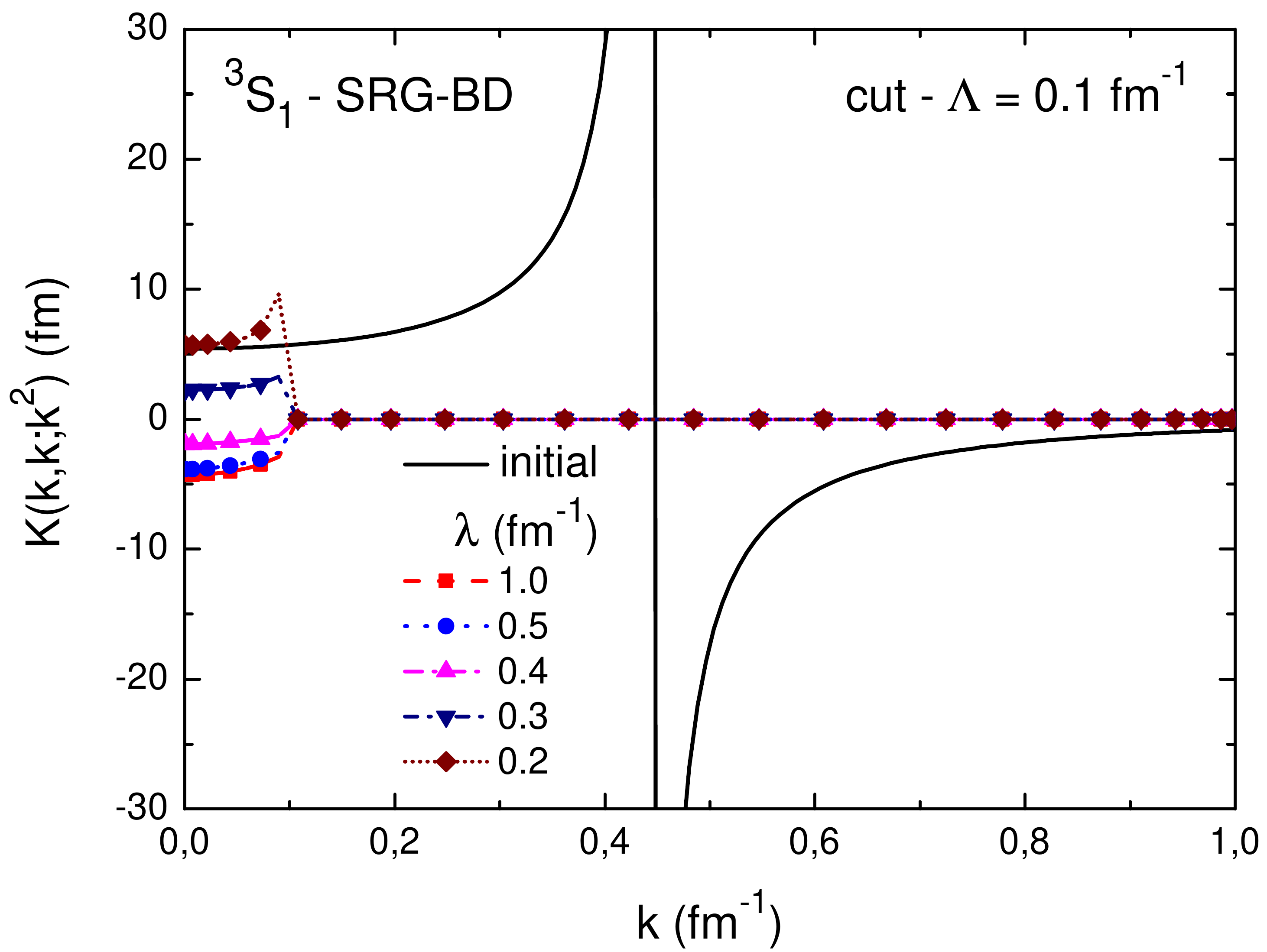}\hspace{0.5cm}
\includegraphics[width=5cm]{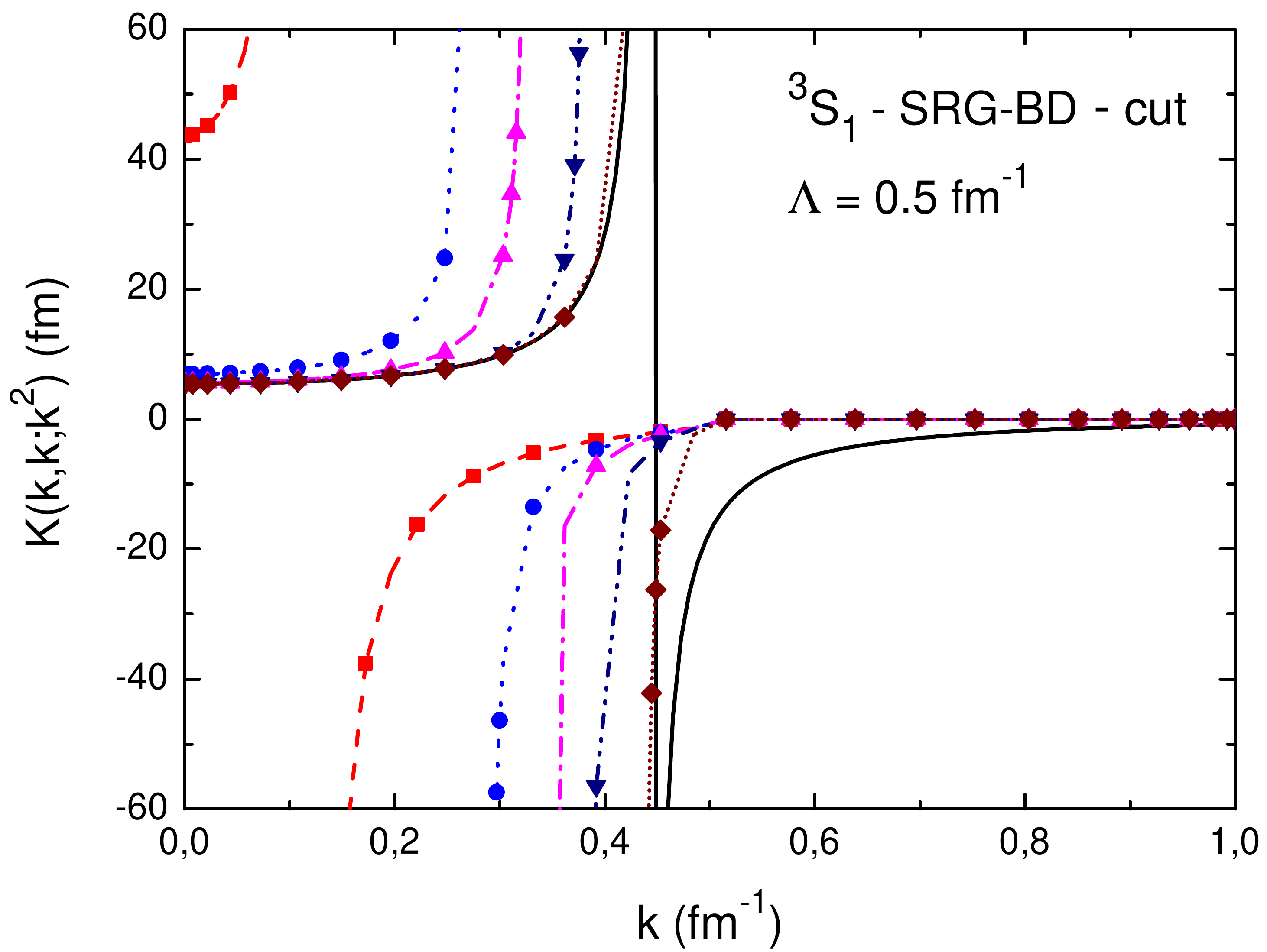}\hspace{0.5cm}
\includegraphics[width=5cm]{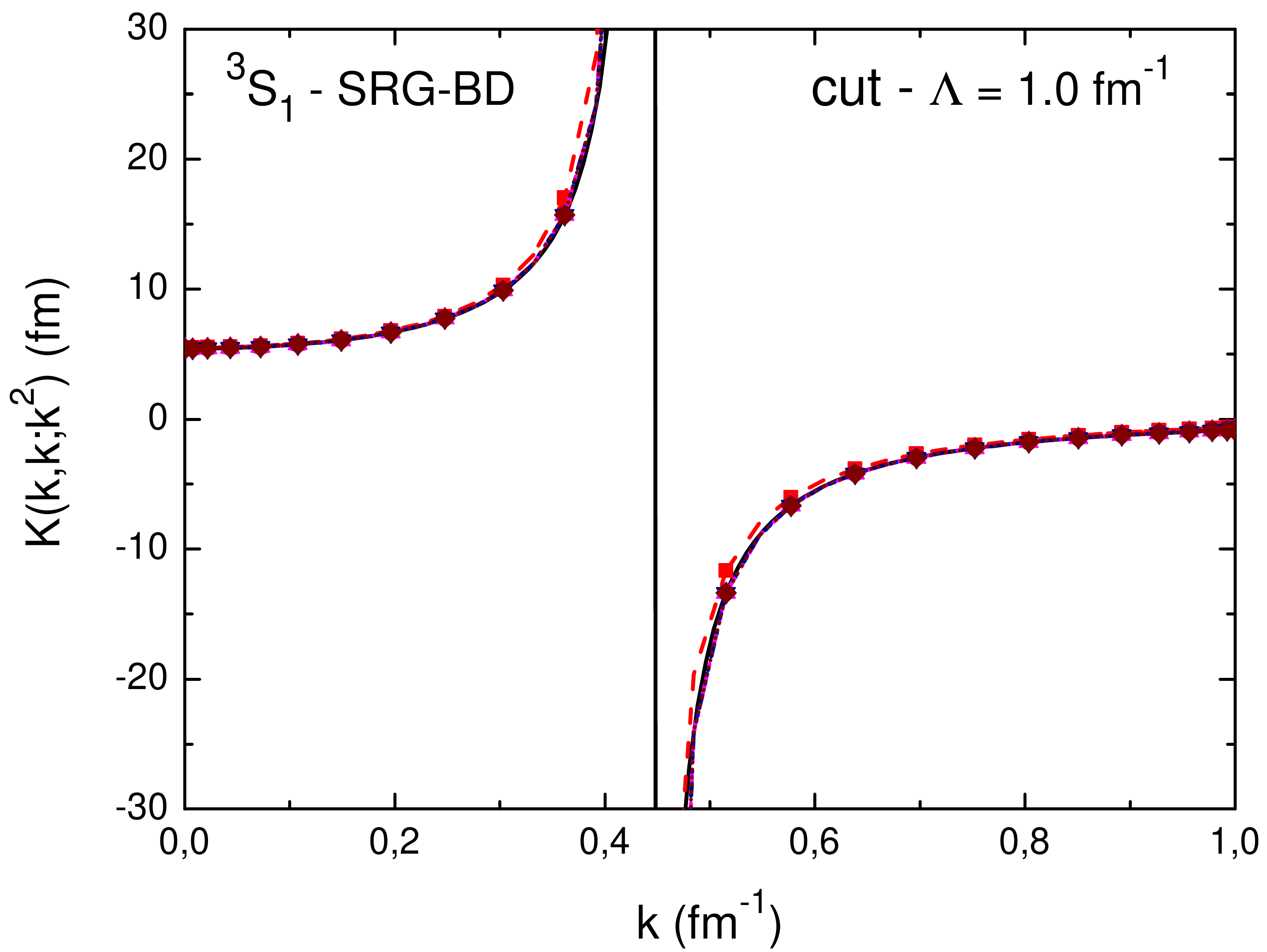}
\end{center}
\caption{On-shell $K$-matrices for the $^1S_0$ (upper panels) and $^3S_1$ (lower panels) channels SRG-evolved toy-model potentials
cut at the block-diagonal cutoff $\Lambda$ (i. e. with the matrix-elements set to zero for momenta above $\Lambda$). For comparison,
we also show the on-shell $K$-matrix for the initial ($\lambda \rightarrow \infty$) toy-model potential.}
\label{fig:13}
\end{figure*}

\begin{figure*}[ht]
\begin{center}
\includegraphics[width=7cm]{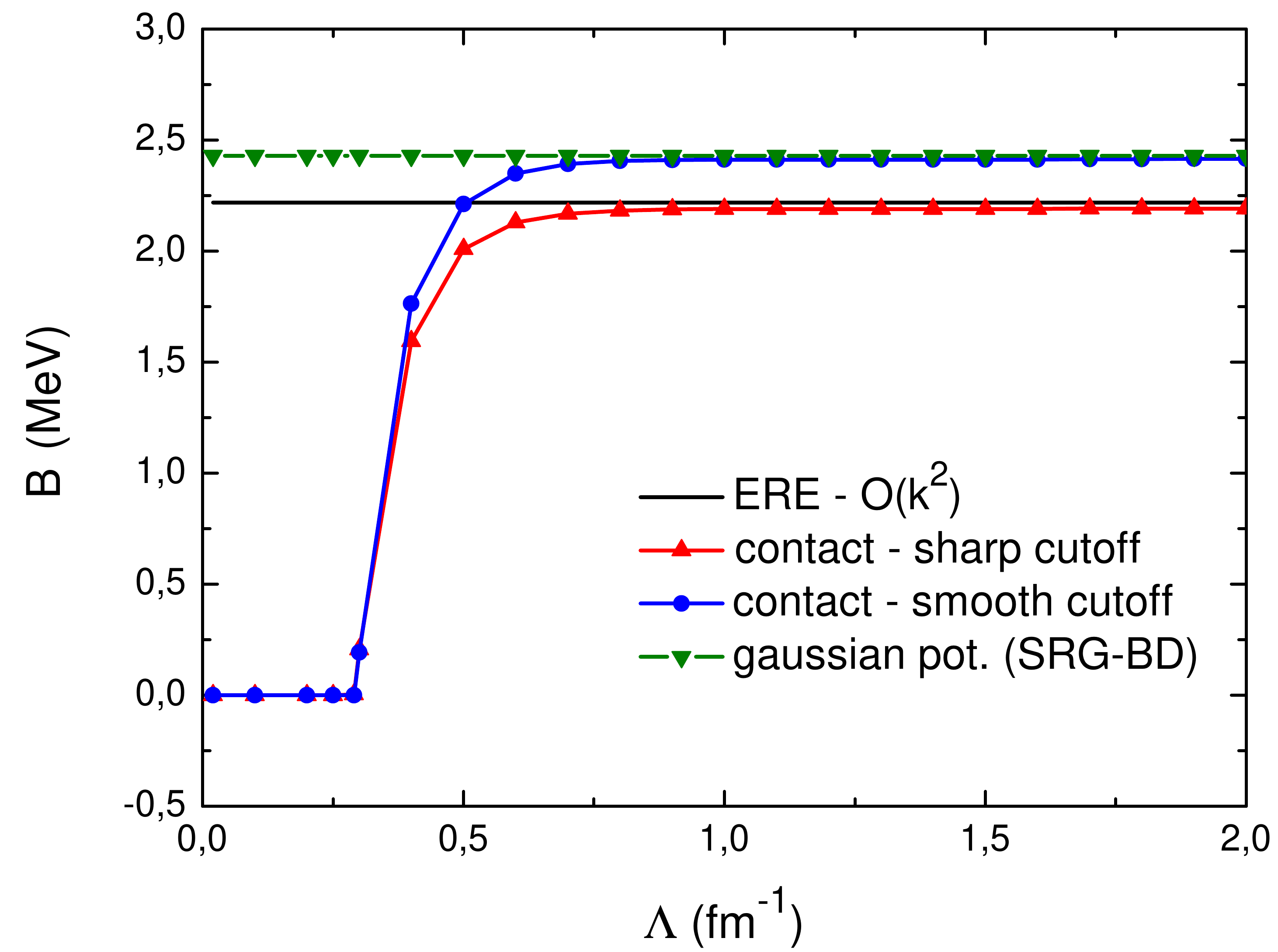}\hspace{0.5cm}
\includegraphics[width=7cm]{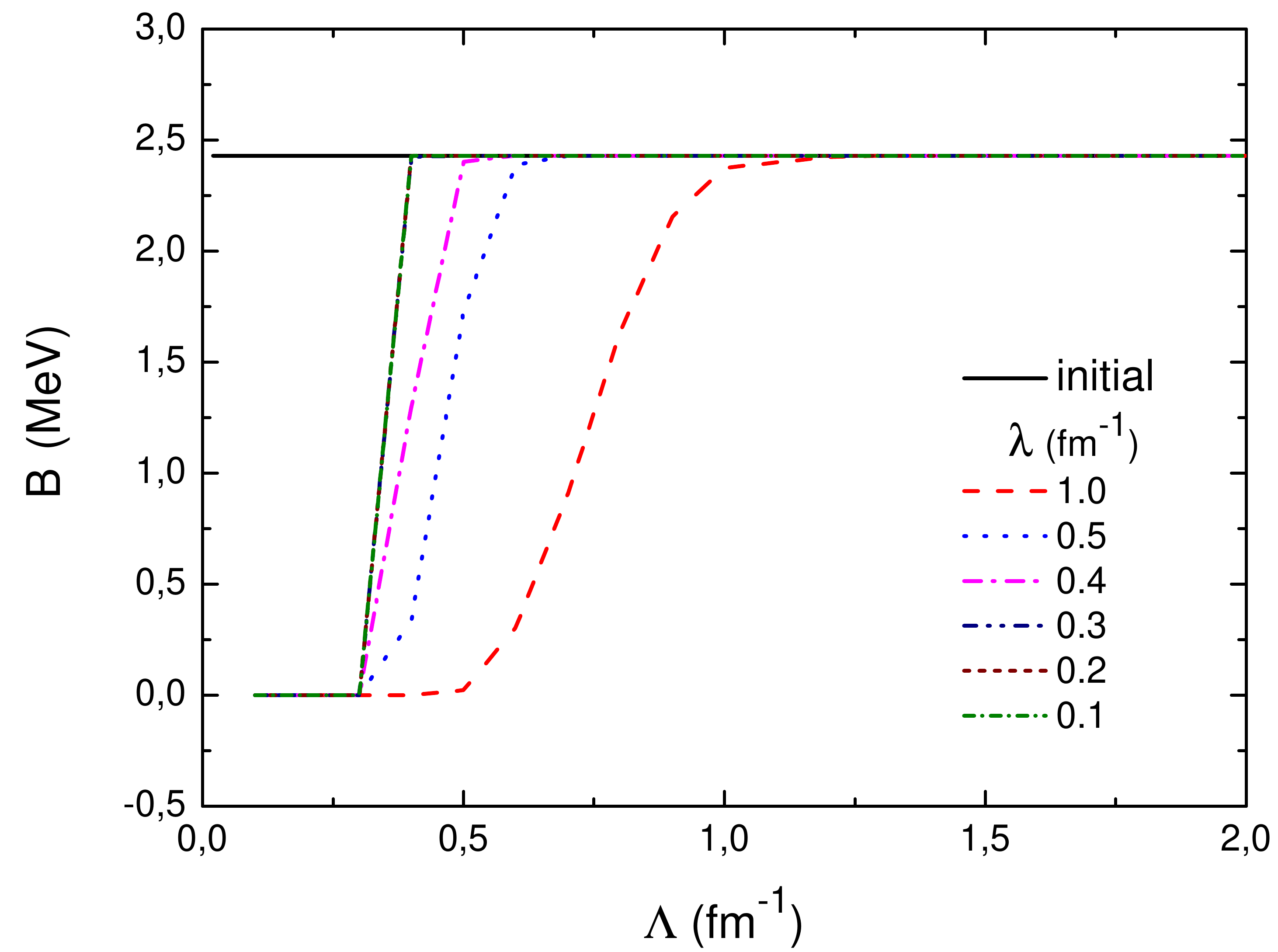}
\end{center}
\caption{Left: deuteron binding-energy as a function of the cutoff scale $\Lambda$ evaluated from the numerical solution of Schr{\"o}dinger's equation with the $^3S_1$ channel potential at $NLO$ for the contact theory in the continuum with a sharp cutoff and for the contact theory on a gaussian grid with a smooth cutoff (with $N=50$ momentum points, $P_{\rm max}=5~{\rm fm}^{-1}$ and $n=16$). For comparison, we also show the binding-energy obtained for the toy-model potential and the binding-energy obtained from the ERE to order ${\cal O}(k^2)$. Right: Deuteron binding-energy as a function of the block-diagonal cutoff $\Lambda$ evaluated from the numerical solution of Schr{\"o}dinger's equation with the SRG-evolved toy-model potential in the $^3S_1$ channel cut at $\Lambda$ (i.e. with the matrix-elements set to zero for momenta above $\Lambda$). For comparison, we also show the binding-energy for the initial ($\lambda \rightarrow \infty$) toy-model potential.}
\label{fig:15}
\end{figure*}

In order to test the decoupling between the $P$-space and the
$Q$-space in the SRG evolution with a block-diagonal generator, we
compute the on-shell $K$-matrices for the $^1S_0$ channel and the
$^3S_1$ channel SRG-evolved toy-model potentials cut at the
block-diagonal cutoff $\Lambda$, i. e. with the matrix-elements set to
zero for momenta above $\Lambda$. As one can observe in
Figs.~\ref{fig:13}, for a given $\Lambda$ the
decoupling improves as $\lambda$ decreases. For $\lambda < \Lambda$,
the results are nearly indistinguishable from those obtained with the
initial ($\lambda \rightarrow \infty$) potential for momenta $k <
\Lambda$, except near the deuteron pole (that vanishes for $\Lambda <
0.3~{\rm fm^{-1}}$) in the case of the $^3S_1$ channel potential. One
should also note that the position of the deuteron pole changes as the
potential evolves and in the limit $\lambda \rightarrow 0$ approaches
that for the initial potential as $\Lambda$ increases, similar to what
happens for the contact theory potential regulated by a sharp or
smooth cutoff. In the left panel of Fig.~\ref{fig:15} we show the results for the
deuteron binding-energy as a function of the cutoff scale $\Lambda$
evaluated from the numerical solution of Schr{\"o}dinger's equation with
the $^3S_1$ channel potential at $NLO$ for the contact theory in the
continuum with a sharp cutoff and for the contact theory on a gaussian
grid with a smooth cutoff (with $N=50$ momentum points, $P_{\rm
  max}=5~{\rm fm}^{-1}$ and $n=16$). As one can observe, in both cases
the deuteron bound-state appears at $\Lambda \sim 0.3~{\rm
  fm^{-1}}$. For the contact theory in the continuum with a sharp
cutoff the binding-energy approaches the value obtained from the ERE
to order ${\cal O}(k^2)$ as $\Lambda$ increases. In the case the
contact theory on a grid with a smooth cutoff the binding-energy
approaches the value obtained for the toy-model potential. One should
note that the results obtained for the toy-model potential evolved
through the SRG transformation with the block-diagonal generator
remain invariant under the change of both the SRG cutoff $\lambda$ and
the block-diagonal cutoff $\Lambda$. In the right panel Fig.~\ref{fig:15} we show the
results for the deuteron binding-energy as a function of the
block-diagonal cutoff $\Lambda$ evaluated from the numerical solution
of Schr{\"o}dinger's equation with the $^3S_1$ channel SRG-evolved
toy-model potential cut at $\Lambda$ (i. e. with the matrix-elements
set to zero for momenta above $\Lambda$). As a consequence of the
decoupling between the $P$-space and the $Q$-space, for a given
$\Lambda ~(> 0.3~{\rm fm^{-1}})$ the binding-energy obtained for the
cut SRG-evolved potential approaches the value obtained for the
initial ($\lambda \rightarrow \infty$) potential provided $\lambda \ll
\Lambda$~\footnote{Similar plots to Fig.~\ref{fig:15} have been
  particularly inspiring in our recent analysis of the infrared fixed
  points of the connection between SRG and Levinson's
  theorem~\cite{Arriola:2014aia}.}.

\section{Fitting the explicitly renormalized potential with the contact interactions from the implicit method}

We want to compare the running of the coefficients $C_{0}^{(2)}$ and $C_{2}^{(2)}$ with the cutoff $\Lambda$
in the contact theory potential at $NLO$ to the running of the corresponding coefficients ${\tilde C_{0}^{(2)}}$
and ${\tilde C_{2}^{(2)}}$ with the $V_{\rm low \, k}$ cutoff ($\equiv \Lambda$) extracted from a polynomial fit of the block-diagonal SRG-evolved toy-model potential,
\begin{equation}
V_{\lambda,\Lambda}(p,p') = {\tilde C_{0}^{(2)}} + {\tilde C_{2}^{(2)}}~(p^2+{p'}^2) + \cdots\; .
\end{equation}

The parameters $(C, L)$ in the initial ($\lambda
\rightarrow \infty$) toy-model potential, given by Eq.~(\ref{gaupot}), and the
coefficients $C_{0}^{(2)}$ and $C_{2}^{(2)}$ in the contact theory potential at
$NLO$ are determined from the numerical solution of the $LS$ equation for the
$K$-matrix by fitting the experimental values of the scattering length
$a_0$ and the effective range $r_e$. The coefficients ${\tilde C_{0}^{(2)}}$
and ${\tilde C_{2}^{(2)}}$ are determined by fitting the diagonal
matrix-elements of the block-diagonal SRG-evolved toy-model potential for the lowest momenta
with the polynomial form.

\begin{figure*}[ht]
\begin{center}
\includegraphics[width=5.2cm]{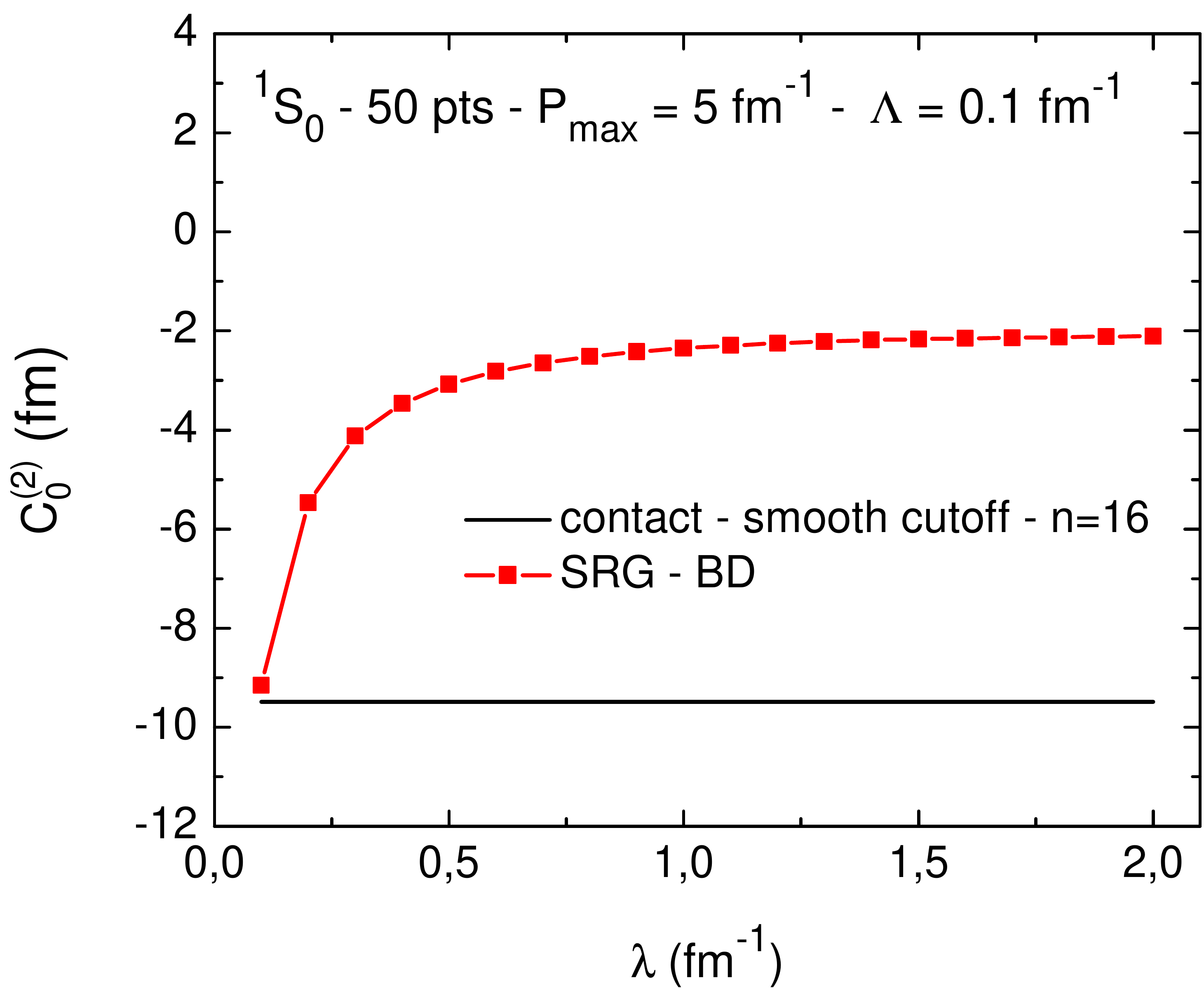}\hspace{0.2cm}
\includegraphics[width=5.2cm]{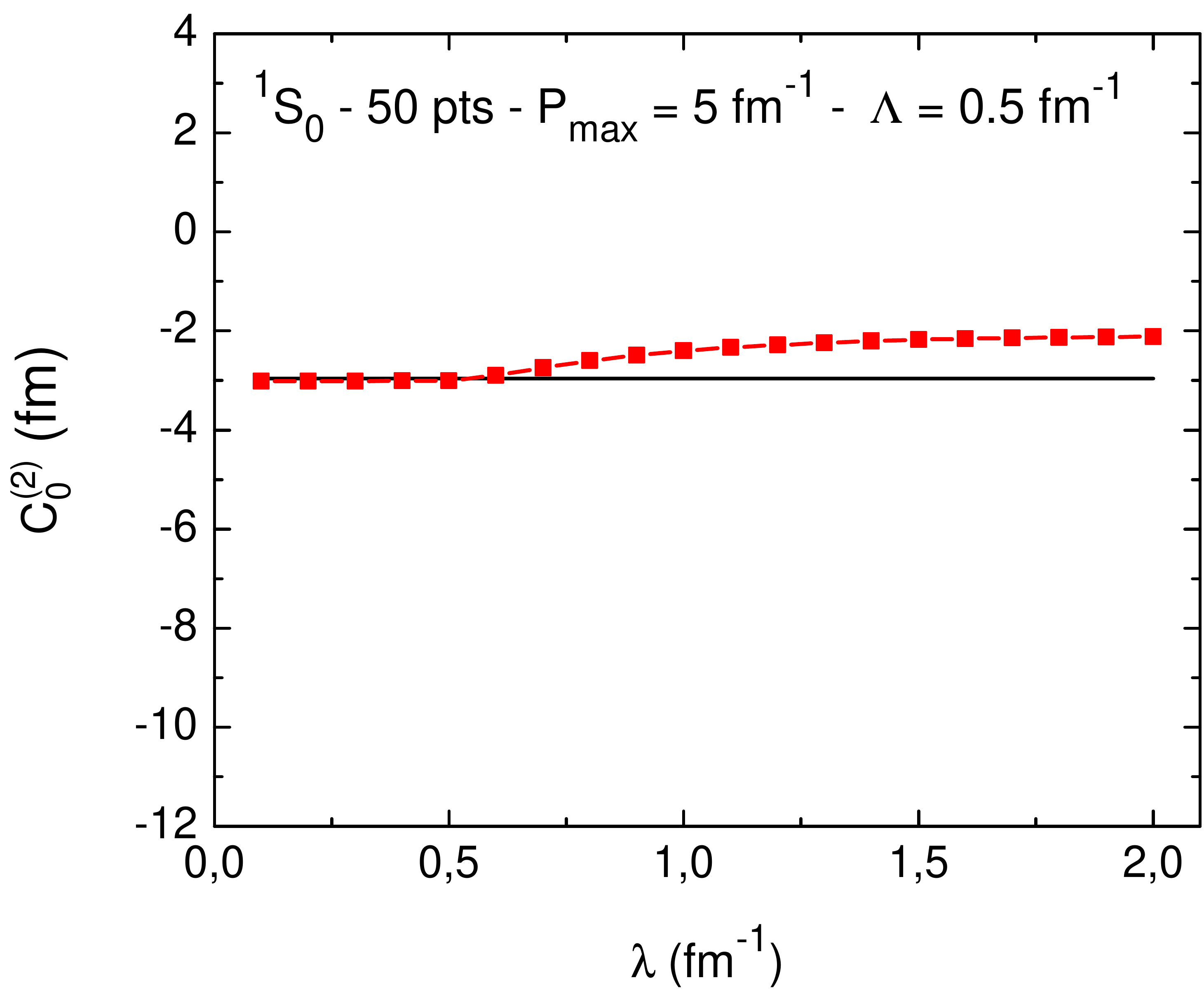}\hspace{0.2cm}
\includegraphics[width=5.2cm]{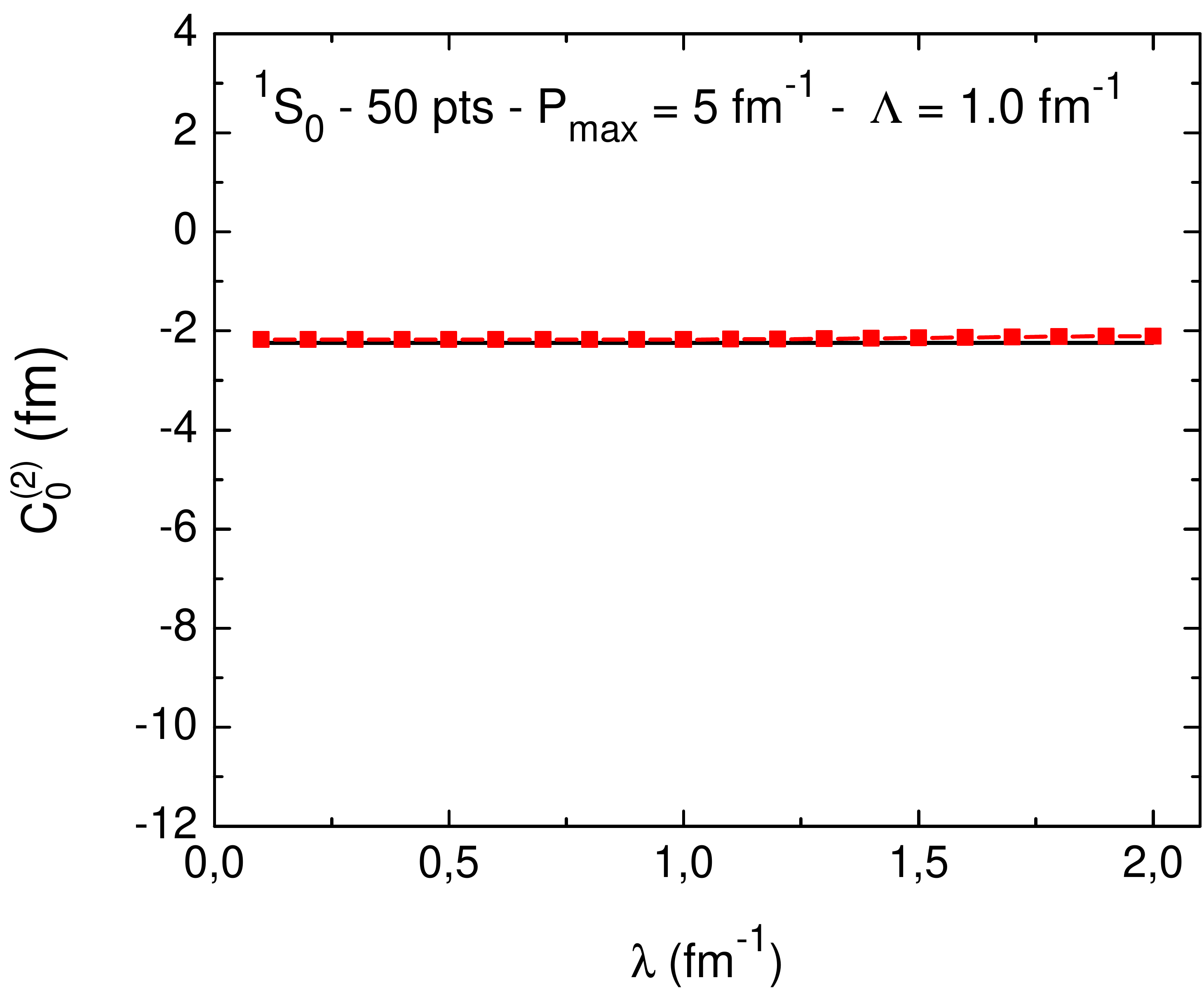}\\\vspace{0.5cm}
\includegraphics[width=5.2cm]{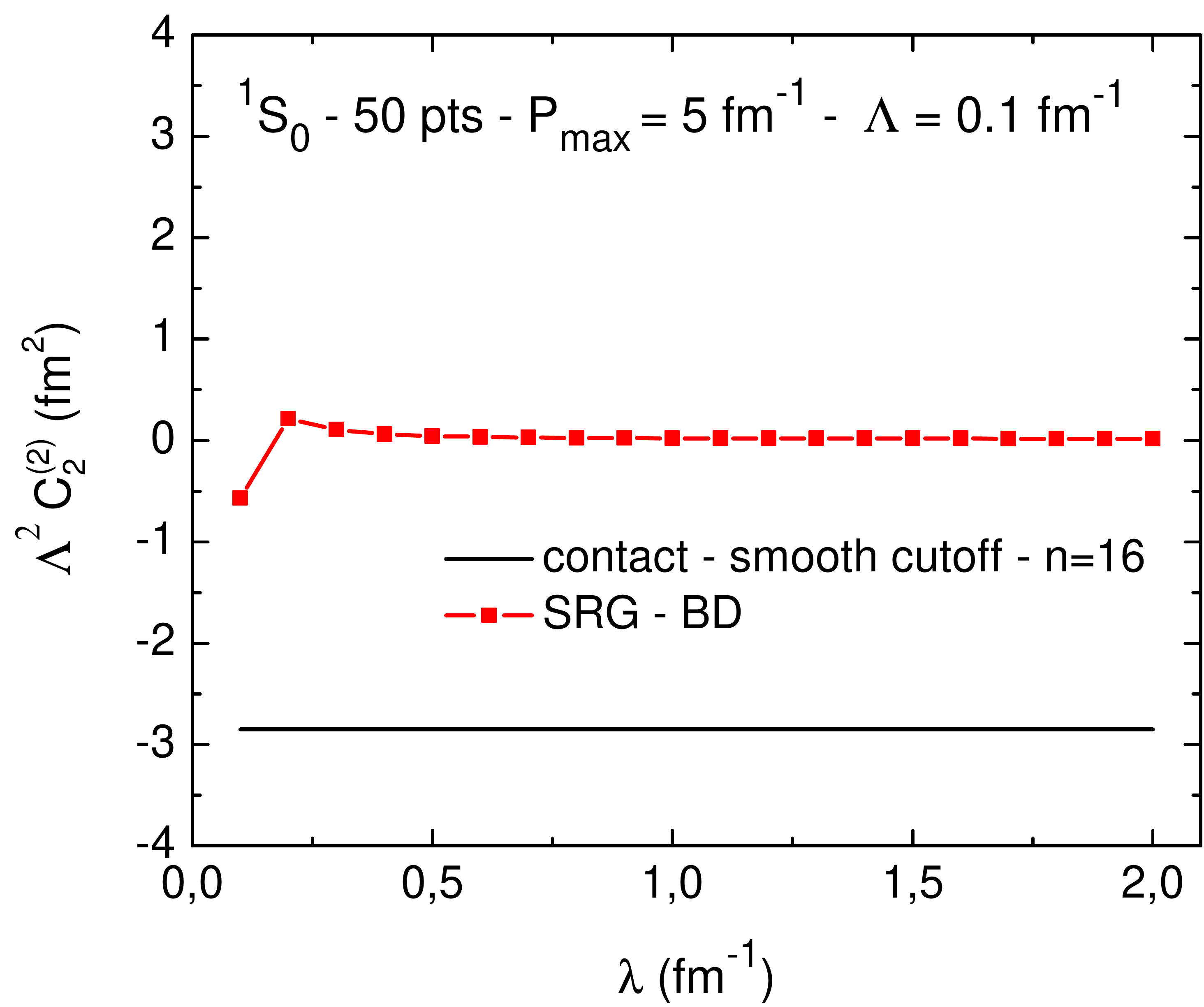}\hspace{0.2cm}
\includegraphics[width=5.2cm]{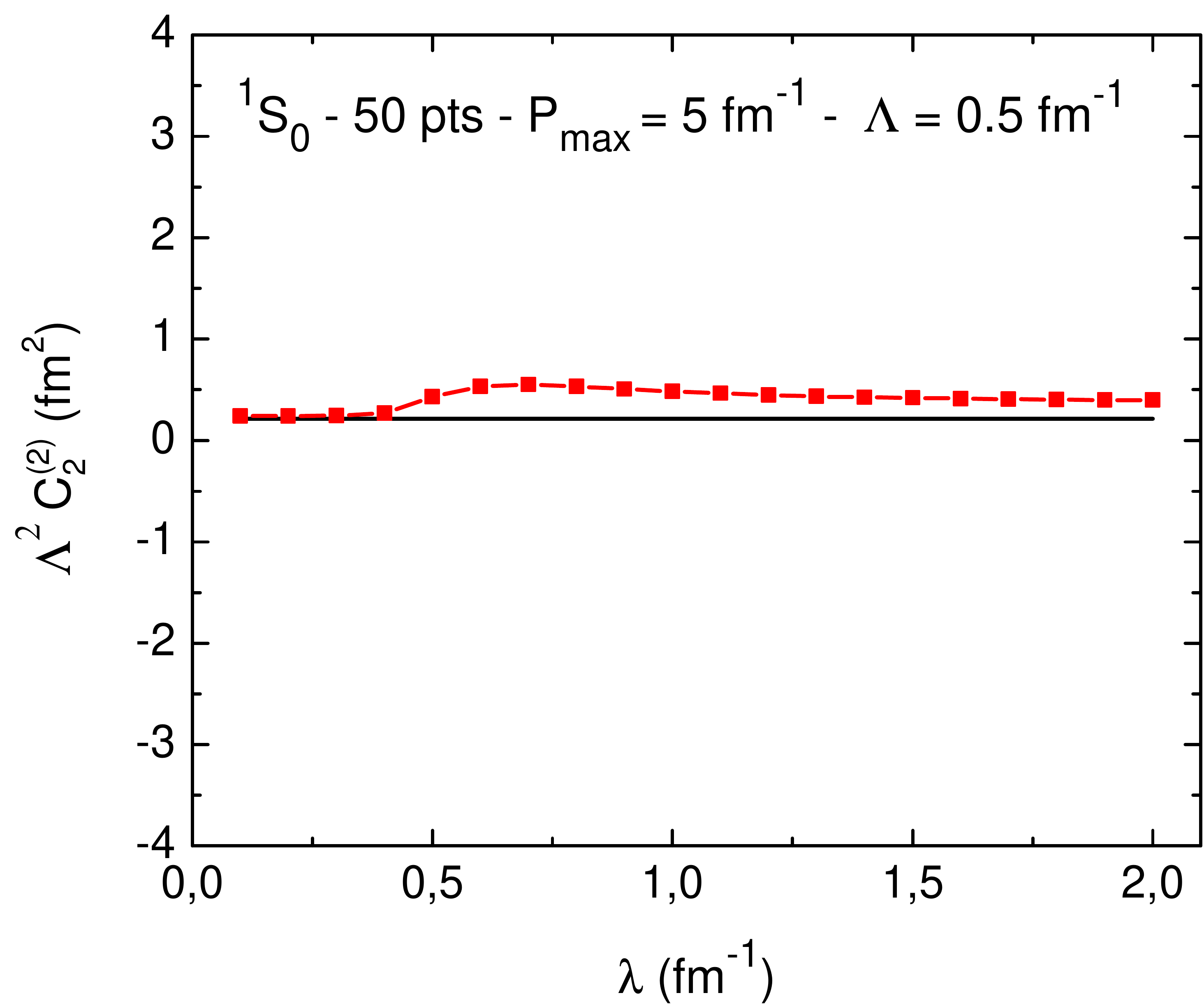}\hspace{0.2cm}
\includegraphics[width=5.2cm]{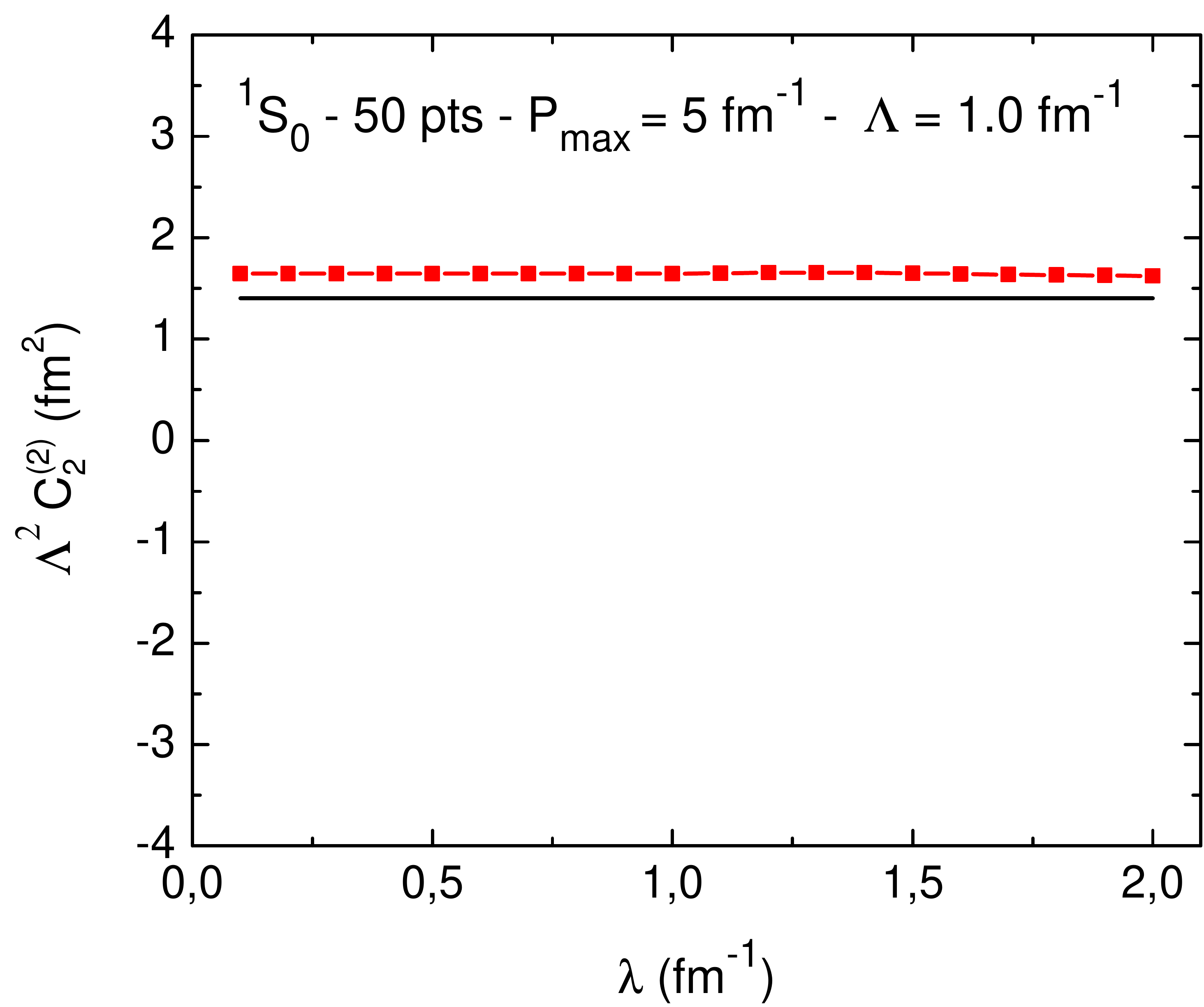}
\end{center}
\caption{${\tilde C_{0}^{(2)}}$ (upper panels) and ${\tilde C_{2}^{(2)}}$ (lower panels) as a function of the SRG cutoff
$\lambda$ extracted from the $^1S_0$ channel toy-model potential on a gaussian grid (with $N=50$ momentum points
and $P_{\rm max}=5~{\rm fm}^{-1}$) evolved through the SRG transformation with the block-diagonal generator for some
values of the block-diagonal cutoff $\Lambda$. For comparison, we also show $C_{0}^{(2)}$ for the $^1S_0$ channel contact
theory potential at $NLO$ (on the same grid) regulated by a smooth exponential momentum cutoff with $n=16$. }
\label{fig:17}
\end{figure*}

\begin{figure*}[ht]
\begin{center}
\includegraphics[width=5.2cm]{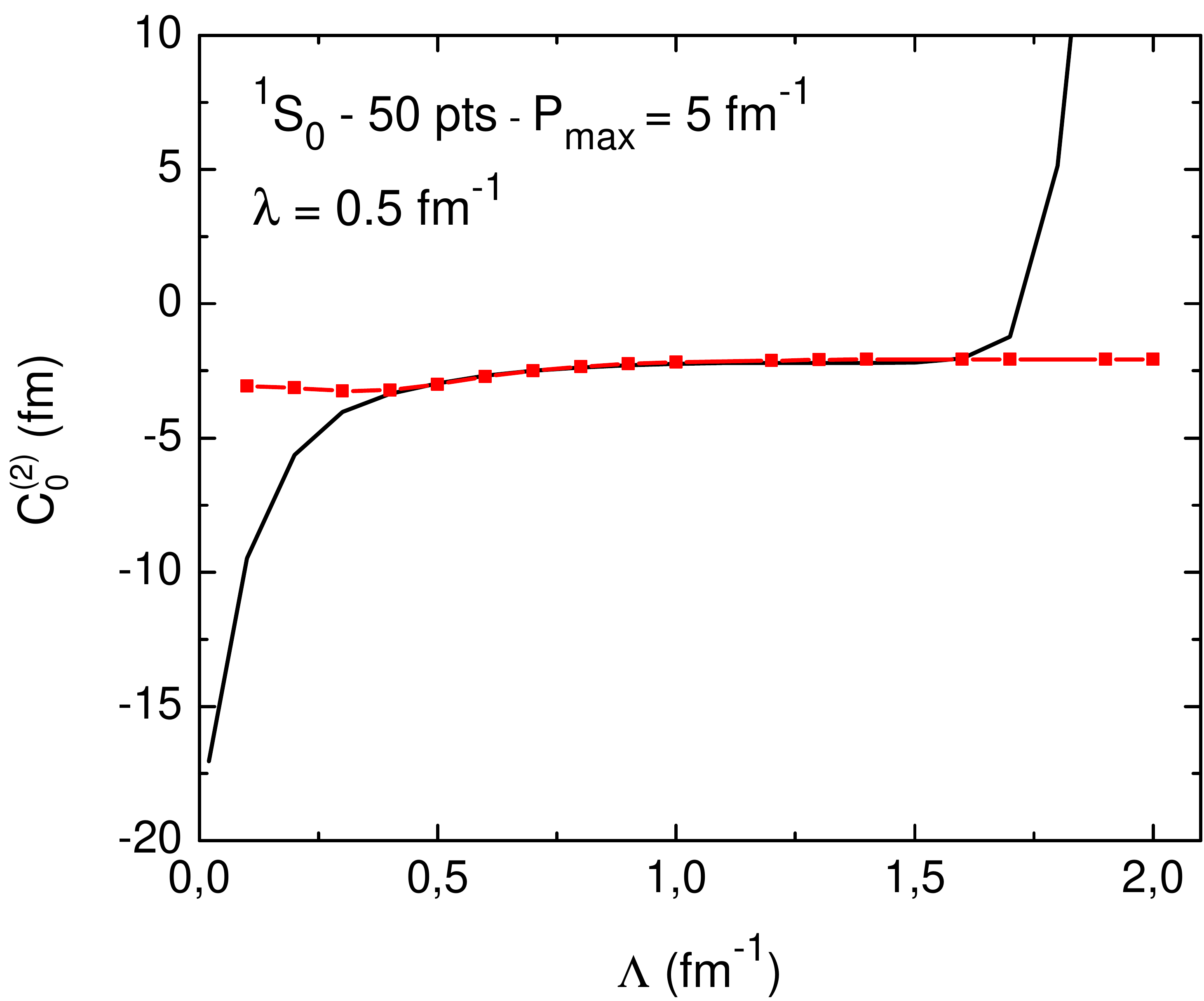}\hspace{0.5cm}
\includegraphics[width=5.2cm]{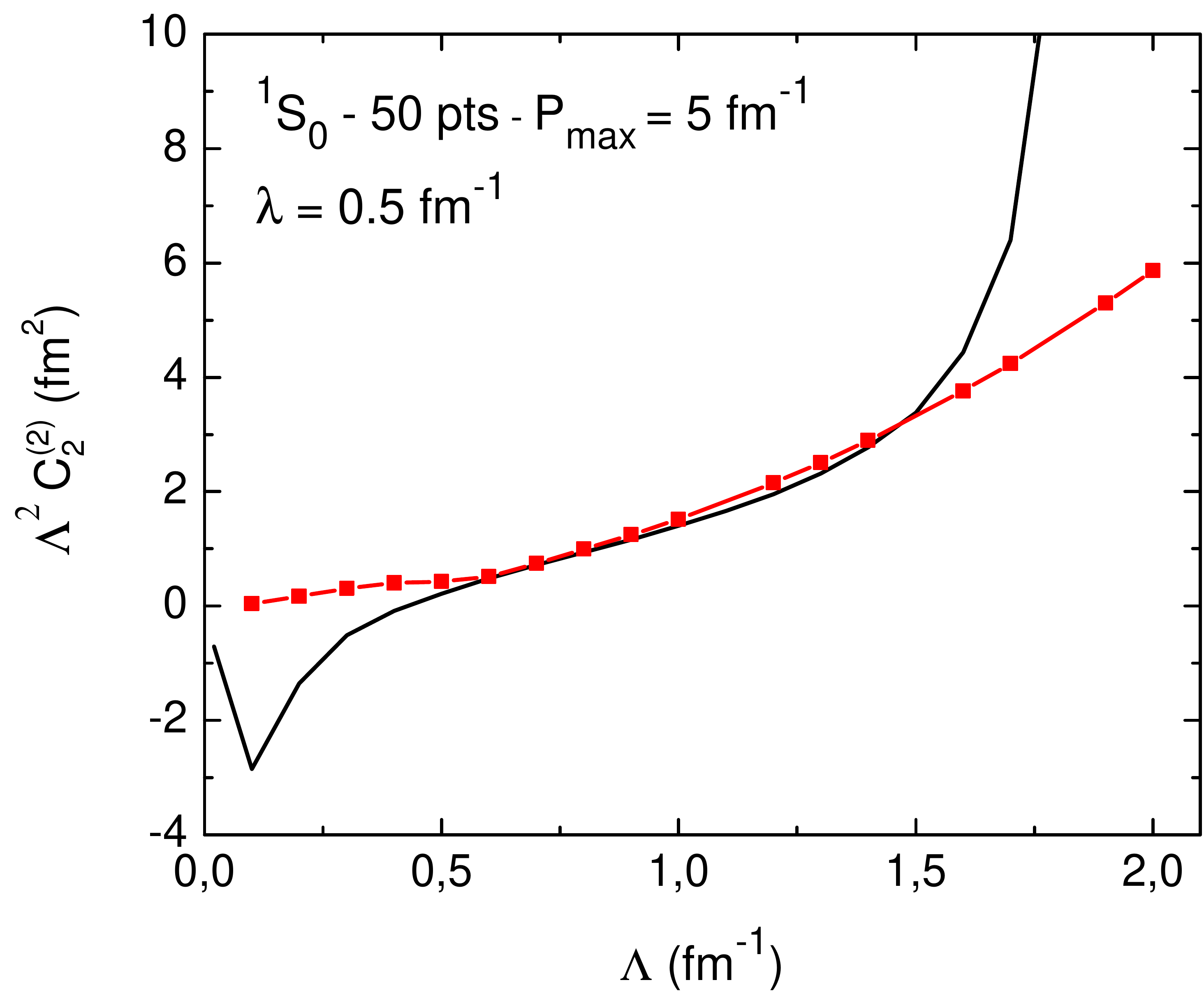} \\\vspace{0.2cm}
\includegraphics[width=5.2cm]{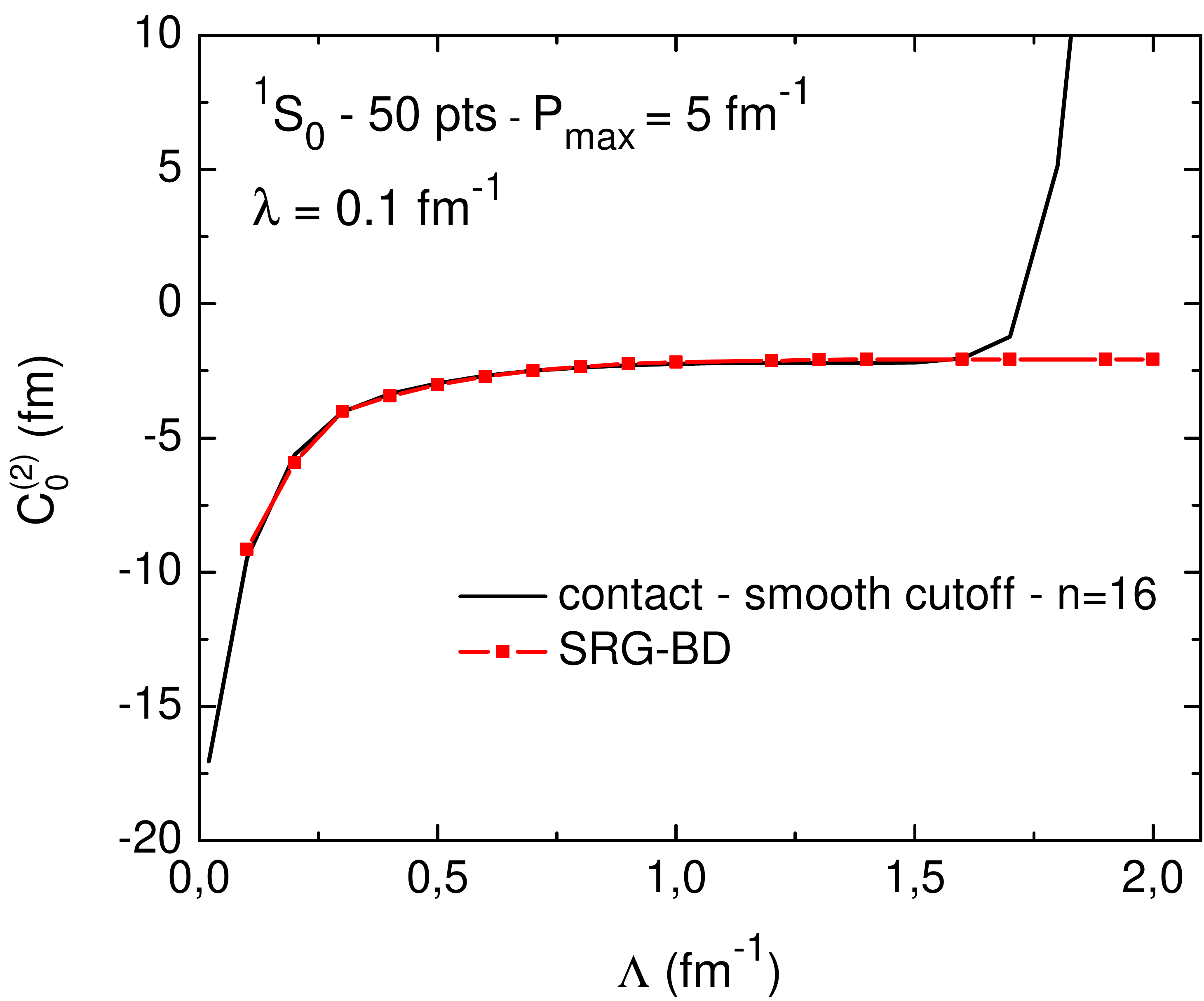}\hspace{0.5cm}
\includegraphics[width=5.2cm]{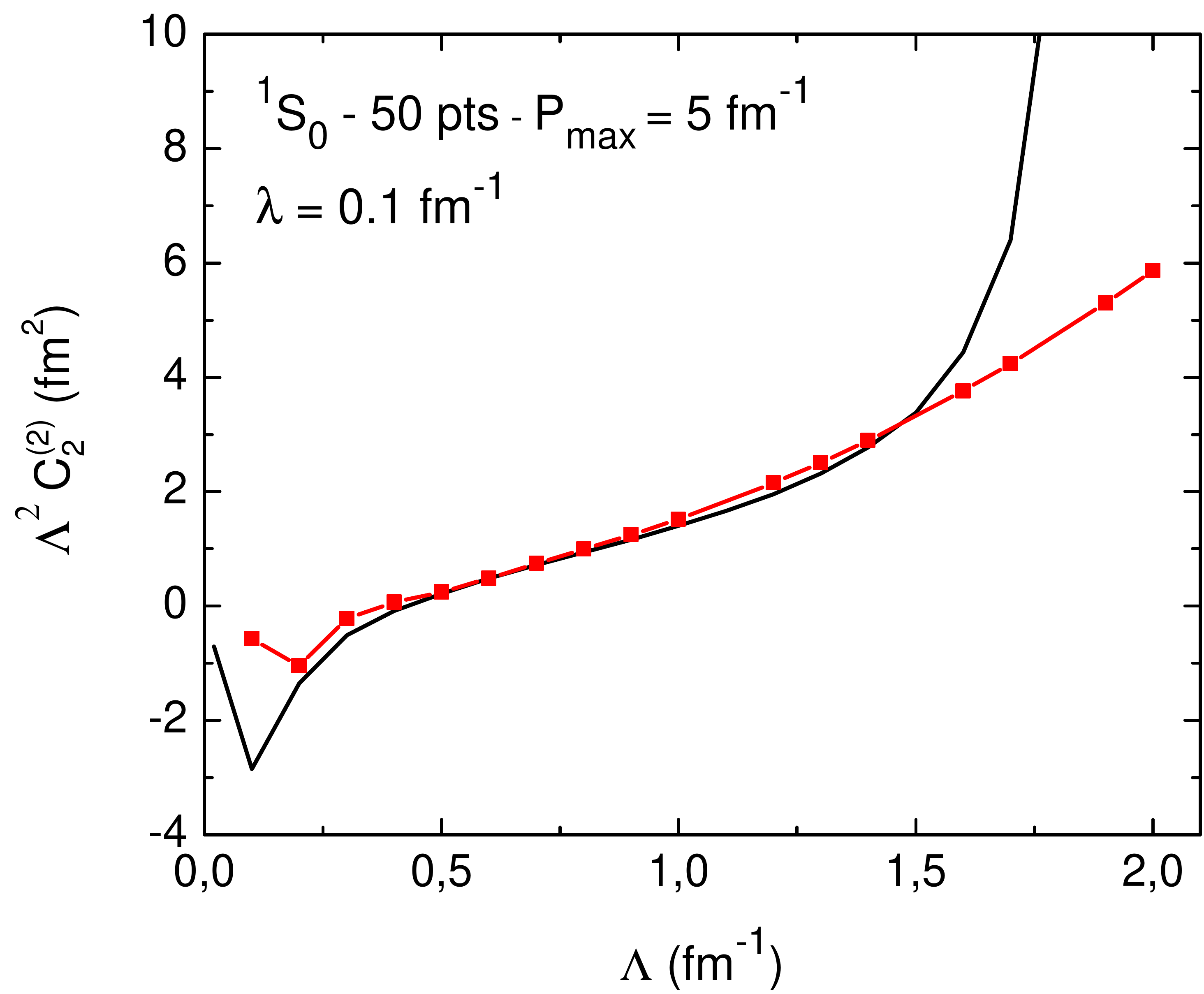}
\end{center}
\caption{${\tilde C_{0}^{(2)}}$ and ${\tilde C_{2}^{(2)}}$ as a function of the block-diagonal cutoff
$\Lambda$ extracted from the $^1S_0$ channel toy-model potential on a gaussian grid
(with $N=50$ momentum points and $P_{\rm max}=5~{\rm fm}^{-1}$) evolved through the SRG
transformation with the block-diagonal generator for two values of the SRG cutoff $\lambda$.
For comparison, we also show $C_{0}^{(2)}$ and $C_{2}^{(2)}$ for the $^1S_0$ channel contact
theory potential at $NLO$ (on the same grid) regulated by a smooth exponential momentum cutoff with $n=16$.}
\label{fig:19}
\end{figure*}

\begin{figure*}[ht]
\begin{center}
\includegraphics[width=5.2cm]{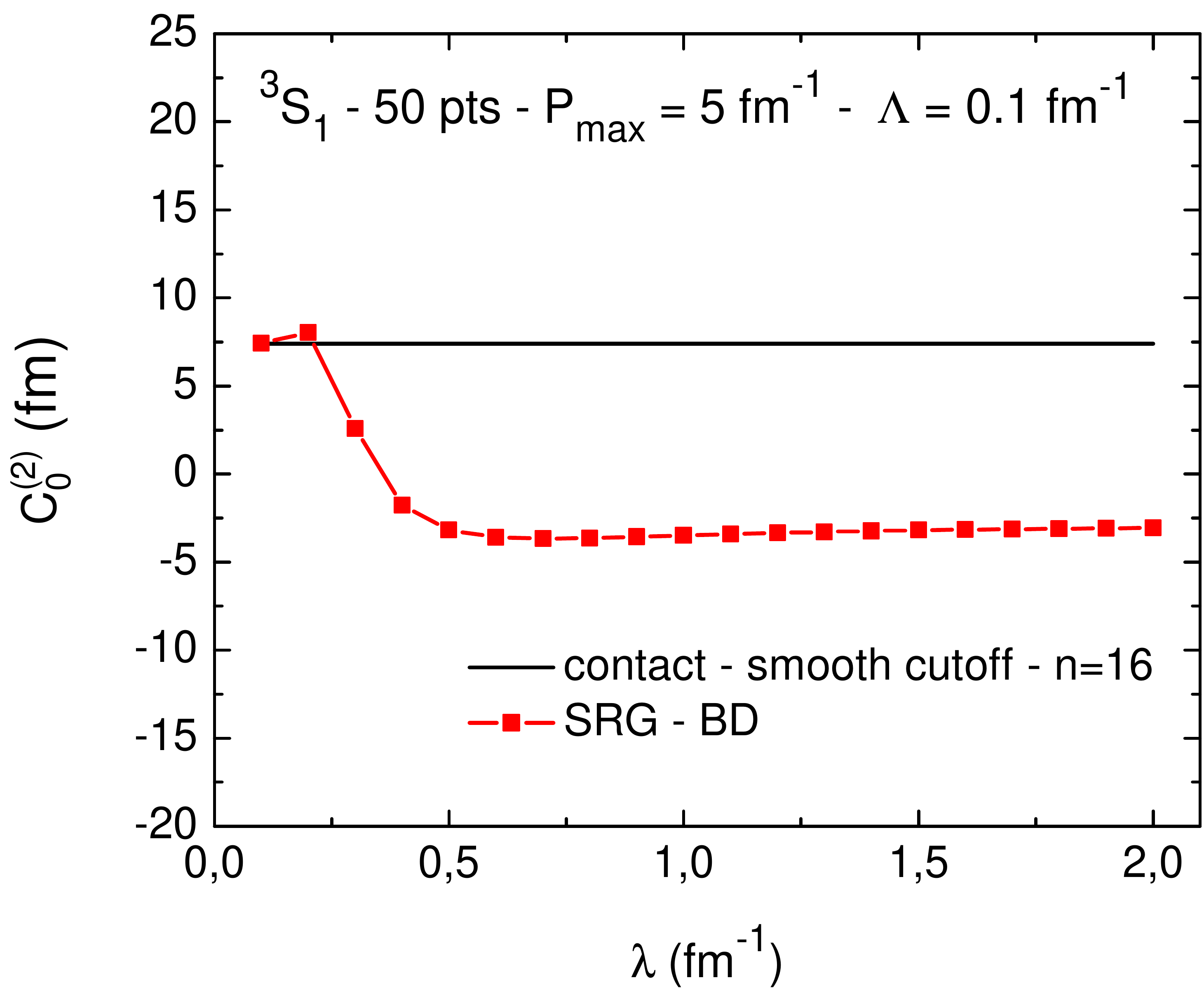}\hspace{0.2cm}
\includegraphics[width=5.2cm]{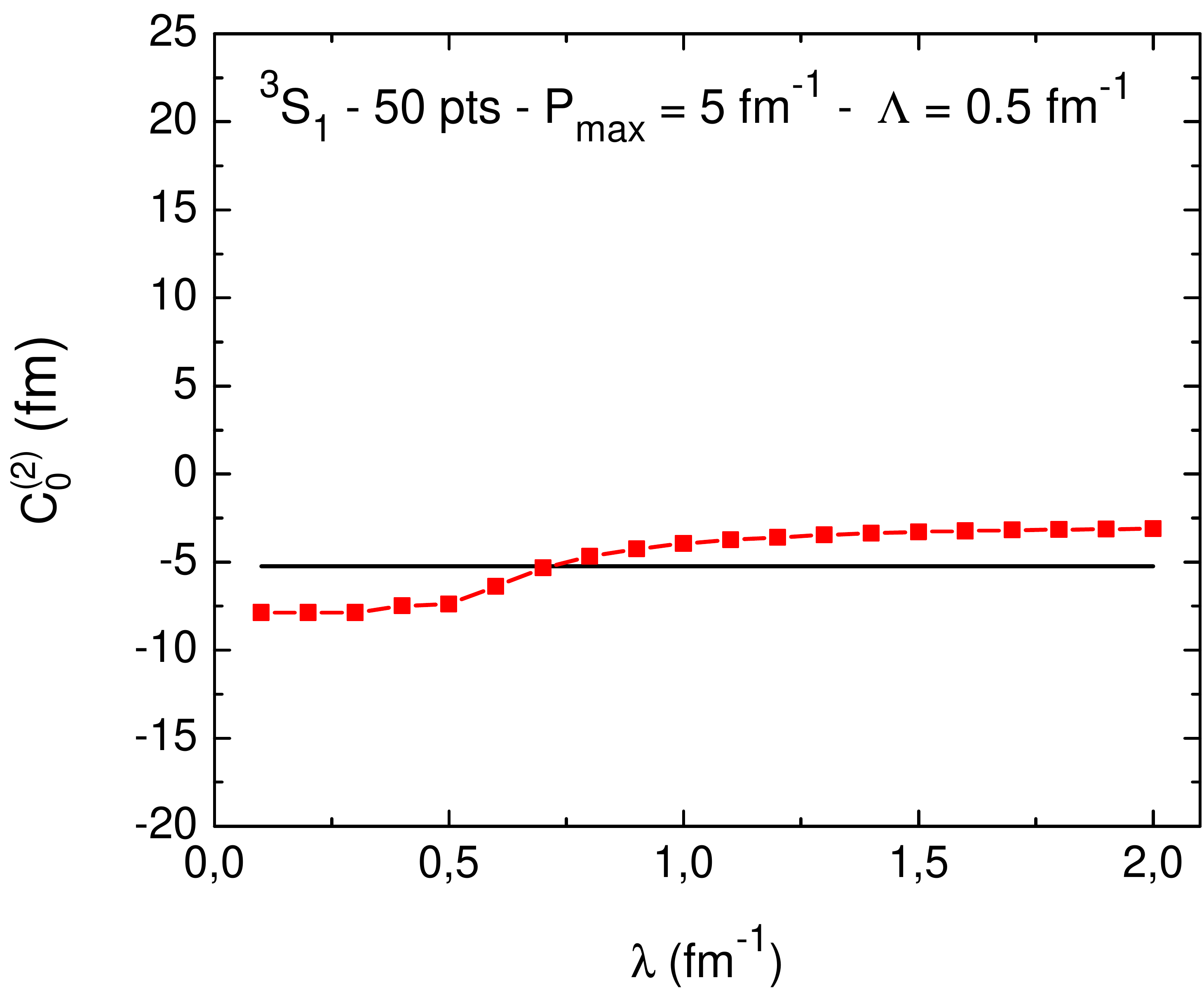}\hspace{0.2cm}
\includegraphics[width=5.2cm]{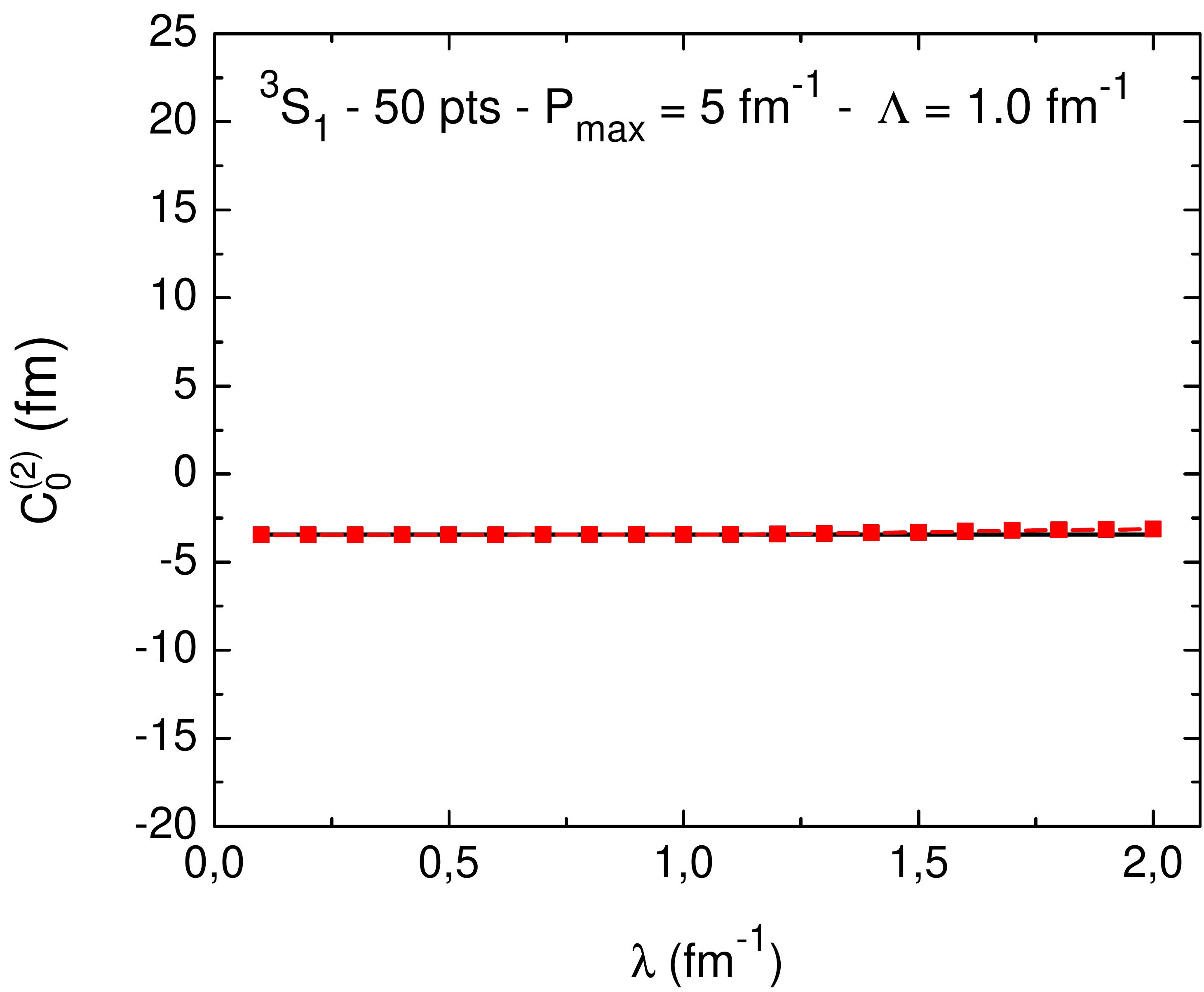}\\\vspace{0.5cm}
\includegraphics[width=5.2cm]{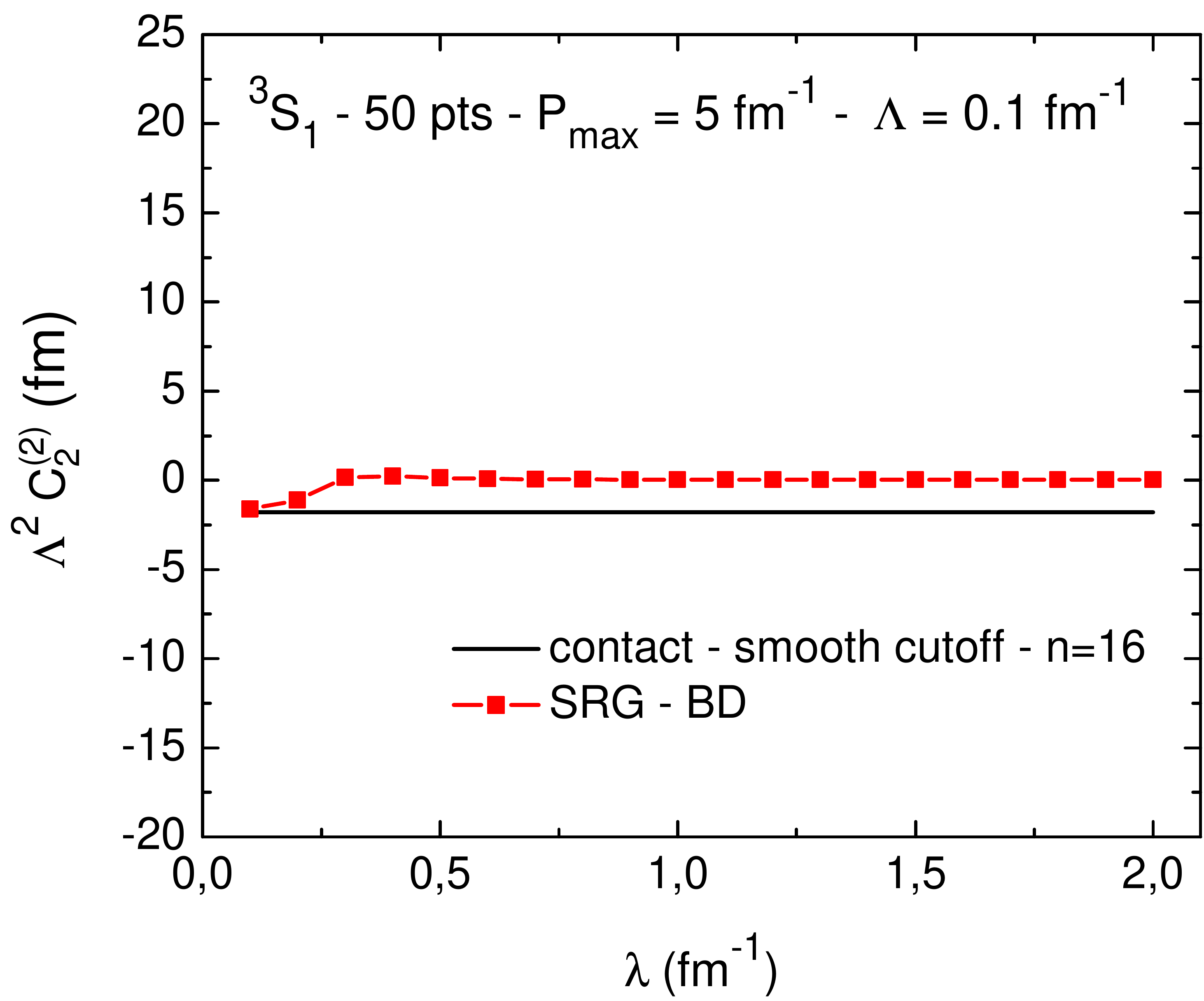}\hspace{0.2cm}
\includegraphics[width=5.2cm]{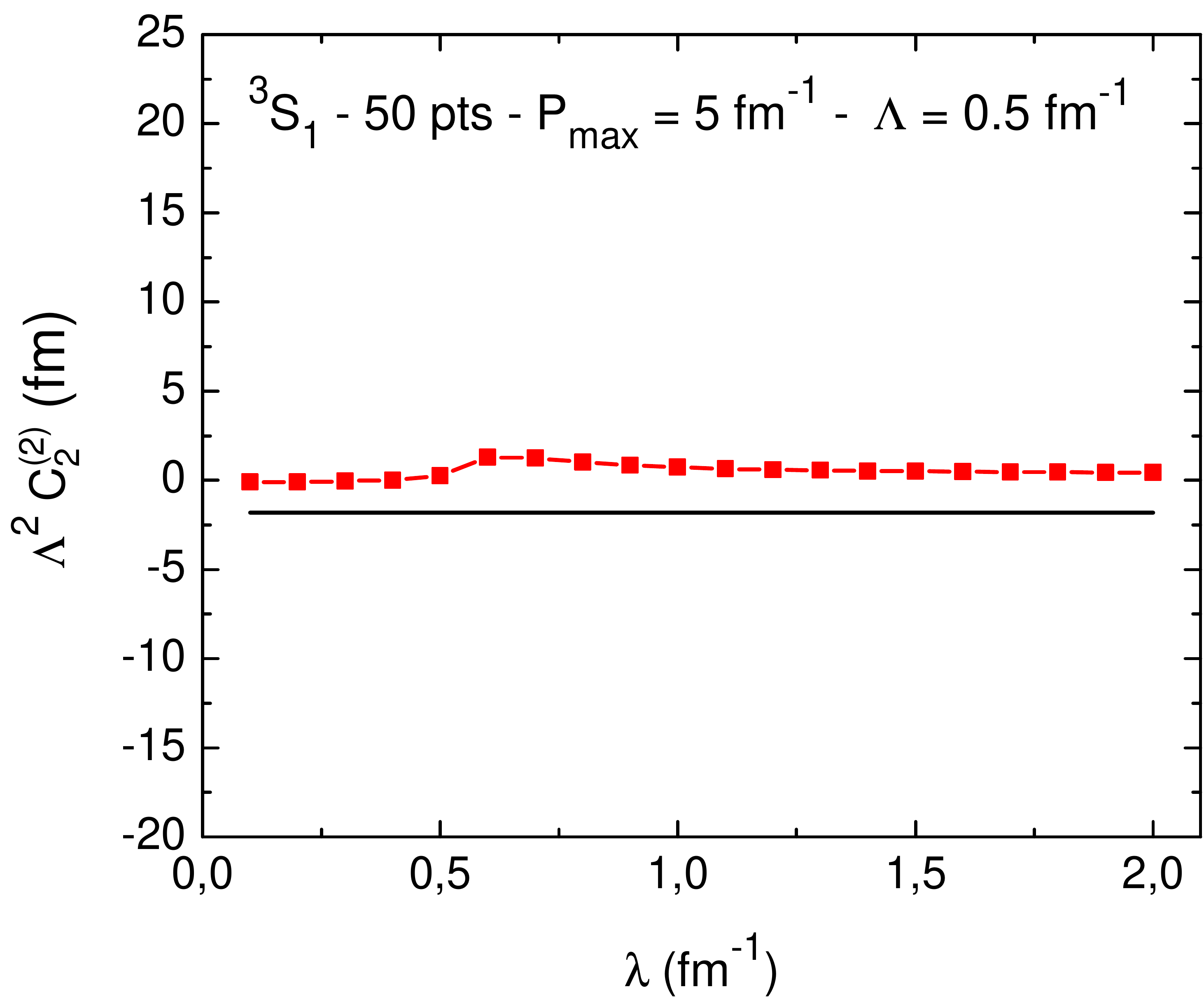}\hspace{0.2cm}
\includegraphics[width=5.2cm]{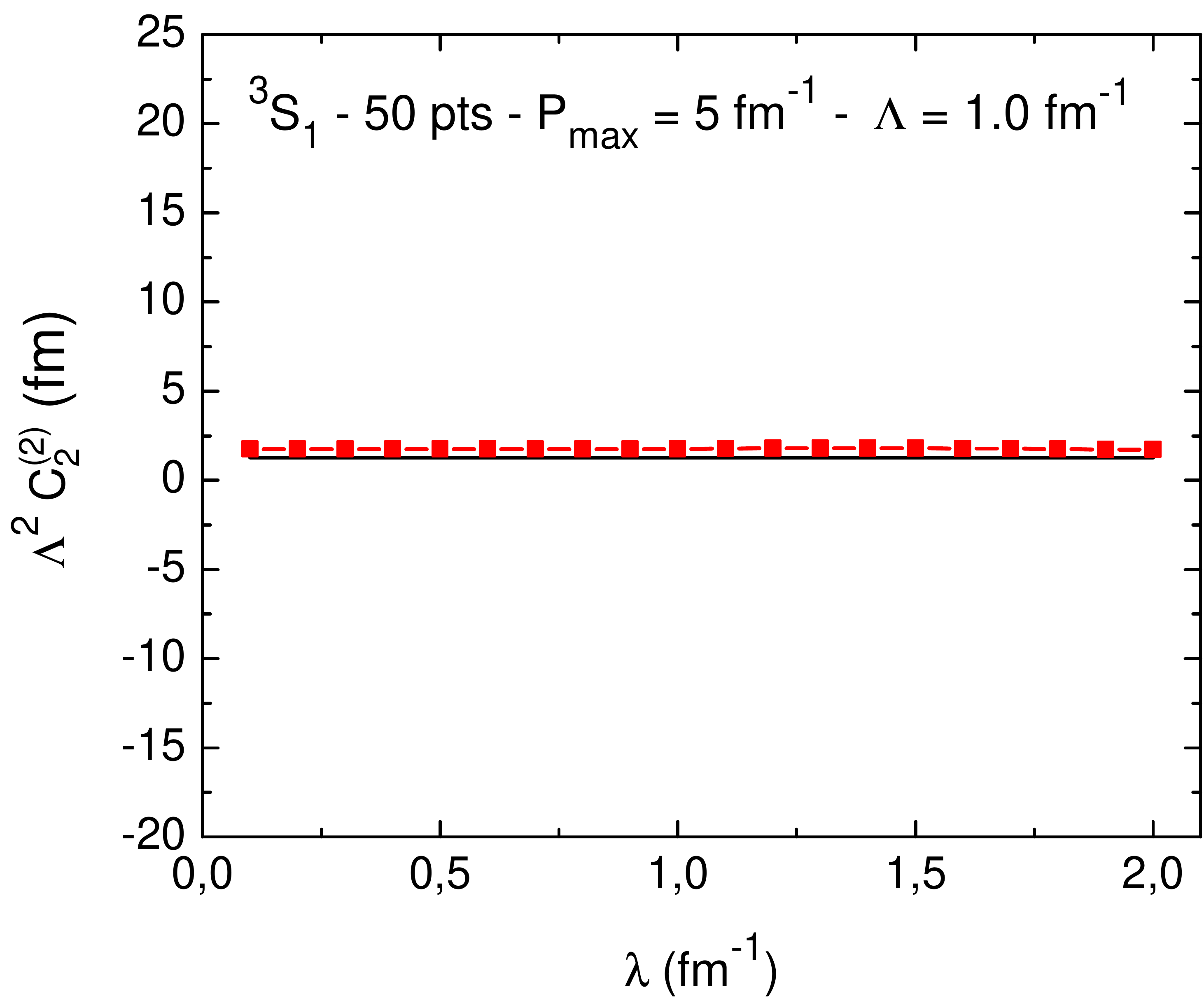}
\end{center}
\caption{${\tilde C_{0}^{(2)}}$ as a function of the SRG cutoff $\lambda$ extracted from the $^3S_1$
channel toy-model potential on a gaussian grid (with $N=50$ momentum points and
$P_{\rm max}=5~{\rm fm}^{-1}$) evolved through the SRG transformation with the block-diagonal
generator for some values of the block-diagonal cutoff $\Lambda$. For comparison, we also show
$C_{0}^{(2)}$ for the $^3S_1$ channel contact theory potential at $NLO$ (on the same grid) regulated
by a smooth exponential momentum cutoff with $n=16$.}
\label{fig:20}
\end{figure*}

\begin{figure*}[ht]
\begin{center}
\includegraphics[width=5.2cm]{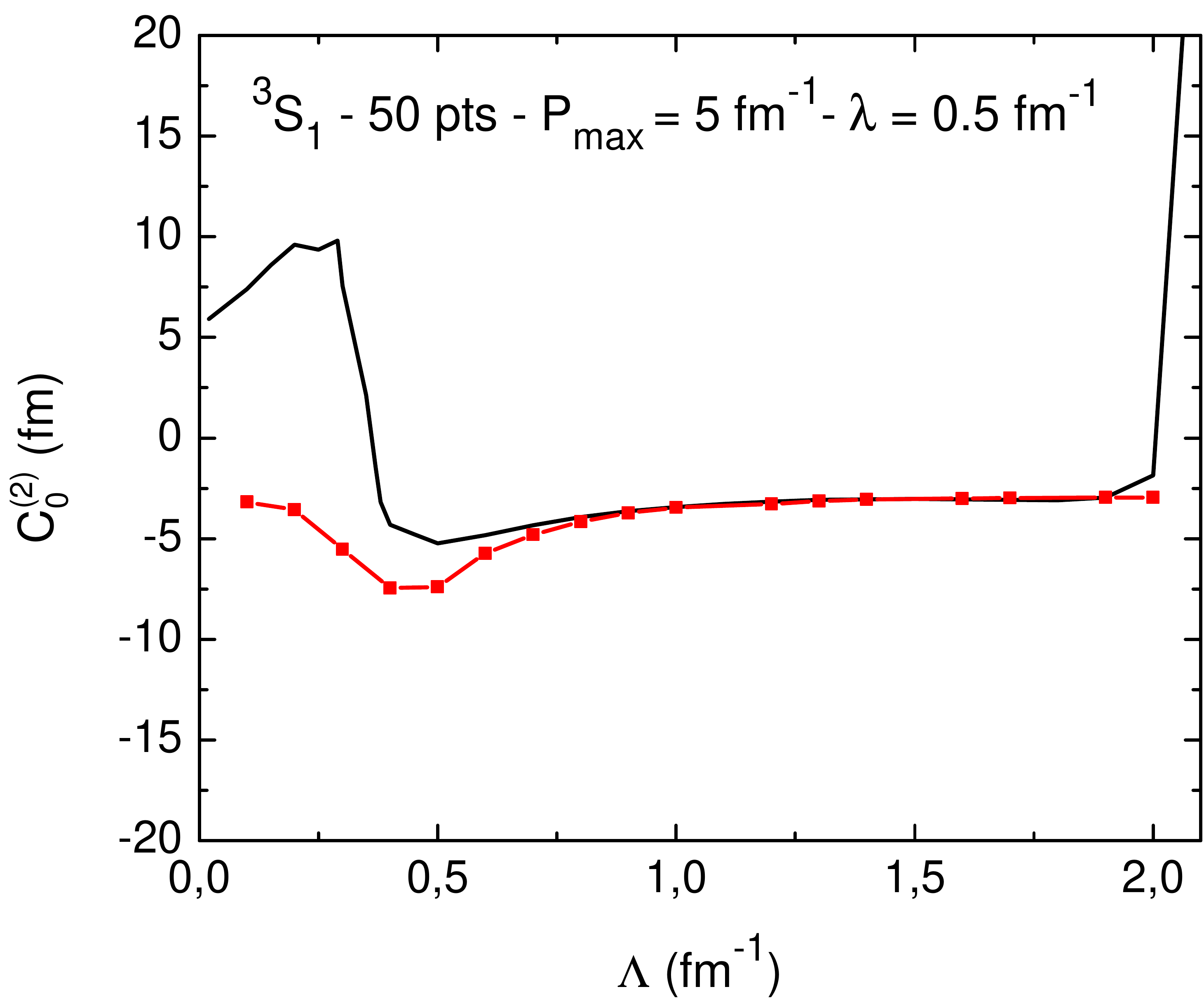}\hspace{0.5cm}
\includegraphics[width=5.2cm]{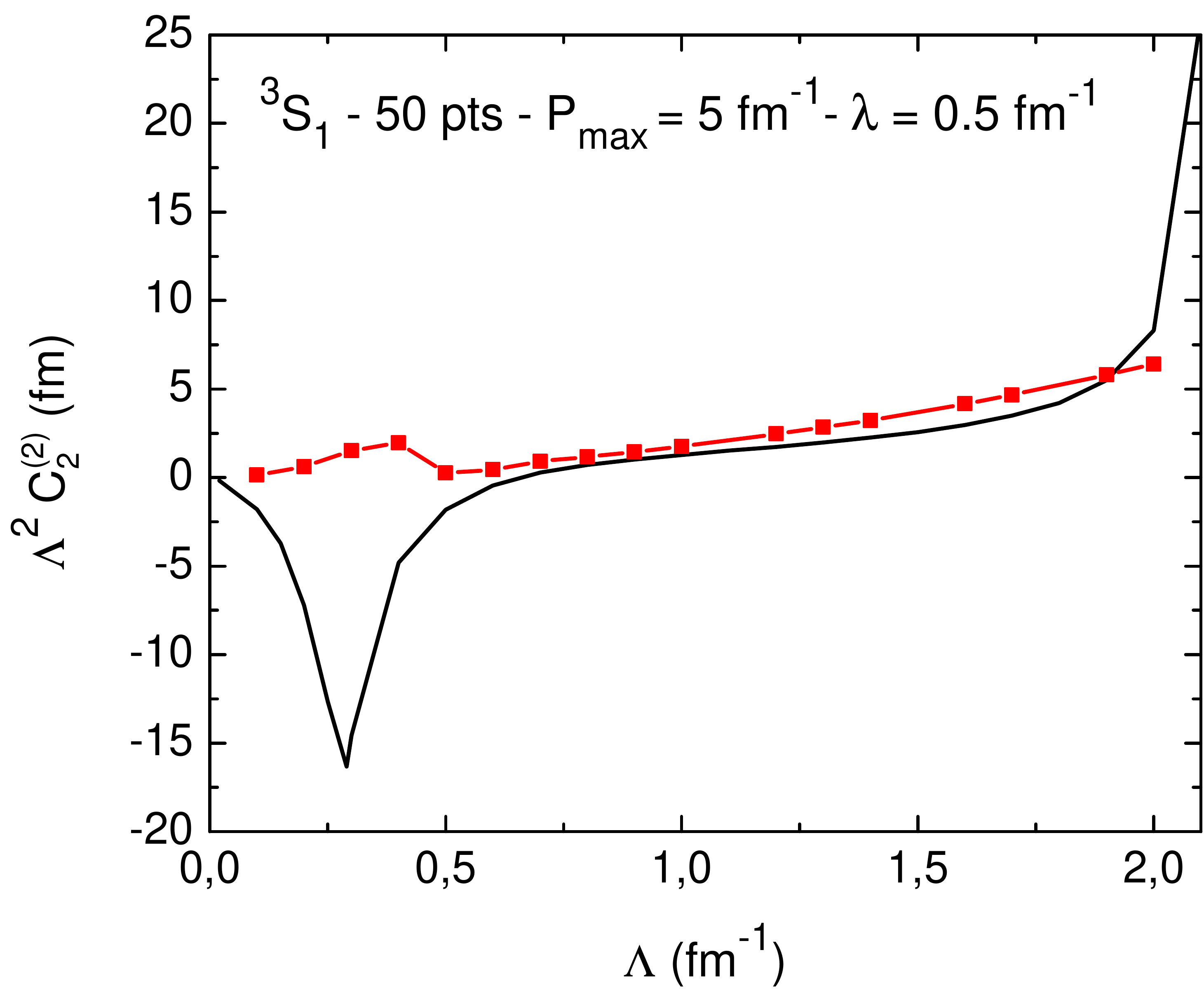} \\\vspace{0.2cm}
\includegraphics[width=5.2cm]{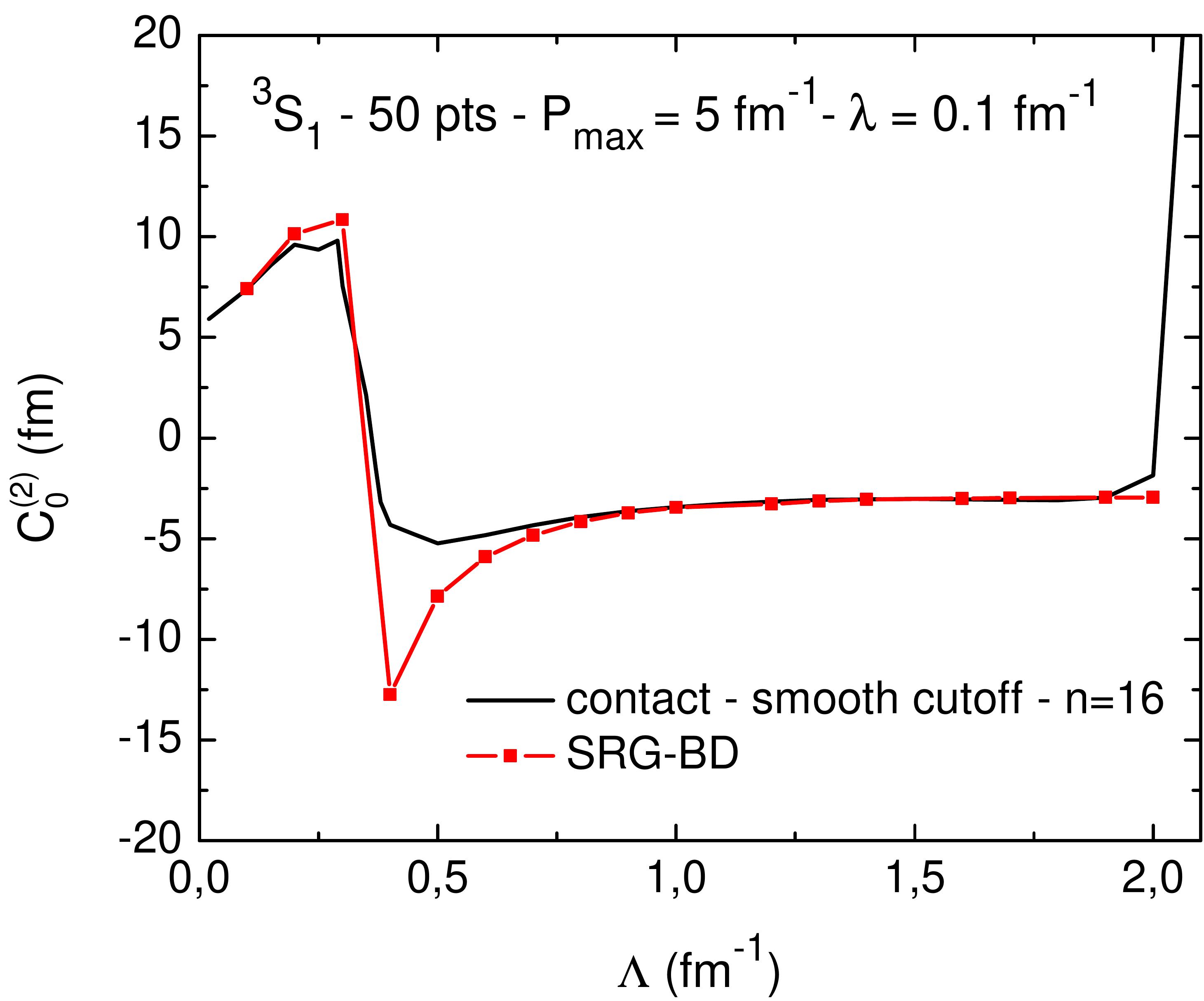}\hspace{0.5cm}
\includegraphics[width=5.2cm]{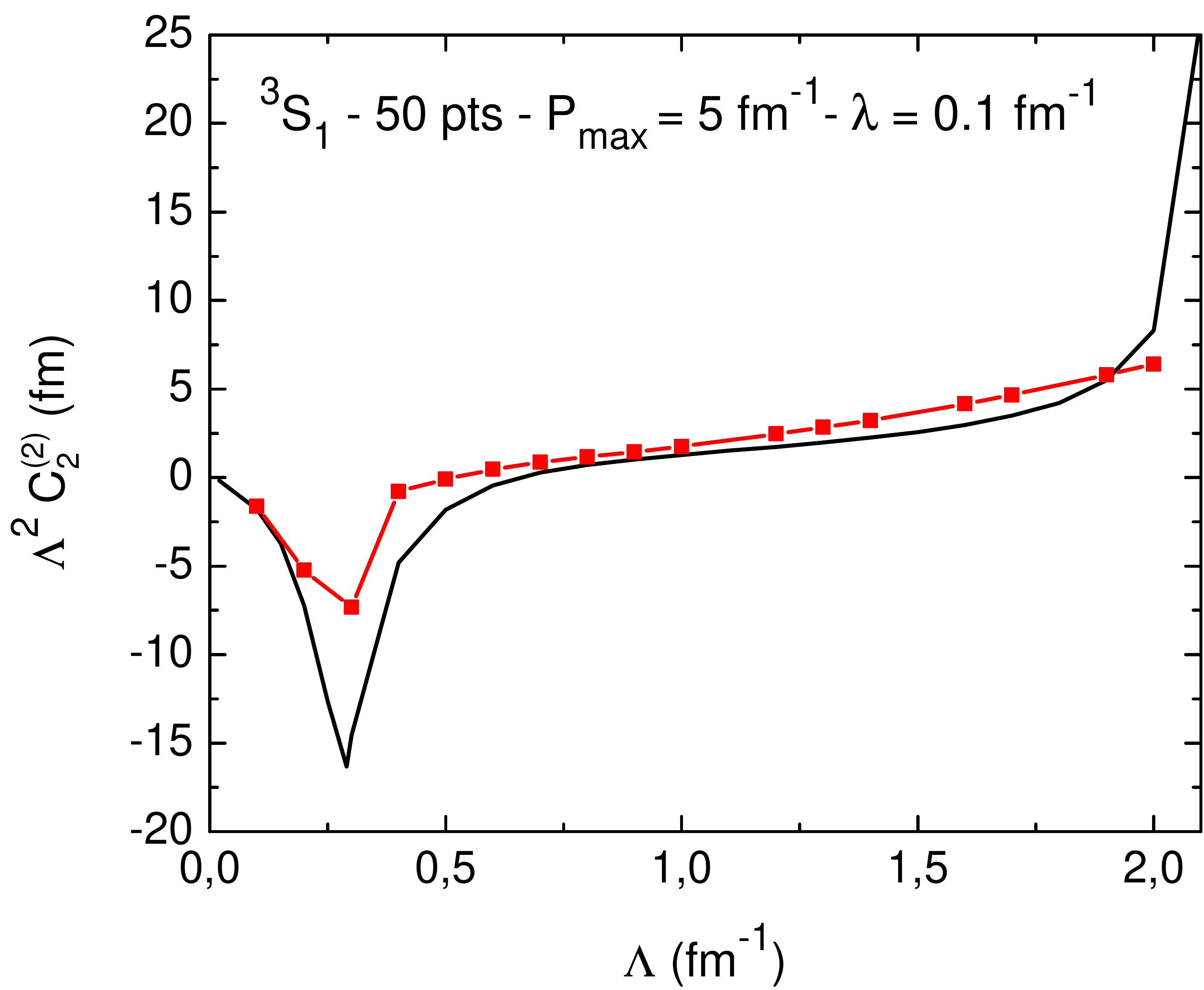}
\end{center}
\caption{${\tilde C_{0}^{(2)}}$ and ${\tilde C_{2}^{(2)}}$ as a function of the block-diagonal cutoff $\Lambda$
extracted from the $^3S_1$ channel toy-model potential on a gaussian grid (with $N=50$ momentum points and
$P_{\rm max}=5~{\rm fm}^{-1}$) evolved through the SRG transformation with the block-diagonal generator for
several values of the SRG cutoff $\lambda$. For comparison, we also show $C_{0}^{(2)}$ and $C_{2}^{(2)}$ for the
$^3S_1$ channel contact theory potential at $NLO$ (on the same grid) regulated by a smooth exponential momentum cutoff with $n=16$.}
\label{fig:22}
\end{figure*}

In Figs.~\ref{fig:17} to \ref{fig:22} we show the results for
${\tilde C_{0}^{(2)}}$ and $\Lambda^2 {\tilde C_{2}^{(2)}}$ extracted from the $^1S_0$
channel and the $^3S_1$ channel SRG-evolved toy-model potentials on a
gaussian grid (with $N=50$ momentum points and $P_{\rm max}=5~{\rm fm}^{-1}$),
compared to $C_{0}^{(2)}$ and $\Lambda^2 C_{2}^{(2)}$ obtained for the contact theory
potentials at $NLO$ (on the same grid) regulated by a smooth
exponential momentum cutoff with sharpness parameter $n=16$. As one
can see, for a given cutoff $\Lambda$ the coefficients
${\tilde C_{0}^{(2)}}$ and $\Lambda^2 {\tilde C_{2}^{(2)}}$ extracted from the
SRG-evolved toy-model potentials approach the coefficients $C_{0}^{(2)}$ and $\Lambda^2 C_{2}^{(2)}$ obtained for the contact theory potentials as the SRG
cutoff $\lambda$ decreases. In general, the running of the
coefficients becomes more significant as $\lambda$ approaches
$\Lambda$ and nearly saturates when $\lambda \leq \Lambda$. In the
case of the $^1S_0$ channel, the running of $\Lambda^2 {\tilde C_{2}^{(2)}}$
deviates from such a pattern for $\Lambda \leq 0.2~{\rm fm}^{-1}$. As
shown in Fig.~\ref{fig:19}, the coefficient $\Lambda^2 C_2^{(2)}$ for the
contact theory potential has a dip in this region. In the case of the
$^3S_1$ channel, we observe large discrepancies in the running of both
${\tilde C_{0}^{(2)}}$ and $\Lambda^2 {\tilde C_{2}^{(2)}}$ for $\Lambda$ in the
range from $\sim 0.3$ to $\sim 0.5~{\rm fm}^{-1}$, even for small
values of $\lambda$. As shown in Fig.~\ref{fig:22}, this is the region
where the running of the coefficients becomes more significant, which
corresponds to the range from the scale where the deuteron bound-state
appears to the momenta around the pole in the $K$-matrix. Nevertheless, one can see that in the limit
$\lambda \rightarrow 0$ there is a remarkably good agreement between the coefficients extracted from the block-diagonal SRG-evolved
toy-model potentials and those obtained for the contact theory potentials.

It is important to point out that the agreement between the running of
the coefficients $C_{0}^{(2)}$ and $C_{2}^{(2)}$ in the contact theory potential
and the running of the coefficients ${\tilde C_{0}^{(2)}}$ and
${\tilde C_{2}^{(2)}}$ extracted from the block-diagonal SRG-evolved toy-model
potential as the SRG cutoff $\lambda$ decreases below $\Lambda$
can be traced to the decoupling between the $P$-space and the
$Q$-space, which follows a similar pattern. Thus, in the limit
$\lambda \rightarrow 0$ we expect to achieve a high degree of
agreement for cutoffs $\Lambda$ up to $\Lambda_{\rm WB}$ determined by
the Wigner bound for the contact theory potential.


We could also analyze the constants $C_4$ and $C_4'$ appearing in
Eq.~(\ref{eq:pot-mom}). As already mentioned, these constants (unlike
$C_0$ and $C_2$ ) feature specific properties of the bare potential,
and their running cannot be determined from two-body scattering data
alone. This ambiguity was manifest in the implicit method when looking
at $\Lambda \to 0$, but the explicit form of the BD-SRG does not solve
it; it is hidden in the initial condition at $\Lambda \to
\infty$. While it is true that for a given bare NN interaction the
explicit method yields a unique answer, the bare interaction itself is
not uniquely fixed by the scattering data, as the off-shell part of
the interaction cannot be fixed this way. The implicit method just
reflects this fact, and it would be misleading to take this feature as
a genuine disadvantage of the method. Actually, it is the opposite,
the explicit method just inherits the arbitrariness of the original
bare two body interaction.  Of course, this can only be solved
unambiguously by solving implicitly the three- and higher body
problem, using the BD-SRG for the CM Hamiltonian, by imposing binding
energies and scattering data as renormalization conditions. While this
is a possible scheme, it is numerically cumbersome and is left for
future research.~\footnote{ One possible strategy is to obtain the
  constants directly from a reference K-matrix at low energies, which
  can be obtained from a phenomenological potential. In principle, it
  is possible to extend the implicit renormalization approach to
  current high precision forces but the the equations cannot be solved
  as easily (see e.g. Ref.~\cite{Arriola:2010hj} for a complete NLO
  calculation).}

\section{Summary and Outlook}

In the present work we have made a thorough investigation on the
renormalization of effective interactions for the nuclear force.  The
main purpose was to display the complementary views between implicit
and explicit renormalization. For simplicity, we have focused on the
two-nucleon problem and the $^1S_0$ and $^3S_1$ states where the
interaction is non-perturbative and has either none or one bound-state
(the deuteron). To further simplify the analysis, we have considered a
separable gaussian potential toy-model as a bare interaction in order
to reduce the computational effort. Already at this level, this much
assumed complementarity is difficult to test at the desired accuracy.

In the explicit renormalization approach implemented via the SRG method with a block-diagonal generator, the original Hilbert
space is separated in two orthogonal sectors according to a given
energy boundary. For the case of two-body interactions in a given
partial-wave channel this energy can be transformed into a CM momentum cutoff
$\Lambda$, which corresponds to the analogous $V_{\rm low \; k}$ cutoff.
This is carried out by a suitable unitary transformation which can
steadily be constructed by using a uniparametric group family via a
differential equation in operator space in the SRG cutoff parameter
$\lambda$. One problem is that the numerical resolution of these
differential equations in operator space requires reducing the Hilbert
space to a finite number of dimensions, $N$, which makes the
non-linear system of order $N^2$. This introduces a minimum momentum
resolution $\Delta p \sim P_{\rm max} /N$. Generally, as $\lambda \to
0$ the equations become stiff and thus convergence issues prevent an
accurate solution. In practice, when $\lambda \ll \Delta p$ the
differential equations yield numerical spurious results. Thus, in
practice the finite $\Delta p$ prevents an accurate
block-diagonalization for $\Lambda \le \Delta p$.

The implicit renormalization approach determines the potential in the low-momentum region in a model independent way by fixing scattering
information, and more specifically the scattering length and the
effective range. Going beyond these two low-energy parameters or
determining the effective interaction to higher energies depends on
the details of the bare interaction. Moreover, the equations can be
solved analytically in the continuum without any finite momentum grid.
This actually allows to work precisely when the explicit
renormalization approach does not. Actually, we find that the complementarity
of both explicit and implicit approaches is verified in a wide cut-off
range $\Lambda$. For the $^1S_0$ and $^3S_1$ neutron-proton scattering
states this range is within $ 0.5 ~{\rm fm}^{-1} \le \Lambda \le 1.5
~{\rm fm}^{-1} $. This is a welcome feature, since it suggests that the
bulk of the effective interaction operating in finite nuclei can be
directly extracted from low-energy $NN$ data, providing a short-cut
to large-scale calculations.

While the complementarity of both explicit and implicit views of the
renormalization procedure is taken for granted at {\it sufficiently}
low energies, it is fair to say that within the present context of
nuclear interactions it has seldomly been verified to the degree
analyzed in the present work and the relevant scales have never been
clearly quantified.  This requires to pin down the numerical
uncertainties with sufficient accuracy in the explicit method. This of
course suggests that the implicit renormalization approach may be a
better method to determine the effective interaction. This was the
purpose of a previous analysis~\cite{Arriola:2010hj} where the Skyrme
force parameters just deducible from the $NN$ interaction were
determined until a low-energy saturation was observed.  The present
discussion provides an {\it a posteriori} explanation of this
observation.

Perturbative discussions provide the customary example to motivate the
idea of an effective field theory (EFT). Actually, such a perturbative
approach holds true when there are large energy gaps in the theory,
since large mass splittings allow for a perturbative treatment due to
well known decoupling theorems. Moreover, the existence of energy gaps
allows for generous variations of the {\it a priori} cutoff scale. A
prominent example is $\pi\pi$ scattering at CM energies much smaller
than the vector meson masses.  In Nuclear Physics, there is no obvious
a priori energy gap. In the meson exchange picture, multipion
exchanges take place before heavier mesons enter the game. This is a
good reason why the discussion on the Wilsonian aspects of
renormalization becomes considerably more complicated.  Given the
relatively low scales where the effective interaction strength is
saturated, our findings suggest also to critically review the
quantitative role of explicit pions and more specifically the
relevance of chiral two-pion exchange (TPE) in the study of the
structure of light nuclei. This issue was partially adressed in our
previous work~\cite{Arriola:2013gya} and will be discussed
comprehensively in an upcoming publication.

While to achieve exact decoupling one has to evolve from $\lambda=
\infty$ to $\lambda =0$ it is not obvious from the start which value
of $\Lambda$ defining the model space is optimal. Naively it should be
defined by $ {\rm min}_\Lambda || Q H(0) Q || $, whence an
insensitivity region around this value should take place. The
existence of such a scale is not obvious, but it has been known since
many years by the work of Moszkowski and
Scott~\cite{1960AnPhy..11...65M} as already recognized by Brown and
Holt \cite{Holt:2004hp}. The smallest value of the high-momentum
$Q$-space block-diagonal Hamiltonian is obtained when $Q H Q=0$. This
corresponds to a situation where above a certain CM momentum scale
nucleons behave to a large extent as free particles.

One of the advantages of the implicit renormalization method is that
it may be applied in situations where the block-diagonal SRG method
cannot due to long tails which intertwine all momentum scales and make
a direct numerical solution out of reach. An interesting case
corresponds to the van der Waals interactions between neutral
atoms~\cite{Arriola:2010tu} where similar trends are found, but the
block-diagonal SRG explicit solutions are extremely difficult to
obtain.

\section*{Acknowledgements}

E.R.A. would like to thank the Spanish DGI (Grant FIS2011-24149) and
Junta de Andalucia (Grant FQM225). S.S. is partially supported by
FAPESP and V.S.T. thanks FAEPEX (Grant 1165/2014), FAPESP (Grant 2014/04975-9), CNPq (Grant 310980/2012-7) for financial
support.

\section*{References}

\bibliographystyle{elsarticle-num}

\bibliography{srg-symmetries}

\begin{thebibliography}{10}
\expandafter\ifx\csname url\endcsname\relax
  \def\url#1{\texttt{#1}}\fi
\expandafter\ifx\csname urlprefix\endcsname\relax\def\urlprefix{URL }\fi
\expandafter\ifx\csname href\endcsname\relax
  \def\href#1#2{#2} \def\path#1{#1}\fi

\bibitem{Wilson:1973jj}
K.~Wilson, J.~B. Kogut, {The Renormalization group and the epsilon expansion},
  Phys.Rept. 12 (1974) 75--200.
\newblock \href {http://dx.doi.org/10.1016/0370-1573(74)90023-4}
  {\path{doi:10.1016/0370-1573(74)90023-4}}.

\bibitem{Perez:2013mwa}
R.~Navarro~Perez, J.~Amaro, E.~Ruiz~Arriola, {Partial Wave Analysis of
  Nucleon-Nucleon Scattering below pion production threshold}, Phys.Rev.
  C88~(6) (2013) 024002.
\newblock \href {http://arxiv.org/abs/1304.0895} {\path{arXiv:1304.0895}},
  \href {http://dx.doi.org/10.1103/PhysRevC.88.024002,
  10.1103/PhysRevC.88.069902} {\path{doi:10.1103/PhysRevC.88.024002,
  10.1103/PhysRevC.88.069902}}.

\bibitem{Perez:2013jpa}
R.~Navarro~Perez, J.~Amaro, E.~Ruiz~Arriola, {Coarse-grained potential
  analysis of neutron-proton and proton-proton scattering below the pion
  production threshold}, Phys.Rev. C88~(6) (2013) 064002.
\newblock \href {http://arxiv.org/abs/1310.2536} {\path{arXiv:1310.2536}},
  \href {http://dx.doi.org/10.1103/PhysRevC.88.064002}
  {\path{doi:10.1103/PhysRevC.88.064002}}.

\bibitem{Perez:2013oba}
R.~Navarro~Perez, J.~Amaro, E.~R. Arriola, {Coarse grained NN potential with Chiral
  Two Pion Exchange}, Phys.Rev. C89 (2014) 024004.
\newblock \href {http://arxiv.org/abs/1310.6972} {\path{arXiv:1310.6972}},
  \href {http://dx.doi.org/10.1103/PhysRevC.89.024004}
  {\path{doi:10.1103/PhysRevC.89.024004}}.

\bibitem{Perez:2014yla}
R.~Navarro~Perez, J.~Amaro, E.~Ruiz~Arriola, {Statistical Error analysis of
  Nucleon-Nucleon phenomenological potentials}, Phys.Rev. C89 (2014) 064006.
\newblock \href {http://arxiv.org/abs/1404.0314} {\path{arXiv:1404.0314}},
  \href {http://dx.doi.org/10.1103/PhysRevC.89.064006}
  {\path{doi:10.1103/PhysRevC.89.064006}}.

\bibitem{Wiringa:1994wb}
R.~B. Wiringa, V.~Stoks, R.~Schiavilla, {An Accurate nucleon-nucleon potential
  with charge independence breaking}, Phys.Rev. C51 (1995) 38--51.
\newblock \href {http://arxiv.org/abs/nucl-th/9408016}
  {\path{arXiv:nucl-th/9408016}}, \href
  {http://dx.doi.org/10.1103/PhysRevC.51.38}
  {\path{doi:10.1103/PhysRevC.51.38}}.

\bibitem{Kukulin:2013oya}
V.~Kukulin, M.~Platonova, {Short-range components of nuclear forces: Experiment
  versus mythology}, Phys.Atom.Nucl. 76~(12) (2013) 1465--1481.
\newblock \href {http://dx.doi.org/10.1134/S1063778813120120}
  {\path{doi:10.1134/S1063778813120120}}.

\bibitem{Pieper:2001mp}
S.~C. Pieper, R.~B. Wiringa, {Quantum Monte Carlo calculations of light
  nuclei}, Ann.Rev.Nucl.Part.Sci. 51 (2001) 53--90.
\newblock \href {http://arxiv.org/abs/nucl-th/0103005}
  {\path{arXiv:nucl-th/0103005}}, \href
  {http://dx.doi.org/10.1146/annurev.nucl.51.101701.132506}
  {\path{doi:10.1146/annurev.nucl.51.101701.132506}}.

\bibitem{Goldstone:1957zz}
J.~Goldstone, {Derivation of the Brueckner Many-Body Theory},
  Proc.Roy.Soc.Lond.A Math.Phys.Eng.Sci. 239 (1957) 267--279.
\newblock \href {http://dx.doi.org/10.1098/rspa.1957.0037}
  {\path{doi:10.1098/rspa.1957.0037}}.

\bibitem{Moshinsky195819}
M.~Moshinsky, Short range forces and nuclear shell theory, Nuclear Physics 8
  (1958) 19 -- 40.

\bibitem{Skyrme:1959zz}
T.~Skyrme, {The effective nuclear potential}, Nucl. Phys. 9 (1959) 615--634.

\bibitem{1960AnPhy..11...65M}
S.~A. {Moszkowski}, B.~L. {Scott}, {Nuclear forces and the properties of
  nuclear matter}, Annals of Physics 11 (1960) 65--115.
\newblock \href {http://dx.doi.org/10.1016/0003-4916(60)90128-7}
  {\path{doi:10.1016/0003-4916(60)90128-7}}.

\bibitem{Vautherin:1971aw}
D.~Vautherin, D.~M. Brink, {Hartree-Fock calculations with Skyrme's
  interaction. 1. Spherical nuclei}, Phys. Rev. C5 (1972) 626--647.
\newblock \href {http://dx.doi.org/10.1103/PhysRevC.5.626}
  {\path{doi:10.1103/PhysRevC.5.626}}.

\bibitem{Bender:2003jk}
M.~Bender, P.-H. Heenen, P.-G. Reinhard, {Self-consistent mean-field models for
  nuclear structure}, Rev. Mod. PHys. 75 (2003) 121--180.
\newblock \href {http://dx.doi.org/10.1103/RevModPhys.75.121}
  {\path{doi:10.1103/RevModPhys.75.121}}.

\bibitem{Dutra:2012mb}
M.~Dutra, O.~Lourenco, J.~Sa~Martins, A.~Delfino, J.~Stone, et~al., {Skyrme
  Interaction and Nuclear Matter Constraints}, Phys.Rev. C85 (2012) 035201.
\newblock \href {http://arxiv.org/abs/1202.3902} {\path{arXiv:1202.3902}},
  \href {http://dx.doi.org/10.1103/PhysRevC.85.035201}
  {\path{doi:10.1103/PhysRevC.85.035201}}.

\bibitem{Arriola:2010hj}
E.~R. Arriola, {Low scale saturation of Effective NN Interactions and their
  Symmetries}\href {http://arxiv.org/abs/1009.4161} {\path{arXiv:1009.4161}}.

\bibitem{Harada:2005tw}
K.~Harada, K.~Inoue, H.~Kubo, {Wilsonian RG and redundant operators in
  nonrelativistic effective field theory}, Phys.Lett. B636 (2006) 305--309.
\newblock \href {http://arxiv.org/abs/nucl-th/0511020}
  {\path{arXiv:nucl-th/0511020}}, \href
  {http://dx.doi.org/10.1016/j.physletb.2006.03.072}
  {\path{doi:10.1016/j.physletb.2006.03.072}}.

\bibitem{NavarroPerez:2013iwa}
R.~Navarro~Perez, J.~Amaro, E.~Ruiz~Arriola, {Effective Interactions in the
  Delta-Shells Potential}, Few Body Syst. 54 (2013) 1487--1490.
\newblock \href {http://dx.doi.org/10.1007/s00601-012-0537-5}
  {\path{doi:10.1007/s00601-012-0537-5}}.

\bibitem{Perez:2014kpa}
R.~N. Perez, J.~Amaro, E.~R. Arriola, {Error analysis of nuclear forces and
  effective interactions}, J. Phys. G (in press), \href
  {http://arxiv.org/abs/1406.0625} {\path{arXiv:1406.0625}}.

\bibitem{Bogner:2001gq}
S.~K. Bogner, T.~T.~S. Kuo, A.~Schwenk, D.~R. Entem, R.~Machleidt, {Towards a
  unique low momentum nucleon nucleon interaction}, Phys. Lett. B576 (2003)
  265--272.
\newblock \href {http://arxiv.org/abs/nucl-th/0108041}
  {\path{arXiv:nucl-th/0108041}}, \href
  {http://dx.doi.org/10.1016/j.physletb.2003.10.012}
  {\path{doi:10.1016/j.physletb.2003.10.012}}.

\bibitem{Bogner:2003wn}
S.~K. Bogner, T.~T.~S. Kuo, A.~Schwenk, {Model-independent low momentum nucleon
  interaction from phase shift equivalence}, Phys. Rept. 386 (2003) 1--27.
\newblock \href {http://arxiv.org/abs/nucl-th/0305035}
  {\path{arXiv:nucl-th/0305035}}, \href
  {http://dx.doi.org/10.1016/j.physrep.2003.07.001}
  {\path{doi:10.1016/j.physrep.2003.07.001}}.

\bibitem{Bogner:2006pc}
S.~K. Bogner, R.~J. Furnstahl, R.~J. Perry, {Similarity Renormalization Group
  for Nucleon-Nucleon Interactions}, Phys. Rev. C75 (2007) 061001.
\newblock \href {http://arxiv.org/abs/nucl-th/0611045}
  {\path{arXiv:nucl-th/0611045}}, \href
  {http://dx.doi.org/10.1103/PhysRevC.75.061001}
  {\path{doi:10.1103/PhysRevC.75.061001}}.

\bibitem{Bogner:2006vp}
S.~K. Bogner, R.~J. Furnstahl, S.~Ramanan, A.~Schwenk, {Low-momentum
  interactions with smooth cutoffs}, Nucl. Phys. A784 (2007) 79--103.
\newblock \href {http://arxiv.org/abs/nucl-th/0609003}
  {\path{arXiv:nucl-th/0609003}}, \href
  {http://dx.doi.org/10.1016/j.nuclphysa.2006.11.123}
  {\path{doi:10.1016/j.nuclphysa.2006.11.123}}.

\bibitem{Coraggio:2008in}
L.~Coraggio, A.~Covello, A.~Gargano, N.~Itaco, T.~T.~S. Kuo, {Shell-model
  calculations and realistic effective interactions}, Prog. Part. Nucl. Phys.
  62 (2009) 135--182.
\newblock \href {http://arxiv.org/abs/0809.2144} {\path{arXiv:0809.2144}},
  \href {http://dx.doi.org/10.1016/j.ppnp.2008.06.001}
  {\path{doi:10.1016/j.ppnp.2008.06.001}}.

\bibitem{Bogner:2009bt}
S.~K. Bogner, R.~J. Furnstahl, A.~Schwenk, {From low-momentum interactions to
  nuclear structure}, Prog. Part. Nucl. Phys. 65 (2010) 94--147.
\newblock \href {http://arxiv.org/abs/0912.3688} {\path{arXiv:0912.3688}},
  \href {http://dx.doi.org/10.1016/j.ppnp.2010.03.001}
  {\path{doi:10.1016/j.ppnp.2010.03.001}}.

\bibitem{Furnstahl:2012fn}
R.~Furnstahl, {The Renormalization Group in Nuclear Physics},
  Nucl.Phys.Proc.Suppl. 228 (2012) 139--175.
\newblock \href {http://arxiv.org/abs/1203.1779} {\path{arXiv:1203.1779}},
  \href {http://dx.doi.org/10.1016/j.nuclphysbps.2012.06.005}
  {\path{doi:10.1016/j.nuclphysbps.2012.06.005}}.

\bibitem{Furnstahl:2013oba}
R.~Furnstahl, K.~Hebeler, {New applications of renormalization group methods in
  nuclear physics}, Rept.Prog.Phys. 76 (2013) 126301.
\newblock \href {http://arxiv.org/abs/1305.3800} {\path{arXiv:1305.3800}},
  \href {http://dx.doi.org/10.1088/0034-4885/76/12/126301}
  {\path{doi:10.1088/0034-4885/76/12/126301}}.

\bibitem{Holt:2003rj}
J.~D. Holt, T.~T.~S. Kuo, G.~E. Brown, S.~K. Bogner, {Counter Terms for Low
  Momentum Nucleon-Nucleon Interactions}, Nucl. Phys. A733 (2004) 153--165.
\newblock \href {http://arxiv.org/abs/nucl-th/0308036}
  {\path{arXiv:nucl-th/0308036}}, \href
  {http://dx.doi.org/10.1016/j.nuclphysa.2003.12.004}
  {\path{doi:10.1016/j.nuclphysa.2003.12.004}}.

\bibitem{Jurgenson:2010wy}
E.~Jurgenson, P.~Navratil, R.~Furnstahl, {Evolving Nuclear Many-Body Forces
  with the Similarity Renormalization Group}, Phys.Rev. C83 (2011) 034301.
\newblock \href {http://arxiv.org/abs/1011.4085} {\path{arXiv:1011.4085}},
  \href {http://dx.doi.org/10.1103/PhysRevC.83.034301}
  {\path{doi:10.1103/PhysRevC.83.034301}}.

\bibitem{Roth:2013fqa}
R.~Roth, A.~Calci, J.~Langhammer, S.~Binder, {Evolved Chiral NN+3N Hamiltonians
  for Ab Initio Nuclear Structure Calculations}\href
  {http://arxiv.org/abs/1311.3563} {\path{arXiv:1311.3563}}.

\bibitem{Partovi:1969wd}
M.~H. Partovi, E.~L. Lomon, {Field theoretical nucleon-nucleon potential},
  Phys.Rev. D2 (1970) 1999--2032.
\newblock \href {http://dx.doi.org/10.1103/PhysRevD.2.1999}
  {\path{doi:10.1103/PhysRevD.2.1999}}.

\bibitem{Machleidt:1989tm}
R.~Machleidt, {The Meson theory of nuclear forces and nuclear structure},
  Adv.Nucl.Phys. 19 (1989) 189--376.

\bibitem{Cordon:2009pj}
A.~Calle~Cordon, E.~Ruiz~Arriola, {Renormalization vs Strong Form Factors for
  One Boson Exchange Potentials}, Phys. Rev. C81 (2010) 044002.
\newblock \href {http://arxiv.org/abs/0905.4933} {\path{arXiv:0905.4933}},
  \href {http://dx.doi.org/10.1103/PhysRevC.81.044002}
  {\path{doi:10.1103/PhysRevC.81.044002}}.

\bibitem{Furnstahl:2007fr}
R.~J. Furnstahl, {Similarity Renormalization Group for Few-Body Systems}, Few
  Body Syst. 44 (2008) 133--136.
\newblock \href {http://arxiv.org/abs/0711.3846} {\path{arXiv:0711.3846}},
  \href {http://dx.doi.org/10.1007/s00601-008-0274-y}
  {\path{doi:10.1007/s00601-008-0274-y}}.

\bibitem{Anderson:2008mu}
E.~Anderson, et~al., {Block Diagonalization using SRG Flow Equations}, Phys.
  Rev. C77 (2008) 037001.
\newblock \href {http://arxiv.org/abs/0801.1098} {\path{arXiv:0801.1098}},
  \href {http://dx.doi.org/10.1103/PhysRevC.77.037001}
  {\path{doi:10.1103/PhysRevC.77.037001}}.

\bibitem{Hebeler:2012pr}
K.~Hebeler, {Momentum space evolution of chiral three-nucleon forces},
  Phys.Rev. C85 (2012) 021002.
\newblock \href {http://arxiv.org/abs/1201.0169} {\path{arXiv:1201.0169}},
  \href {http://dx.doi.org/10.1103/PhysRevC.85.021002}
  {\path{doi:10.1103/PhysRevC.85.021002}}.

\bibitem{Jurgenson:2009qs}
E.~D. Jurgenson, P.~Navratil, R.~J. Furnstahl, {Evolution of Nuclear Many-Body
  Forces with the Similarity Renormalization Group}, Phys. Rev. Lett. 103
  (2009) 082501.
\newblock \href {http://arxiv.org/abs/0905.1873} {\path{arXiv:0905.1873}},
  \href {http://dx.doi.org/10.1103/PhysRevLett.103.082501}
  {\path{doi:10.1103/PhysRevLett.103.082501}}.

\bibitem{Tsukiyama:2010rj}
K.~Tsukiyama, S.~Bogner, A.~Schwenk, {In-Medium Similarity Renormalization
  Group for Nuclei}, Phys.Rev.Lett. 106 (2011) 222502.
\newblock \href {http://arxiv.org/abs/1006.3639} {\path{arXiv:1006.3639}},
  \href {http://dx.doi.org/10.1103/PhysRevLett.106.222502}
  {\path{doi:10.1103/PhysRevLett.106.222502}}.

\bibitem{Launey:2012zz}
K.~D. Launey, T.~Dytrych, J.~P. Draayer, {Similarity renormalization group and
  many-body effects in multiparticle systems}, Phys.Rev. C85 (2012) 044003.
\newblock \href {http://dx.doi.org/10.1103/PhysRevC.85.044003}
  {\path{doi:10.1103/PhysRevC.85.044003}}.

\bibitem{Hergert:2012nb}
H.~Hergert, S.~Bogner, S.~Binder, A.~Calci, J.~Langhammer, et~al., {In-Medium
  Similarity Renormalization Group with Chiral Two- Plus Three-Nucleon
  Interactions}, Phys.Rev. C87~(3) (2013) 034307.
\newblock \href {http://arxiv.org/abs/1212.1190} {\path{arXiv:1212.1190}},
  \href {http://dx.doi.org/10.1103/PhysRevC.87.034307}
  {\path{doi:10.1103/PhysRevC.87.034307}}.

\bibitem{Tsukiyama:2012sm}
K.~Tsukiyama, S.~Bogner, A.~Schwenk, {In-Medium Similarity Renormalization
  Group for Open-Shell Nuclei}, Phys.Rev. C85 (2012) 061304.
\newblock \href {http://arxiv.org/abs/1203.2515} {\path{arXiv:1203.2515}},
  \href {http://dx.doi.org/10.1103/PhysRevC.85.061304}
  {\path{doi:10.1103/PhysRevC.85.061304}}.

\bibitem{Timoteo:2011tt}
V.~Timoteo, S.~Szpigel, E.~Ruiz~Arriola, {Symmetries of the Similarity
  Renormalization Group for Nuclear Forces}, Phys.Rev. C86 (2012) 034002.
\newblock \href {http://arxiv.org/abs/1108.1162} {\path{arXiv:1108.1162}},
  \href {http://dx.doi.org/10.1103/PhysRevC.86.034002}
  {\path{doi:10.1103/PhysRevC.86.034002}}.

\bibitem{Launey:2012mda}
K.~Launey, T.~Dytrych, J.~Draayer, {Importance of symmetries in the similarity
  renormalization group approach}, Bulg.J.Phys. 39 (2012) 271--281.

\bibitem{Arriola:2013nja}
E.~Ruiz~Arriola, V.~Timoteo, S.~Szpigel, {Nuclear Symmetries of the similarity
  renormalization group for nuclear forces}, PoS CD12 (2013) 106.
\newblock \href {http://arxiv.org/abs/1302.3978} {\path{arXiv:1302.3978}}.

\bibitem{Bogner:2007qb}
S.~K. Bogner, R.~J. Furnstahl, R.~J. Perry, {Three-Body Forces Produced by a
  Similarity Renormalization Group Transformation in a Simple Model}, Annals
  Phys. 323 (2008) 1478--1501.
\newblock \href {http://arxiv.org/abs/0708.1602} {\path{arXiv:0708.1602}},
  \href {http://dx.doi.org/10.1016/j.aop.2007.09.001}
  {\path{doi:10.1016/j.aop.2007.09.001}}.

\bibitem{Jurgenson:2008jp}
E.~D. Jurgenson, R.~J. Furnstahl, {Similarity Renormalization Group Evolution
  of Many-Body Forces in a One-Dimensional Model}, Nucl. Phys. A818 (2009)
  152--173.
\newblock \href {http://arxiv.org/abs/0809.4199} {\path{arXiv:0809.4199}},
  \href {http://dx.doi.org/10.1016/j.nuclphysa.2008.12.007}
  {\path{doi:10.1016/j.nuclphysa.2008.12.007}}.

\bibitem{Szpigel:1999gf}
S.~Szpigel, R.~J. Perry, {Simple Applications of Effective Field Theory and
  Similarity Renormalization Group Methods}\href
  {http://arxiv.org/abs/nucl-th/9906031} {\path{arXiv:nucl-th/9906031}}.

\bibitem{Arriola:2013gya}
E.~Ruiz~Arriola, S.~Szpigel, V.~Timoteo, {Fixed points of the Similarity
  Renormalization Group and the Nuclear Many-Body Problem}, Few Body Syst. 55
  (2014) 971--975.
\newblock \href {http://arxiv.org/abs/1310.8246} {\path{arXiv:1310.8246}},
  \href {http://dx.doi.org/10.1007/s00601-014-0858-7}
  {\path{doi:10.1007/s00601-014-0858-7}}.

\bibitem{Arriola:2013yca}
E.~Ruiz~Arriola, S.~Szpigel, V.~S. Timoteo, {Implicit Versus Explicit
  Renormalization of the $NN$ Force: An S-Wave Toy Model}, Few Body Syst. 55
  (2014) 989--992.
\newblock \href {http://arxiv.org/abs/1310.8526} {\path{arXiv:1310.8526}},
  \href {http://dx.doi.org/10.1007/s00601-014-0811-9}
  {\path{doi:10.1007/s00601-014-0811-9}}.

\bibitem{Entem:2007jg}
D.~R. Entem, E.~Ruiz~Arriola, M.~Pavon~Valderrama, R.~Machleidt,
  {Renormalization of chiral two-pion exchange NN interactions. Momentum vs.
  coordinate space}, Phys. Rev. C77 (2008) 044006.
\newblock \href {http://arxiv.org/abs/0709.2770} {\path{arXiv:0709.2770}},
  \href {http://dx.doi.org/10.1103/PhysRevC.77.044006}
  {\path{doi:10.1103/PhysRevC.77.044006}}.

\bibitem{Szpigel:2010bj}
S.~Szpigel, V.~S. Timoteo, F.~d.~O. Duraes, {Similarity Renormalization Group
  Evolution of Chiral Effective Nucleon-Nucleon Potentials in the Subtracted
  Kernel Method Approach}, Annals Phys. 326 (2011) 364--405.
\newblock \href {http://arxiv.org/abs/1003.4663} {\path{arXiv:1003.4663}},
  \href {http://dx.doi.org/10.1016/j.aop.2010.11.007}
  {\path{doi:10.1016/j.aop.2010.11.007}}.

\bibitem{Steele:1998un}
J.~V. Steele, R.~Furnstahl, {Regularization methods for nucleon-nucleon
  effective field theory}, Nucl.Phys. A637 (1998) 46--62.
\newblock \href {http://arxiv.org/abs/nucl-th/9802069}
  {\path{arXiv:nucl-th/9802069}}, \href
  {http://dx.doi.org/10.1016/S0375-9474(98)00219-X}
  {\path{doi:10.1016/S0375-9474(98)00219-X}}.

\bibitem{Stoks:1994wp}
V.~G.~J. Stoks, R.~A.~M. Klomp, C.~P.~F. Terheggen, J.~J. de~Swart,
  {Construction of high quality N N potential models}, Phys. Rev. C49 (1994)
  2950--2962.
\newblock \href {http://arxiv.org/abs/nucl-th/9406039}
  {\path{arXiv:nucl-th/9406039}}, \href
  {http://dx.doi.org/10.1103/PhysRevC.49.2950}
  {\path{doi:10.1103/PhysRevC.49.2950}}.

\bibitem{Stoks:1993tb}
V.~G.~J. Stoks, R.~A.~M. Kompl, M.~C.~M. Rentmeester, J.~J. de~Swart, {Partial
  wave analysis of all nucleon-nucleon scattering data below 350-MeV}, Phys.
  Rev. C48 (1993) 792--815.
\newblock \href {http://dx.doi.org/10.1103/PhysRevC.48.792}
  {\path{doi:10.1103/PhysRevC.48.792}}.

\bibitem{Glazek:1993rc}
S.~D. Glazek, K.~G. Wilson, {Renormalization of Hamiltonians}, Phys. Rev. D48
  (1993) 5863--5872.
\newblock \href {http://dx.doi.org/10.1103/PhysRevD.48.5863}
  {\path{doi:10.1103/PhysRevD.48.5863}}.

\bibitem{Glazek:1994qc}
S.~D. Glazek, K.~G. Wilson, {Perturbative renormalization group for
  Hamiltonians}, Phys. Rev. D49 (1994) 4214--4218.
\newblock \href {http://dx.doi.org/10.1103/PhysRevD.49.4214}
  {\path{doi:10.1103/PhysRevD.49.4214}}.

\bibitem{Wegner200177}
F.~J. Wegner, Flow equations for hamiltonians, Physics Reports 348~(1-2) (2001)
  77 -- 89.

\bibitem{Kehrein:2006ti}
S.~Kehrein, {The flow equation approach to many-particle systems}, Springer,
  2006.

\bibitem{Arriola:2014aia}
E.~Ruiz~Arriola, S.~Szpigel, V.~Timoteo, {The infrared limit of the Similarity
  Renormalization Group evolution and Levinson's theorem}, Phys.Lett. B735
  (2014) 149--156.
\newblock \href {http://arxiv.org/abs/1404.4940} {\path{arXiv:1404.4940}},
  \href {http://dx.doi.org/10.1016/j.physletb.2014.06.032}
  {\path{doi:10.1016/j.physletb.2014.06.032}}.

\bibitem{Holt:2004hp}
J.~W. Holt, G.~E. Brown, {Separation of Scales in the More Effective Field
  Theory and Moszkowski-Scott Methods}\href
  {http://arxiv.org/abs/nucl-th/0408047} {\path{arXiv:nucl-th/0408047}}.

\bibitem{Arriola:2010tu}
E.~Ruiz~Arriola, {Van der Waals forces and Photon-less Effective Field
  Theories}, Few Body Syst. 50 (2011) 399--402.
\newblock \href {http://arxiv.org/abs/1012.2284} {\path{arXiv:1012.2284}},
  \href {http://dx.doi.org/10.1007/s00601-010-0203-8}
  {\path{doi:10.1007/s00601-010-0203-8}}.

\end{thebibliography}

\end{document}